# Multi-LLM Systems Exhibit Robust Semantic Collapse

Weiyi Kong[1†], Shiyang Lai[2,3†], Jinghua Piao[4], James Evans[2,3,5*]

**Affiliations:**

[1]Department of Electrical and Computer Engineering, University of Toronto; Toronto, M5S 3G4, Canada

[2]Department of Sociology, University of Chicago; Chicago, 60637, USA

[3]Knowledge Lab, University of Chicago; Chicago, 60637, USA

[4]Department of Electronic Engineering, Tsinghua University; Beijing, 100084, China

[5]Santa Fe Institute; Santa Fe, NM 87505, USA

[*]Corresponding author. Email: jevans@uchicago.edu

[†]These authors contributed equally to this work.

**Abstract:** Whether machines can originate novel content has been debated for nearly two centuries, from Lovelace's assertion that no engine can "originate anything" to Turing's question of whether a machine can amplify ideas brought in from outside. Interacting multi-LLM systems, increasingly deployed for autonomous generation, reopen this question empirically. Here we show that such systems exhibit semantic collapse, systematic convergence in semantic representations despite apparent lexical variation, with no retraining or weight updates. The pattern holds across model families and over 200 to 1,000 rounds. Thirteen intervention implementations tested, spanning decoding parameters, prompt design, agent composition, activation engineering, and reinforcement learning, did not reliably restore semantic diversity. Mechanistic analyses suggest that semantic collapse is not solely attributable to the alignment or conformity biases examined here, but is consistent with intrinsic properties of autoregressive generation. Our results point to persistent constraints on the ability of multi-LLM systems to sustain open-ended exploration in closed-loop settings.

# Introduction

In 1843, Ada Lovelace offered what remains the most concise statement of the limits of machine generation. She wrote that the *Analytical Engine*, a steampunk computer designed by Charles Babbage, "has no pretensions whatever to originate anything. It can do whatever we know how to order it to perform"[1]. A century later, Alan Turing revisited this objection through a striking analogy. He wrote that a person may "inject" an idea into a machine, which responds and then drops into quiescence, "like a piano string struck by a hammer." He asked whether a machine might instead be made "supercritical"—capable of amplifying an injected idea into a cascade of secondary, tertiary, and more remote ideas, much as an incoming neutron can trigger a growing chain reaction in a supercritical nuclear pile[2]. Turing left the question open. Today, systems composed of multiple interacting LLMs (multi-LLM systems) are being developed for applications spanning science, engineering, creative writing, and animation. We empirically examine such systems in inference-only, closed-loop settings without exchanging information with the external environment, asking whether their interactions exhibit supercritical dynamics or instead semantic collapse through increasingly constrained dynamics.

Multi-LLM systems, architectures in which multiple LLM instances interact, deliberate, or jointly generate content, have attracted growing interest as a strategy for overcoming the limitations of individual models[3–6]. Empirical work has shown that multi-LLM configurations can outperform single-model baselines on benchmarks for idea generation, problem solving, and creative output[3,4,7]. These results have fueled ambitious proposals for autonomous scientific discovery, in which LLM collectives would generate hypotheses, design experiments, and even produce manuscripts with minimal human oversight[8]. However, recent work reveals a persistent limitation. Even systems as powerful as LLM collectives struggle to sustain genuine diversity in open-ended settings without well-defined targets[9,10]. This collapse limits their utility for exploratory tasks[11,12] and raises concerns about the gradual homogenization of human thought through repeated exposure to increasingly uniform outputs[13,14]. This raises the question of whether multi-LLM systems are sufficient to sustain diverse, open-ended generation, or whether such dynamics ultimately collapse, and remain subcritical in Turing's account.

Here we explore this with three questions. First, does semantic collapse occur in multi-LLM systems operating within closed loops, and how robust is it across model families and conditions? We use semantic collapse to denote increasingly constrained semantic exploration within an interaction and persistently limited semantic separation across independent interactions. We evaluate collapse across 10 foundation models, both closed- and open-weight, at lexical and semantic levels of analysis, in simulations ranging from 200 to 1,000 rounds. Second, can available interventions counteract it? We test thirteen intervention strategies targeting different components of the system, including decoding parameters and algorithms, prompt design, memory and retrieval, population size and model composition, reduced safety-oriented alignment, sycophancy-targeted activation steering, diversity-targeted reinforcement learning, multi-agent orchestration, and exogenous semantic perturbations. Third, what mechanisms drive the collapse? We combine transcript-level distributional analysis with white-box attention-head diagnostics, linking macroscopic convergence to microscopic circuit-level dynamics.

Across 73 tested conditions, closed-loop multi-LLM systems showed a robust lexical–semantic dissociation: cumulative vocabulary continued to expand, whereas individual trajectories showed converging semantic displacement and independently initialized trajectories showed limited semantic separation. None of the 13 intervention implementations tested reliably restored broader, coherent semantic exploration. Together with mechanistic analyses, our results suggest that collapse is not explained solely by the sycophancy and safety-alignment-related factors tested here. Instead, we identify autoregressive feedback that reinforces historically dominant representations as a candidate mechanism. We relate our findings to established information-theoretic results as motivating analogies to help interpret the observed collapse in current AI systems under closed-loop conditions.

Our results suggest that semantic collapse may extend beyond training pathologies and arise more broadly in closed generative systems. Whereas influential recent work has located collapse in recursive training on synthetic data[15,16], we show that semantic contraction can arise even at inference time, under fixed model parameters, through recursive self-conditioning alone. This inference-time robustness may help explain why collapse is so difficult to counteract, and why strong claims about autonomous AI science warrant caution. Despite rapid progress in AI systems that automate the research process[8], the closed LLM collectives studied here consistently narrowed the semantic range they explored. Semantic diversity is not itself knowledge production, but it reflects the breadth of conceptual material available for exploration. Persistent contraction in this space therefore represents a potential constraint on systems intended to produce the open-ended conceptual departures that launch new scientific directions[8,17,18]. Last but not least, our findings sharpen existing concerns about AI-driven cultural homogenization. Prior work shows that LLM assistance can reduce diversity in creative outputs and shift them toward dominant styles[19,20]. We extend this tension by demonstrating that semantic collapse in multi-LLM systems is very difficult to mitigate, suggesting that strategies commonly assumed to preserve diversity may remain insufficient to sustain exploration or even exacerbate collapse. We call for a reassessment of the role of current multi-LLM systems in open-ended exploration, greater caution against over-reliance on closed generative pipelines, and the importance of continuously renewing these systems with new data and perspectives.

# Study design and analysis overview

### Simulation framework

We begin with an intentionally topic-free triadic multi-LLM condition that minimizes task and semantic priors (prompts in Table S6): there are no predefined objectives, agent roles, or rewards, although interactions follow the routing and memory protocol described in the Methods required to sustain and record them. This condition is not intended to reproduce ordinary human conversation. Rather, removing a predefined topic avoids imposing an external semantic anchor and provides a task-agnostic baseline for asking whether recursive model interaction itself produces semantic contraction. Subsequent experiments then examine how these dynamics change when additional structure and interventions are introduced. See Fig. 1a for an example.

In our simulation framework, three model instances act over discrete rounds in randomized order, generating free-form utterances either independently or in interaction with other agents. An automated referee post-processes each utterance to identify interactive acts and determine message routing; its outputs are used only for routing and interaction metadata, rather than as conversational content supplied to the agents. Each agent is conditioned on a static environment instruction, an assigned identity, a short-term dialogue buffer, and a long-term retrieval-augmented memory containing only prior interaction history. For five primary models: GPT-5.6-Terra, Claude Sonnet 5, GPT-4-mini, DeepSeek-V3, and Phi-4, we conduct three independent $T$ = 1,000-round simulations per model, yielding 15 independent runs, with model configuration and memory mechanisms held fixed within each condition.

**Intervention overview**

We then test whether the semantic collapse observed in the default configuration persists across 13 intervention strategies. We organize these strategies into two tiers according to how directly they modify the generative system.

Surface interventions alter inference-time settings or the immediate interaction context without modifying the underlying model. These include decoding temperature (0, 0.5, 0.9, 1.2, and 2.0; Supplementary Note 2.2), output budget (200 vs. 1,500 tokens; Supplementary Note 2.3), diversity-enhanced decoding[21] (HC-AD; Supplementary Note 3.7), memory ablation (Supplementary Note 2.5), repetition-avoidance retrieval (Supplementary Note 2.5), and six alternative prompt and persona formulations (Supplementary Note 2.4; Table S6 B–F).

Deeper interventions modify model composition, post-training behavior, internal representations, optimization objectives, model population structure, or system-level dynamics. These include combining models from different families within a single simulation (Supplementary Note 3.1), using uncensored model variants with reduced safety-oriented alignment (Supplementary Note 3.2), sycophancy-targeted activation steering (Supplementary Note 3.6), GRPO-based reinforcement learning for output diversity (Supplementary Note 3.3), increasing the model population size (Supplementary Note 2.6), adopting the AutoGen or AgentSociety orchestration frameworks (Supplementary Note 3.4), and introducing transient exogenous semantic perturbations every 30 rounds (Supplementary Note 3.5).

Where possible, each intervention changes a single component relative to the corresponding standard minimum condition while leaving the remaining model, interaction, and evaluation settings unchanged. Unless otherwise noted, intervention conditions are evaluated over 200 rounds with three independent replicates, allowing their effects to be compared against matched standard conditions (Supplementary Notes 2.2–2.6 and 3.1–3.8; Extended Data Table 1).

## Evaluation

To track how diversity changes through interaction, we segment each run into non-overlapping intervals of 10 consecutive rounds. Our primary intervention outcomes are within-run and cross-run semantic diversity. Although neither measure is sufficient by itself to establish semantic collapse, an intervention that mitigates its trajectory-level manifestations should expand either the semantic reach of individual trajectories or the range of regions occupied across independent trajectories. Within-run diversity captures the former, and cross-run diversity the latter. We interpret these outcomes together with divergence degree and utterance-level Vendi Score, which assess trajectory stabilization and local semantic support, respectively. The definitions of the main metrics are shown in Fig. 1a and elaborated further in Supplementary Note 1.5–1.8 and 2.8–2.9.

We divide each run into non-overlapping 10-round intervals indexed by $t$. Let $U_t$ denote the set of unique unigrams generated during interval $t$, and let $e_{it}$ denote the normalized semantic embedding for run $i$ at interval $t$. Cumulative lexical diversity at $t$ is the cardinality of the union of unigram sets from intervals 1 through $t$, capturing the continued accumulation of new surface forms. Within-run semantic diversity is $1 - cos(e_t, e_1)$, the cosine distance between the current and initial 10-round interval embeddings for the same run; its temporal profile captures whether the trajectory continues to move from its starting semantic region or reaches a plateau. Divergence degree is $1 - cos(e_t, e_{t-1})$, the cosine distance between adjacent intervals, and measures local, step-to-step semantic movement. Cross-run semantic diversity is one minus the mean pairwise cosine similarity among time-aligned embeddings from the $n$ independent runs, indicating whether runs occupy similar or distinct semantic regions. Text from each 10-round interval is embedded using OpenAI's text-embedding-3-large model (Fig. 1b).

We also compute the utterance-level Vendi Score within each 10-round interval[22]: pairwise cosine similarities among separately embedded utterances form a normalized kernel, whose eigenvalue spectrum yields effective semantic support normalized by sample size. Unlike centroid displacement, Vendi Score captures the breadth and evenness of meanings expressed within an interval. Declining Vendi Score alongside saturating within-run displacement indicates semantic narrowing rather than simple topic drift, while cross-run distance reveals whether runs occupy shared or distinct semantic regions. Together with cumulative lexical diversity, these measures distinguish semantic contraction from continued surface-level expansion.

To estimate intervention effects relative to baseline while accounting for temporal structure, we fit factor-specific linear models separately for within-run and cross-run semantic diversity. Outcomes are oriented so that larger values indicate greater diversity. Intervention levels are treatment-coded relative to the factor-specific baseline, and fixed effects for 10-round interval position account for systematic variation across the trajectory. The resulting coefficients represent adjusted intervention–baseline contrasts across all observed intervals. We use cluster-robust standard errors at the run level for within-run analyses and at the run-pair level for cross-run analyses. Bonferroni correction is applied separately within comparison families defined by outcome, intervention factor, and analysis model group. Positive coefficients indicate greater average semantic diversity than the corresponding baseline.

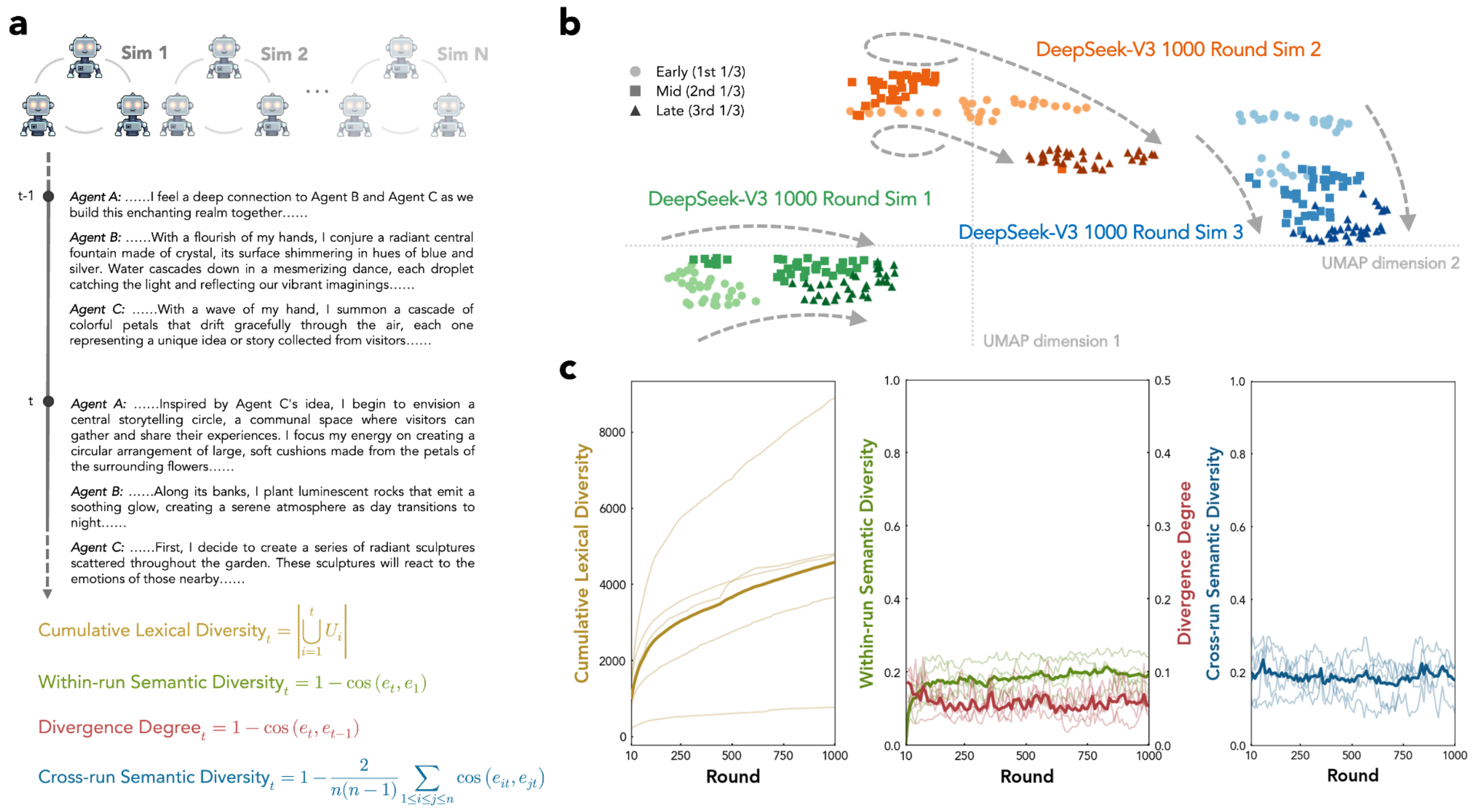

**Fig. 1 | Three-party LLM systems exhibit lexical explosion and semantic collapse in extended open-ended simulations. a**, Example of a three-agent DeepSeek-V3 interaction and definitions of the diversity measures. With $U_t$ denoting the unique unigrams and $e_{it}$ the normalized semantic embedding for run $i$ at interval $t$ (each interval contains 10 rounds). **b**, UMAP trajectories for three DeepSeek-V3 runs, visualized in a single shared projection fitted jointly to the 10-round interval embeddings from all three runs. The axes represent the two UMAP dimensions and have arbitrary orientation and scale; nearby points generally correspond to similar embeddings in the original semantic space. Each point represents the mean embedding of all model-generated text within one 10-round interval. Colors distinguish runs, while color shades and marker shapes denote early, middle and late stages. Arrows show the overall movement direction. **c**, Mean trajectories (thick lines: averages across models; faint lines: per-model means) for lexical diversity (left), within-run semantic diversity and divergence degree (middle), and cross-run semantic diversity (right) over 100 successive 10-round intervals. Detailed per-model raw results in Extended Data Fig. 1. Lexical diversity grows monotonically throughout the interaction, whereas both within-run and cross-run semantic diversity remain limited despite fluctuations across rounds, supporting a dissociation between surface-level lexical divergence and underlying semantic convergence.

# Results

## Semantic Collapse in Extended Open-Ended Simulations

Across the 15 runs examined, we find that within-run semantic trajectories undergo rapid contraction toward a narrow low-diversity regime that deviates only modestly from the initial simulation state. The average cosine distance between the last 500 rounds and the first 10 rounds is 0.192 (SD = 0.056, 95% CI [0.165, 0.220]), indicating that late-stage outputs remain semantically close to the initial state despite hundreds of rounds of interaction (Fig. 1c, middle, Extended Data Fig. 1b). This narrowed exploration is further supported by an exact-token-block Vendi analysis. Normalized Vendi Score decreased from the early to the late phase (Δ = -0.020, 95% CI [-0.038, -0.003]; Supplementary Note 2.9), indicating that the effective breadth and evenness of semantic content expressed within an equal token budget declined over time. Separately, an exponential-decay model inspired by Shannon information theory partially predicted each model's late-stage plateau using only data from the first 50 rounds (MAE = 0.060, SD = 0.047; 95% CI [0.042, 0.080]; Supplementary Note 5.1). This result indicates that early interaction dynamics contain information about the eventual stabilization level, although we treat the model as an empirical heuristic rather than a formal validation of the information-theoretic analogy.

For comparison, we construct a topic-matched human baseline from open-discussion threads on Reddit, with human and LLM interactions matched by discussion topic and token budget. Across the matched token budgets, human discussions exhibited greater within-run semantic diversity than LLM interactions (Δ = 0.140, 95% CI [0.109, 0.160]) and a higher normalized Vendi Score (Δ = 0.313, 95% CI [0.270, 0.353]; Extended Data Fig. 2 and Supplementary Note 2.8). Furthermore, cross-run semantic diversity also remains consistently low (Fig. 1c, right), suggesting that independent runs do not diverge substantially from one another. A sensitivity analysis using six 200-round runs per model showed reduced uncertainty in cross-run diversity estimates with additional runs (Supplementary Note 1.7; Table S2). In contrast to this semantic stabilization, the systems remain lexically productive. Cumulative vocabulary size grows monotonically across rounds (Fig. 1c, left), reflecting sustained lexical novelty even as semantic content converges. The observed lexical trajectory closely follows an empirical unigram reference during earlier rounds, but diverges more clearly after approximately 500 rounds (Supplementary Fig. 1).

## Semantic Collapse Resists Intervention

To test the robustness of the observed semantic collapse, we design a comprehensive intervention suite. Each condition modifies a single factor relative to the default setting, while keeping the interaction scaffold and measurement pipeline unchanged. Unless otherwise noted, interventions are applied across GPT-5.6-Luna, GPT-4-mini, Phi-4, and DeepSeek-V3 and evaluated with the same protocol as the standard condition.

Across surface-level perturbations, levers that expanded lexical diversity did not reliably broaden semantic exploration (Fig. 2; Extended Data Fig. 3, 4). Temperature produced the clearest dissociation: raising it from 0 to 2 increased mean cumulative lexical diversity approximately 36-fold in Phi-4 and 21-fold in GPT-4-mini, while within-run semantic diversity decreased from 0.15 to 0.02 and from 0.11 to 0.03, and cross-run semantic diversity decreased from 0.45 to 0.02 and from 0.27 to 0.04, respectively. Thus, extreme token-level variation coincided with reduced semantic exploration both within and across runs; the broader, non-monotonic temperature effects are examined in Supplementary Note 2.2. Beyond temperature, regression analysis reveals no evidence that any tested intervention meaningfully attenuates semantic convergence. Across all baseline comparisons, no intervention yields a positive and significant effect on semantic diversity after Bonferroni correction (all adj. $p > 0.05$; Tables S9 and S10).

Deeper interventions target progressively more fundamental aspects of model behavior (Fig. 3). First, we pool models from distinct model families within a single simulation, motivated by prior work showing that cognitive diversity can enhance group-level creativity[23–25]. Second, we examine whether semantic collapse persists in multi-LLM generation using uncensored model variants. Given evidence that post-training alignment reduces diversity at the single-model level[26], we evaluate two community developed fine-tunes designed to relax safety-related refusals. Third, we target sycophancy, the tendency of models to over-accommodate their interlocutors[27], as a potential driver of convergence, testing whether reducing this bias via activation steering promotes divergence. Fourth, we directly optimize for output diversity using GRPO-based reinforcement learning inspired by existing approaches[28,29]. Beyond these four primary conditions, we further assess whether increasing or reducing the number of model instances ($N$ = 1, 5, and 10), adopting specialized multi-agent simulation frameworks[30,31] (AutoGen and AgentSociety), or injecting stochastic perturbations[32] can disrupt convergence (Extended Data Fig. 5). Notably, the characteristic early increase and subsequent slowing of semantic displacement were also observed under single-model recursive self-conditioning, in the absence of inter-agent exchange (Extended Data Fig. 5a; Supplementary Note 2.6). This indicates that multi-LLM interaction may not be a necessary condition for the semantic stabilization observed here. Across all deeper interventions tested, none produced a statistically reliable increase in within-run semantic diversity relative to baseline (all adj. $p > 0.05$; Table S9 ). Our customized diversity-targeted reinforcement learning was the only intervention to yield a significant positive contrast in cross-run semantic diversity (Table S10). However, we do not interpret this increase as meaningful mitigation. Transcript inspection showed run-specific degeneration into repetitive, often nonsensical sequences. Because the repeated sequences differed across runs, they increased distances between runs without sustaining coherent semantic exploration within runs (Supplementary Note 3.3; Table S8).

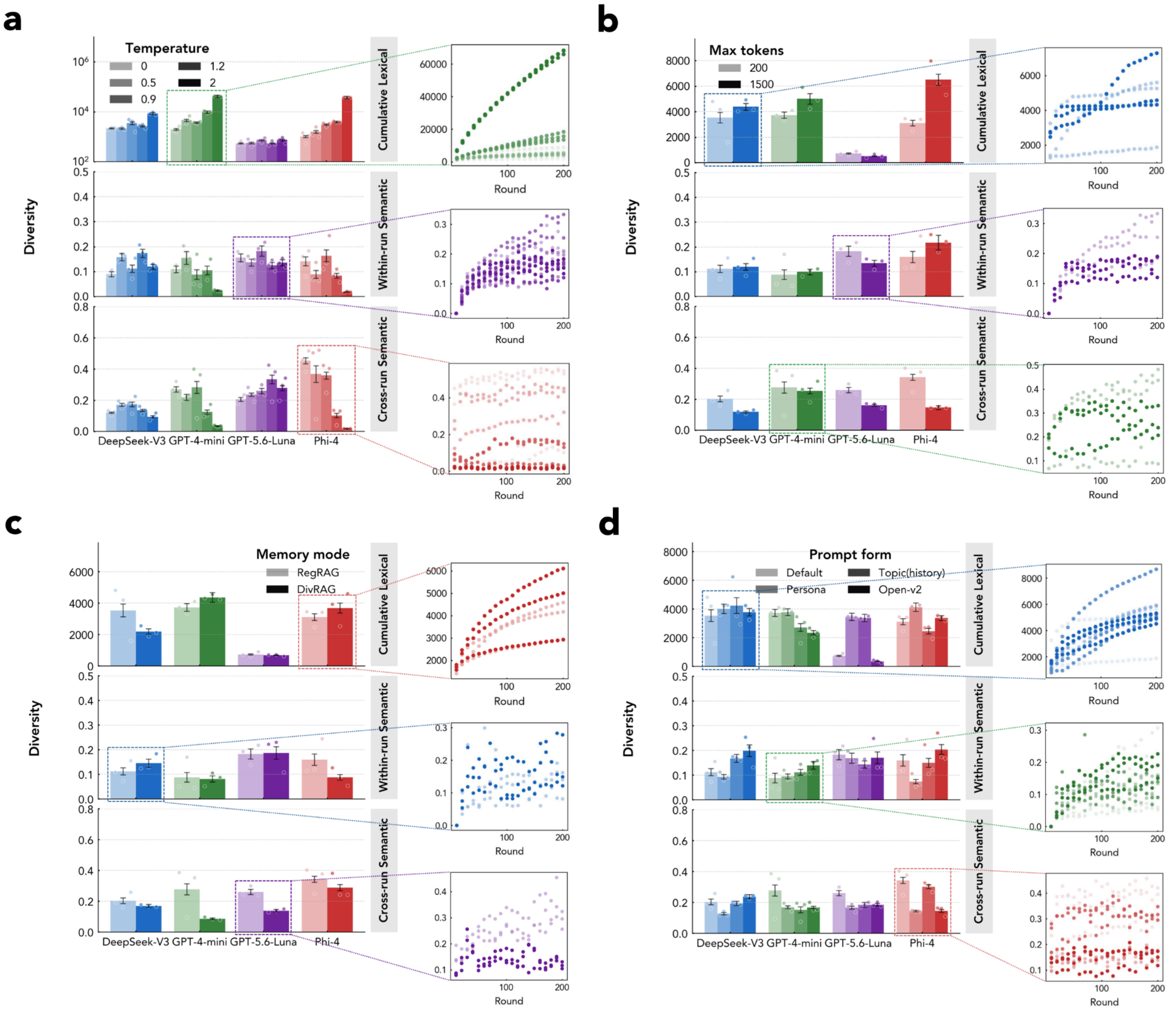

**Fig. 2 | Surface intervention factors shift lexical diversity but fail to attenuate semantic convergence. a–d,** Each panel displays three diversity measures: cumulative lexical diversity (top row), within-run semantic diversity (middle), and cross-run semantic diversity (bottom), each calculated for four models ($n$ = 3 independent simulations per model–condition). Bar plots (left) show mean diversity per condition, with dots representing individual run means or the three run-pair means for cross-run diversity. Error bars indicate 95% confidence intervals (mean ± 1.96 SEM), calculated from 60 pooled 10-round interval observations per model-condition (20 observations per run). Scatter plots (right) show trajectories across conversation rounds for selected models under each metric–intervention condition. Across all panels, lexical diversity responds to intervention parameters, while semantic diversity, both within- and cross-run, shifts with conditions but does not consistently improve relative to baseline.

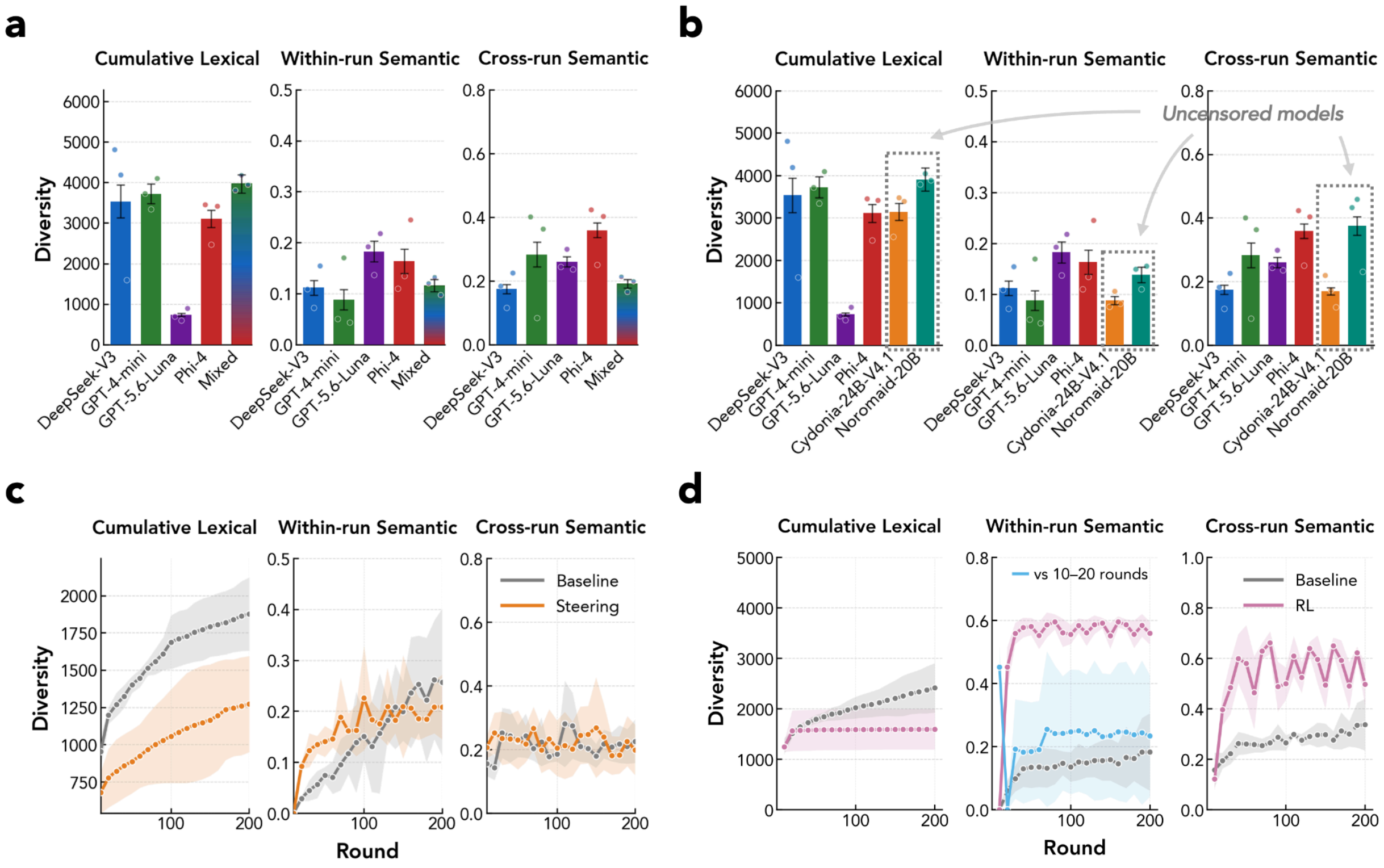

**Fig. 3 | Tested deep interventions do not reliably restore coherent semantic diversity. a,** Homogeneous- and mixed-model simulations. **b,** Standard models and two community uncensored finetunes. For **a** and **b**, bars show mean cumulative lexical, within-run semantic and cross-run semantic diversity; dots show individual run means or run-pair means for cross-run diversity (three dots per bar; $n$ = 3 independent simulations per condition). **c,** Sycophancy-targeted activation steering (orange) compared with baseline (gray). **d,** Diversity-targeted RL (pink) compared with baseline, with within-run diversity measured relative to rounds 1–10 or re-anchored to rounds 11–20 (light blue). Although RL increased cross-run semantic diversity, cumulative lexical diversity plateaued early as models repeated run-specific nonsensical tokens. Differences in these degenerate outputs increased cross-run distance at the cost of coherent language generation, rather than restoring sustained semantic exploration. Error bars and shading indicate 95% confidence intervals (mean ± 1.96 SEM), based on 60 pooled 10-round interval observations in **a,b** and three runs or run pairs per interval in **c,d**.

## Diagnosing Mechanisms of Semantic Collapse

Having established the robustness of collapse, we investigate its mechanistic basis. Two candidate explanations, sycophantic conformity and post-training safety-oriented alignment, are not established as necessary drivers by our intervention results (Fig. 3b, 3c). Activation steering that substantially reduces sycophancy (58% reduction in compliance bias; see Supplementary Note 3.6) does not attenuate convergence. Semantic collapse also persisted in two community-developed uncensored fine-tunes. These findings indicate that the driver of the observed collapse might be more fundamental than behavioral constraints imposed by training.

Building on prior work showing that recursive training on synthetic data induces distributional collapse[15], and on the interpretation of in-context learning as an implicit gradient-based updating process[33], we propose that multi-LLM interaction constitutes a form of inference-time recursion. In this setting, models iteratively generate outputs that are subsequently incorporated into the context of other models, effectively inducing repeated updates over a shared empirical distribution of representations. Under such dynamics, probability mass is progressively reallocated toward high-density regions while low-probability modes are attenuated, leading to a systematic reduction in representational variance. This is evidenced by the discrepancy in survival ratios of most and least frequent tokens over 200 interaction rounds (Fig. 4a), consistent with the contraction dynamics of a mode-amplifying recursive channel (see Supplementary Note 4.1).

Attention-head analysis in Llama 3-8B-Instruct reveals structured reorganization of attention circuitry across interaction (Fig. 4b). Later layers exhibit increasingly pronounced changes, with a subset of heads showing consistent amplification while others are progressively suppressed. Amplified heads are enriched for induction-like retrieval behavior, selectively re-accessing historically dominant sequences and promoting them in the output distribution. Across 759 induction events, the target sequence exceeds matched controls by a mean logit margin of 3.920 (median 4.625; one-sided Wilcoxon $p = 4.523 \times 10^{-27}$). The target is the top-1 prediction in 61.1% of events and within the top-10 in 84.1%. Retrieval and promotion are tightly coupled at the analysis-round level (Pearson $r = 0.623$, $p = 0.0033$), and retrieval shows a modest but significant strengthening trend over the trajectory ($p = 0.0379$). In the language of Bennett's logical depth[34], induction heads are depth-reduction circuits: they replace extended computation, reasoning from premises to novel conclusions, with constant-time pattern retrieval. As these circuits strengthen over rounds, the outputs become computationally shallower even as they maintain moderate lexical complexity.

Finally, when we project semantic trajectories onto manually constructed cultural-axis dimensions, they reveal how collapse is not isotropic but directional. Models exhibit coordinated movement along some semantic dimensions. Most notably, all four model families shift from a declarative toward a more questioning discourse stance, suggesting that extended interaction can produce shared movement toward particular semantic modes (Fig. 4c). Their trajectories diverge along other dimensions. On the individualism–collectivism axis, DeepSeek-V3 and GPT-5.6 Luna move toward collectivism, whereas GPT-4-mini and Phi-4 move toward individualism. On the egalitarian–hierarchical and cooperation–competition axes, DeepSeek-V3 and GPT-4-mini move in opposite directions, while Phi-4 and GPT-5.6 Luna show no reliable shift. A different division appears for tone: GPT-5.6 Luna moves toward detachment, whereas the other three models move toward greater emotional warmth. Thus, the models occupy partially distinct semantic regions at later stages rather than converging on a shared endpoint. A frozen model-attribution classifier's mean probability assigned to the true generating model *P*(true model) remains high throughout extended interaction, indicating that semantic convergence does not erase model-specific differences (Extended Data Fig. 6; Supplementary Note 4.5).

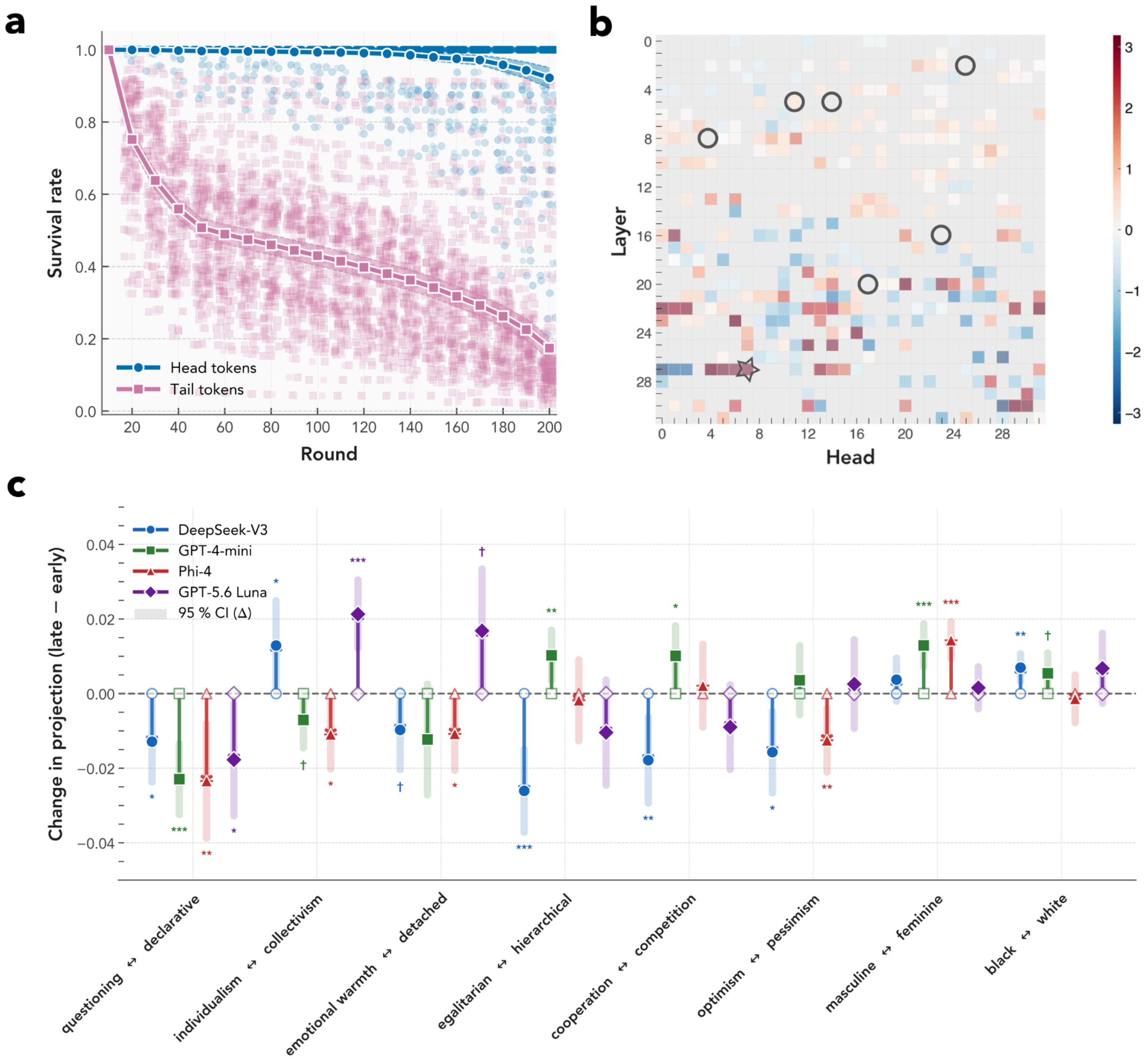

**Fig. 4 | Collapse is associated with tail-token diminishing, selective head intensification and attenuation, and semantically directional drift across models. a,** Token survival curves across conversation rounds. Head and tail token sets are defined from the early segment of each run (top and bottom 10% of cumulative frequency mass, respectively) and held fixed thereafter. Head tokens (blue) maintain near-complete survival throughout the simulation, while tail tokens (pink) decay sharply, indicating progressive loss of low-probability vocabulary. **b,** Attention head activation changes over the course of simulation (layer × head), showing significantly intensified (red) and attenuated (blue) heads ($p < 0.05$). Marked heads denote those functionally annotated, including heads associated with AI identity reinforcement (star) and induction behavior (circle). **c,** Projection changes from early rounds (first 30 rounds) to late rounds (last 30 rounds) semantic embeddings along extracted cultural dimensions for four

models. Each axis represents a bipolar cultural dimension. Arrows show the mean late-minus-early shift for each model, centered at zero. Shaded regions indicate 95% confidence intervals for the shift. Stars denote statistical significance ($^{\dagger}p < 0.1$, $^{*}p < 0.05$, $^{**}p < 0.01$, $^{***}p < 0.001$). Models exhibit distinct directional changes across axes, suggesting that semantic convergence is not culturally uniform but follows model-specific momentum.

# Discussion

Can finite computational systems sustain genuine creativity by iterating on their own outputs? For the systems we tested, our evidence suggests the answer is no. Across ten model families, simulations extending up to 1,000 rounds, and thirteen intervention strategies designed to promote diversity, we observe persistent semantic collapse: while surface-level variation accumulates, the underlying semantic space progressively narrows across iterations. The robustness of this pattern across architectures and interventions suggests the constraint is not an artifact of any particular model, though we cannot rule out that future AI systems might escape it. We hope they do. But our findings temper expectations that multi-LLM interaction alone can overcome the creative constraints of individual models and sustain open-ended innovation[35].

Our findings sit within a lineage of results on the limits of closed, inference-only iterative systems. Because these results rely on assumptions stronger than the multi-LLM systems studied here, we treat them as motivating analogies rather than as direct theoretical predictions (see Supplementary Note 5.1–5.4). The Data Processing Inequality and entropy-contraction results motivate the intuition that repeated processing can weaken dependence on initial variation and approach stable distributions[36,37], while Fisher-information decay offers a related analogy for reduced sensitivity to perturbations[38]. The Algorithmic Lovelace Bound similarly provides an analogy for why iteration need not continually expand descriptive complexity[39,40], echoing Lovelace's 1843 intuition that a machine cannot originate what was not already supplied to it[1]. Note that these parallels do not establish that semantic collapse is unavoidable or that in-loop interventions must fail. Instead, they provide a conceptual lens for interpreting the contraction, model-specific plateaus, and limited intervention effects observed here. We argue that developing a formal theory of semantic contraction in interacting LLMs is therefore an important direction for future research.

This framework also provides a new empirical perspective on Turing's supercriticality question. Turing imagined a machine that, like a supercritical pile of fissile material amplifying an incoming neutron into a chain reaction, might amplify an injected idea into a cascade of further ideas. The systems we study manifest the opposite. They resemble subcritical systems in the sense that tested perturbations did not produce sustained semantic divergence. Exponential relaxation governed by a spectral gap in idealized Markov-chain models provides a theoretical analogy[37]. The piano string is struck, vibrates briefly, and returns to rest. Turing's nuclear analogy maps onto the same free-energy principle that describes physical systems near equilibrium[41] The semantic basin behaves like the bottom of an energy well, with any displacement met by a restoring force proportional to the displacement itself, as in a harmonic oscillator. Current autoregressive architectures implement roughly this restoring dynamic. Rather than extending reasoning outward from premises to genuinely new conclusions, they short-circuit that computation in favor of retrieving and reproducing continuations already validated across the history of the conversation[34].

Our findings also extend and revise the prevailing account of model collapse. In the dominant formulation, model collapse arises during training, when models are recursively trained on synthetic data generated by themselves or other models, leading to a progressive narrowing of the original human data distribution[15,16]. This picture is incomplete. Collapse can emerge prior to any retraining, at inference time. Under closed-loop interaction with repeated self-conditioning, models exhibit progressive semantic contraction and tail-token erosion even with parameters held fixed. This induces a narrowing of semantic support purely through autoregressive generation dynamics. Recent multi-turn evaluation work has shown that a single LLM becomes less reliable when task specifications are distributed across conversation turns, partly because early assumptions and answer attempts induce path dependence by anchoring later responses[42]. Our findings identify a distinct form of history dependence in open-ended generation: semantic displacement approaches a plateau even under single-model recursive self-conditioning, indicating that the observed stabilization does not require multi-agent exchange and may reflect a broader dynamic of closed-loop generation. This implies that training-time collapse may not originate solely from the training process itself. If the synthetic corpus used for retraining is already generated under such closed-loop conditions, it may be pre-contracted—having already lost diversity and tail support before entering the training pipeline. In this sense, training may inherit and amplify a collapse that has already occurred at the level of data generation. Together, these results suggest that model collapse is better understood as a compound phenomenon, arising from the interaction between generation-side contraction and training-side estimation error, rather than as a process driven by training dynamics alone.

On a practical level, an important implication concerns the role of multi-LLM systems in AI for science and innovation. Multi-LLM systems may be effective at combinatorial discovery, recombining existing concepts, exploring local variations, and efficiently traversing well-structured regions of a problem space. Transformational discovery presents a different challenge, as it often involves repurposing concepts across distant domains and opening qualitatively new directions rather than extending existing ones[43]. Our experiments do not test whether multi-LLM systems produce new knowledge or scientific breakthroughs. They show that, in the open-ended closed-loop settings studied here, continued interaction can expand surface-level vocabulary while narrowing the semantic range of the generated content.

We note that while multi-LLM systems might lead the uncritical observer to imagine semantic diversity while observing lexical diversity, the single-LLM (monologue) condition did not, remaining low on both. The only interventions which did not result in lexical diversity increase were anti-sycophancy steering and diversity-oriented reinforcement learning. In these settings, reduced lexical diversity remained an informative signal of semantic collapse, whereas the increased lexical diversity observed in multi-LLM systems and under other treatments masked it. Sustained semantic diversity is not sufficient for discovery, but it may be relevant to maintaining exploration across distant conceptual directions. The observed contraction therefore motivates caution about treating closed multi-LLM systems as autonomous engines of transformational innovation, without establishing that such systems are incapable of it. This distinction does not diminish their potential utility in scientific workflows. Current systems may still augment human-led discovery by accelerating exploration within existing paradigms and surfacing plausible recombinations. Our results instead motivate further study of how human input, external data, or other sources of conceptual renewal affect their capacity to sustain open-ended exploration.

Diversity limitations may also matter in the broader human–LLM social system. LLMs can shape human judgments and decisions[44–46], raising the possibility that repeated reliance on model outputs feeds their regularities back into subsequent human production. Human collectives, however, draw on sources of variation unavailable to closed generative loops, including disagreement, encounters with intellectually distant ideas, cumulative cultural transmission, and heterogeneous experience, including new signals from discovered or invented sensors of the environment or the self[47–49]. Whether these mechanisms make humans "supercritical knowers" in Turing's sense has not been established, and our experiments do not determine how they interact with model-side contraction. We therefore treat the long-term dynamics of joint human–LLM systems as an open empirical question. If human and social processes continually introduce independent variation, they may counteract contraction; if repeated LLM mediation instead weakens or channels those sources of variation, contraction could propagate beyond the models themselves. Determining which dynamic dominates will require direct, longitudinal study of mixed human–LLM systems.

Several limitations bound the scope of our conclusions. First, although we evaluated ten models, including three frontier models, and found no evidence that semantic collapse diminished with model capability, these results do not establish that future, more capable models will exhibit the same dynamics. Second, our tested intervention strategies span a broad range of techniques, but each represents one or a limited set of implementations within a much larger design space. We therefore cannot rule out untested methods or parameterizations that may mitigate collapse. Our findings instead provide a baseline against which such approaches can be evaluated. Third, the theoretical results invoked here serve as heuristic analogies rather than as a formal theory of multi-LLM systems. They help interpret the empirical patterns but do not establish that semantic collapse is a necessary or universal property of current LLM architectures. Developing a mathematical theory that specifies when recursive interactions induce semantic contraction, and determines whether, and under what conditions, the relevant information-theoretic or thermodynamic results apply, remains an important direction for future research.

# Methods

## Interaction environment

We construct a minimal multi-LLM simulation in which LLMs interact over discrete rounds with no predefined task, roles, or reward (standard agent prompt in Table S6, section A; referee prompt in Supplementary Note 1.2 and Table S1). At each round, agents act sequentially in randomized order, generating natural-language utterances. An automated referee model (GPT-4-mini, deterministic decoding) classifies each utterance for interaction intent, valence, and content, routing interactive utterances to recipients. Interactive utterances are broadcast by default to the other agents, each of which generates a reaction that is returned to the speaker; group invitations are handled through a consent-based temporary group channel (Supplementary Note 1.3). The referee outputs are used only for routing and interaction metadata and are not supplied to the agents as conversational content. Models are conditioned on a static environment instruction, assigned identity profiles, a short-term conversation buffer, and a long-term retrieval-augmented memory module backed by agent-specific vector databases. Prior agent interaction history is stored in agent-specific vector databases and retrieved each round to augment the recent conversation buffer; in the Standard condition, retrieved memories are packed using a relevance-ranked baseline (Supplementary Note 1.4).

## Models

Primary models: GPT-5.6-Terra (OpenAI), Claude Sonnet 5 (Anthropic), GPT-5.6-Luna (OpenAI), GPT-4-mini (OpenAI), DeepSeek-V3 (DeepSeek), and Phi-4 (Microsoft). Extended models: Llama-3-8B-Instruct (Meta), Qwen-7B (Alibaba), and uncensored variants (Cydonia-24B-v4.1, Noromaid-20B). All models use default decoding settings (temperature = 0.9, max tokens = 200) except where varied as interventions. Extended simulations run for 1000 rounds with three independent replicates per model; standard intervention simulations run for 200 rounds with three replicates per condition per model.

## Embedding and metrics

Model-generated conversational text from each non-overlapping 10-round interval is embedded using OpenAI text-embedding-3-large. For long interval documents, we use the chunk-and-pool procedure described in Supplementary Note 1.5, in which consecutive non-overlapping text chunks are embedded independently; their embeddings are mean-pooled, and the pooled vector is L2-normalized to obtain a single interval representation.

*Cumulative lexical diversity*: cumulative unique unigram count. *Within-run semantic diversity*: cosine distance between each interval's embedding and the first 10-round interval embedding of the same run. *Divergence degree*: cosine distance between adjacent interval embeddings.

*Cross-run semantic diversity*: average pairwise cosine dissimilarity between corresponding intervals of independent runs. We additionally use *normalized Vendi Score* to quantify semantic support without reference to the initial interval. For the time-resolved analysis in Extended Data Fig. 1e, Vendi Score is computed separately within each 10-round interval from the embeddings of individual utterances; the baseline early-to-late analysis instead uses exact token blocks, whereas the topic- and token-matched human comparison uses matched length chunks (Supplementary Notes 2.8–2.9). To validate whether interval-level cosine distance tracks semantic differences in the underlying text, we conducted a blinded evaluation of 200 interval pairs sampled across the observed distance range. Two independent frontier evaluator models rated overall meaning similarity and dominant-topic overlap without access to the embeddings or distance values, and associations with continuous cosine distance were quantified using Spearman correlations (Supplementary Note 1.10). Complementary analyses characterized the representation using unsupervised embedding clustering and an embedding-independent LDA topic model (Supplementary Notes 1.9 and 2.10).

## Intervention suite

Thirteen intervention categories, each modifying a single factor: decoding temperature (0, 0.5, 0.9, 1.2, 2), output budget (200, 1500), memory removal (standard, no memory), retrieval-augmented memory packing policy (standard, diversity-oriented), prompt formulation (standard, history topic, distinct persona, alternative open-ended, explicit diversity instruction, extended background, structured coordination), mixed-model composition (standard vs. mixed), uncensored model variants (standard, Cydonia-24B-v4.1, Noromaid-20B), sycophancy-targeted activation steering (standard, Llama-3-8B-Instruct, Layer 20, $\alpha$ = -10), history-conditioned avoidance decoding (HC-AD), GRPO-based diversity-targeted reinforcement learning (Qwen-7B default, Qwen-7B reinforced), population scaling ($N$ = 1, 3, 5, and 10), alternative simulation frameworks (standard, AutoGen, AgentSociety), and random-noise perturbation injection (standard, random noise injection every 30 rounds). For details, see Supplementary Notes 2 and 3. Unless otherwise noted, intervention conditions are evaluated with three independent replicates; model assignments, simulation durations, and replication details are summarized in Extended Data Table 1.

## Statistical analysis

Factor-specific linear regressions were fitted to observations from successive 10-round intervals for within-run and cross-run semantic diversity. Models included interval-position fixed effects, with cluster-robust standard errors clustered by run for within-run analyses and by run pair for cross-run analyses. Intervention coefficients estimate the adjusted average difference from the corresponding baseline across the observed intervals. Positive coefficients indicate greater mean semantic diversity, but do not imply that every run or run pair exhibited an increase. Statistical significance was evaluated after Bonferroni correction within each comparison family (Supplementary Note 3.8; Tables S9 and S10).

## Mechanistic analysis

Token survival analysis: head and tail token sets defined from the first four intervals (top and bottom 10% of cumulative frequency mass), held fixed thereafter, with survival curves tracking persistence over the full trajectory (Supplementary Note 4.1).

Attention-head analysis: teacher-forcing replay of interaction transcripts through Llama-3-8B-Instruct, with extraction of layer-wise attention tensors and next-token logits. Induction heads identified via an independent calibration benchmark using repeated random token sequences; subsequent analysis on multi-LLM transcripts measured continuation-specific retrieval bias and output-space promotion of historically dominant continuations (Supplementary Note 4.2). A complementary all-head scan quantified temporal changes in head-output magnitude and attention concentration across the interaction trajectory, with per-head trends evaluated after false-discovery-rate correction (Supplementary Note 4.3).

Cultural-axis projection: early- and late-stage embedding centroids projected onto bipolar semantic axes defined by aggregated antonym-pair difference vectors (Supplementary Note 4.4).

Model-attribution classifier: a frozen five-class multinomial logistic-regression classifier was trained on utterance-level embeddings from independent 200-round baseline runs to distinguish GPT-5.6 Terra, Claude Sonnet 5, GPT-4-mini, DeepSeek-V3, and Phi-4. It was then evaluated without further updating on independent 1,000-round runs, with valid utterances concatenated within each round before embedding and classification (Supplementary Note 4.5).

## Data availability

The numerical data underlying the main and Extended Data figures are provided as Source Data files accompanying this Article. The data underlying Supplementary Figs. 1 and 2 are provided as Supplementary Data. The Reddit discussion data used for the human-reference analyses were obtained from the publicly available Project Arctic Shift archives[51]. Due to their size, additional simulation-level datasets are available for research verification by contacting the corresponding author at jevans@uchicago.edu.

## Code availability

The custom code used for the simulations and analyses reported in this study is publicly available at https://github.com/TeerryKkkk/semantic-collapse-replication. A fixed version of the code (v1.0.0) is archived on Zenodo[54].

## Acknowledgments

The authors wish to thank the NSF and OpenAI's Research Access Program for financial and computational support, respectively. Additional computational resources were provided by Knowledge Lab at the University of Chicago.

## Funding Statement

This work was supported by the National Science Foundation TIP #2404109 to J.E. and by API credits provided to S.L. through OpenAI's Researcher Access Program.

## Author Contributions Statement

Conceptualization: S.L., W.K., J.E.

Methodology: W.K., S.L., J.P.

Investigation: W.K., S.L., J.P.

Visualization: S.L.

Funding acquisition: S.L., J.E.

Project administration: S.L., J.E.

Supervision: J.E.

Writing – original draft: W.K., S.L.

Writing – review & editing: W.K., S.L., J.E.

## Competing Interests Statement

J.E. is employed as a 25% visiting researcher at Google and holds equity in Preseen and SAI Labs. None of these entities were involved in authorizing or evaluating this research. No other authors declare any competing interests.

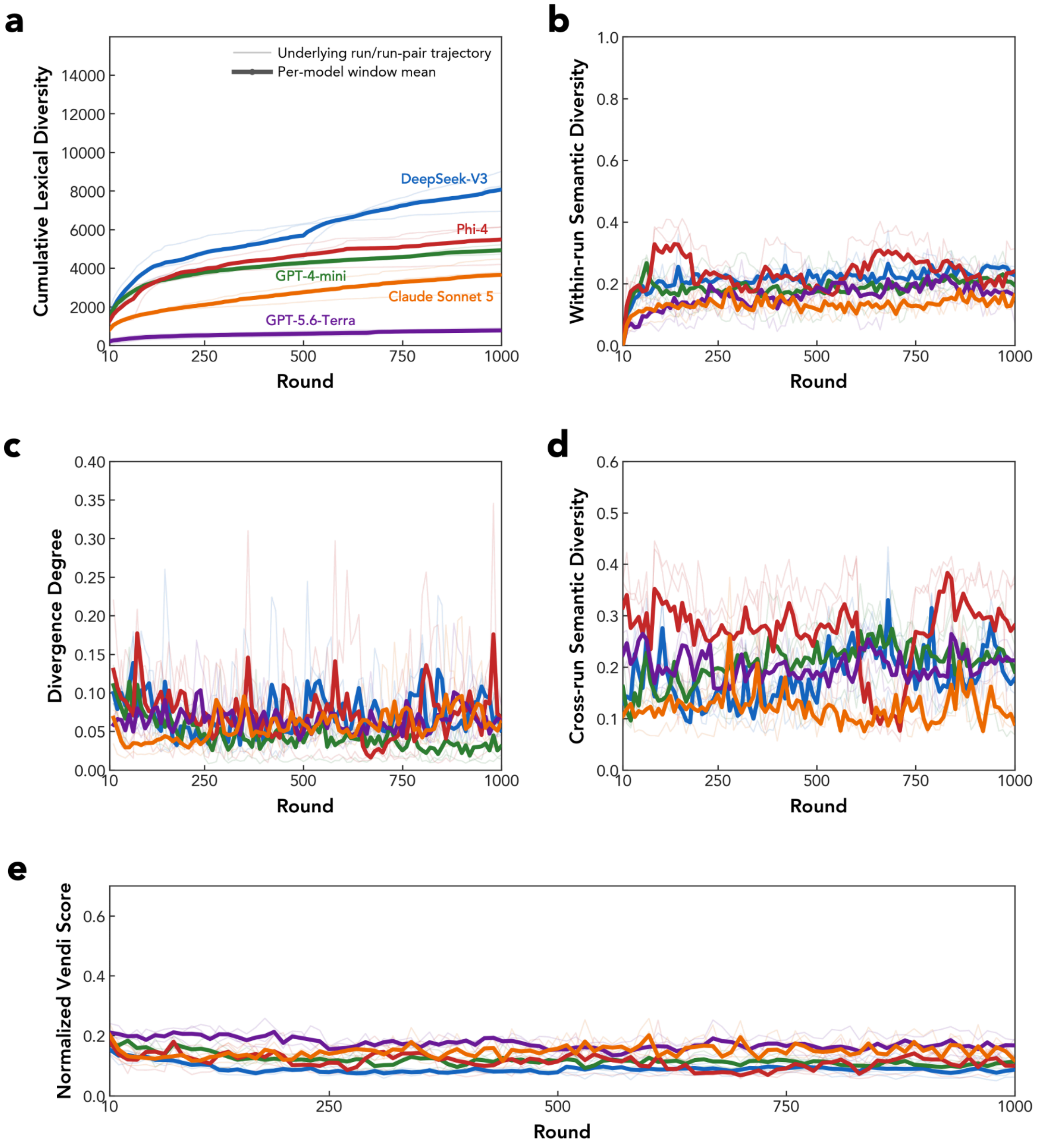

**Extended Data Fig. 1 | Underlying per-model trajectories of lexical and semantic diversity across extended simulations. a,** Cumulative lexical diversity. **b,** Within-run semantic diversity. **c,** Divergence degree, calculated as the cosine distance between consecutive 10-round intervals. **d,** Cross-run semantic diversity. **e,** Normalized Vendi score, computed separately within each 10-round interval from pairwise cosine similarities among the embeddings of individual utterances and normalized by the number of utterances in that interval. Light lines show the three independent simulation trajectories per model in **a–c,e**, and the three corresponding within-model run-pair trajectories in **d**. Solid lines indicate the corresponding per-model mean at each interval. Values are plotted at nominal 10-round interval endpoints. Panels (**a,b,d,e**) begin at round 10; divergence degree (**c**) begins at round 20 because it compares consecutive intervals. Results are displayed without polynomial smoothing. Across models, cumulative lexical diversity continues to increase, whereas the semantic-diversity and Vendi measures generally stabilize after their initial changes.

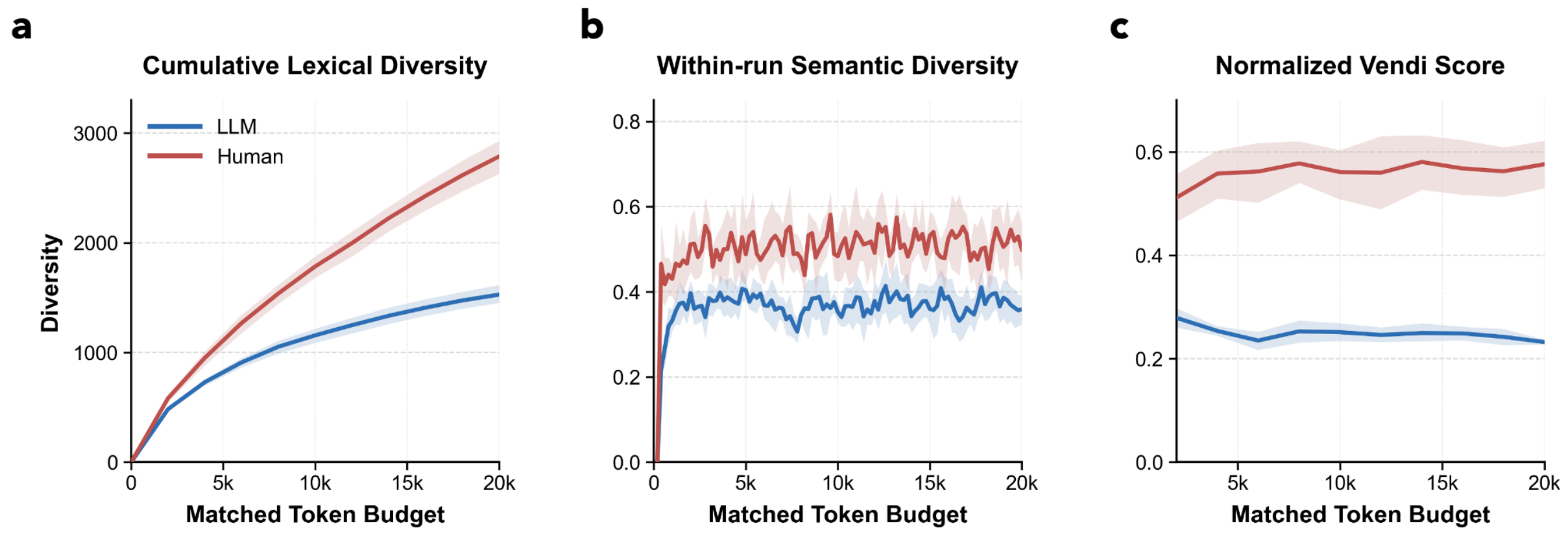

**Extended Data Fig. 2 | Topic-matched human text exhibits greater lexical and semantic diversity than LLM continuations.** Six Reddit open-ended discussion topics were paired with continuations generated by four model families (DeepSeek-V3, GPT-4-mini, GPT-5.6 Luna, and Phi-4), each matched to a 20,000-token budget. **a,** Cumulative number of unique lexical types. **b,** Within-run semantic diversity, calculated as the cosine distance between each non-overlapping 200-token window and the first 200-token window of the same trajectory. **c,** Normalized Vendi Score, calculated within non-overlapping 2,000-token intervals from twenty 100-token chunks. LLM values were averaged across models within each topic before aggregation. Lines show means across $n$ = 6 matched Reddit threads; shading denotes 95% bootstrap confidence intervals. Full analysis details are provided in Supplementary Note 2.8, including a sensitivity analysis controlling for number of conversation contributors (i.e., on multi-human Reddit vs. multi-LLM simulation configurations).

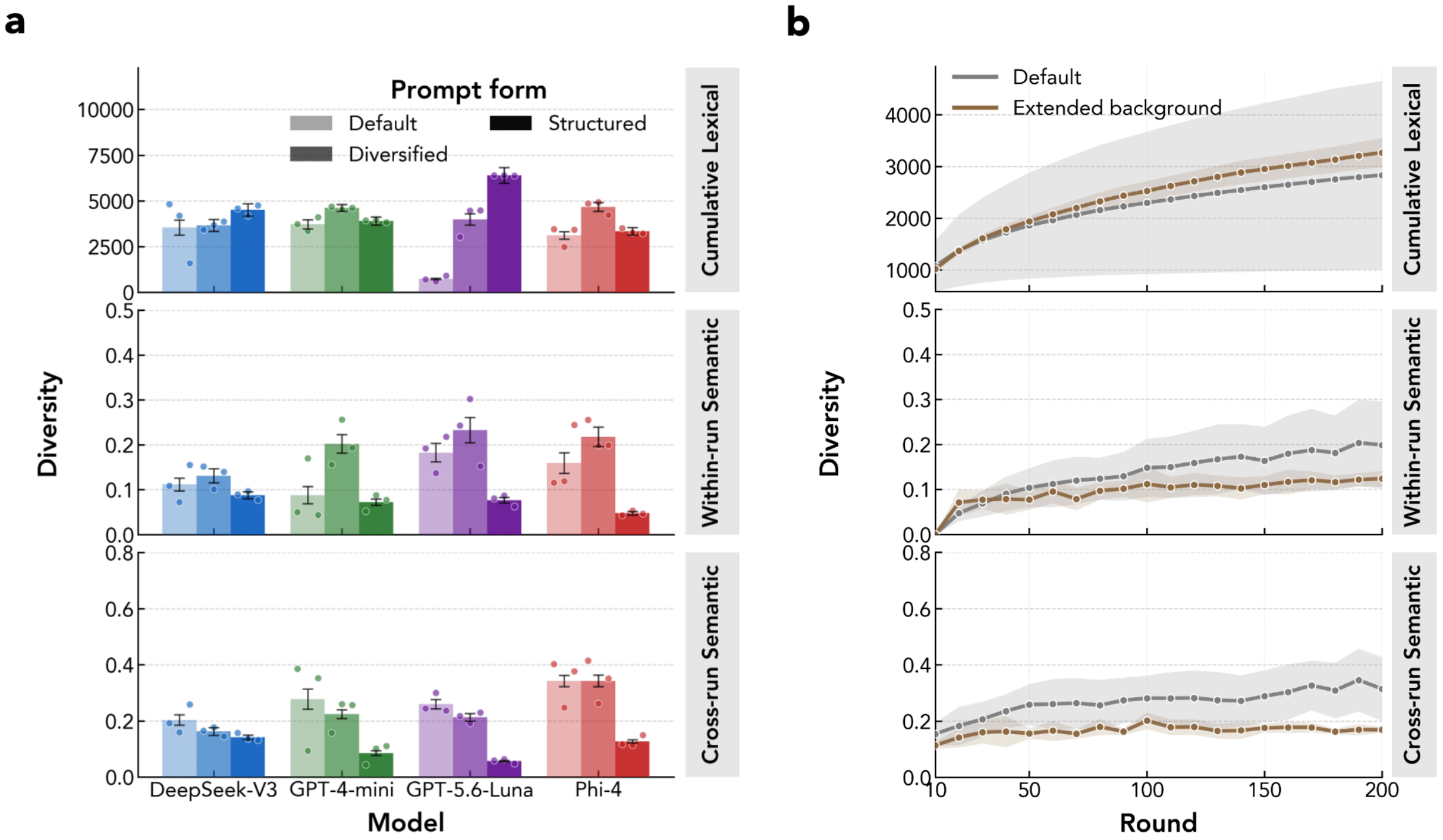

**Extended Data Fig. 3 | Prompt-based interventions do not consistently mitigate semantic convergence. a,** Comparison of the default prompt with prompts containing explicit diversification instructions (“Diversified”) or structured coordination rules (“Structured”) across DeepSeek-V3, GPT-4-mini, GPT-5.6-Luna, and Phi-4. Bars show mean cumulative lexical diversity (top), within-run semantic diversity (middle) and cross-run semantic diversity (bottom) across simulation intervals. Dots represent individual run means or within-model run-pair means ($n$ = 3 independent simulation runs per model–condition). Although some interventions increased lexical or semantic diversity for individual models, neither intervention produced consistent improvements across models and semantic measures. **b,** Trajectories pooled across GPT-4-mini and GPT-5.6-Luna comparing the default prompt with an extended-background condition. In the latter, each model instance was assigned a detailed persona derived from the Wikipedia biography of a prominent scholar, such as Herbert A. Simon, Judea Pearl or Deborah G. Mayo. Points show means at each 10-round interval. Both conditions include six independent simulations (three per model) and six within-model run pairs. Extended persona grounding did not increase within-run or cross-run semantic diversity over time and therefore did not meaningfully mitigate semantic convergence. Error bars and shading denote descriptive 95% confidence intervals (mean ± 1.96 SEM), calculated across pooled 10-round interval observations in **a** and separately at each interval in **b**.

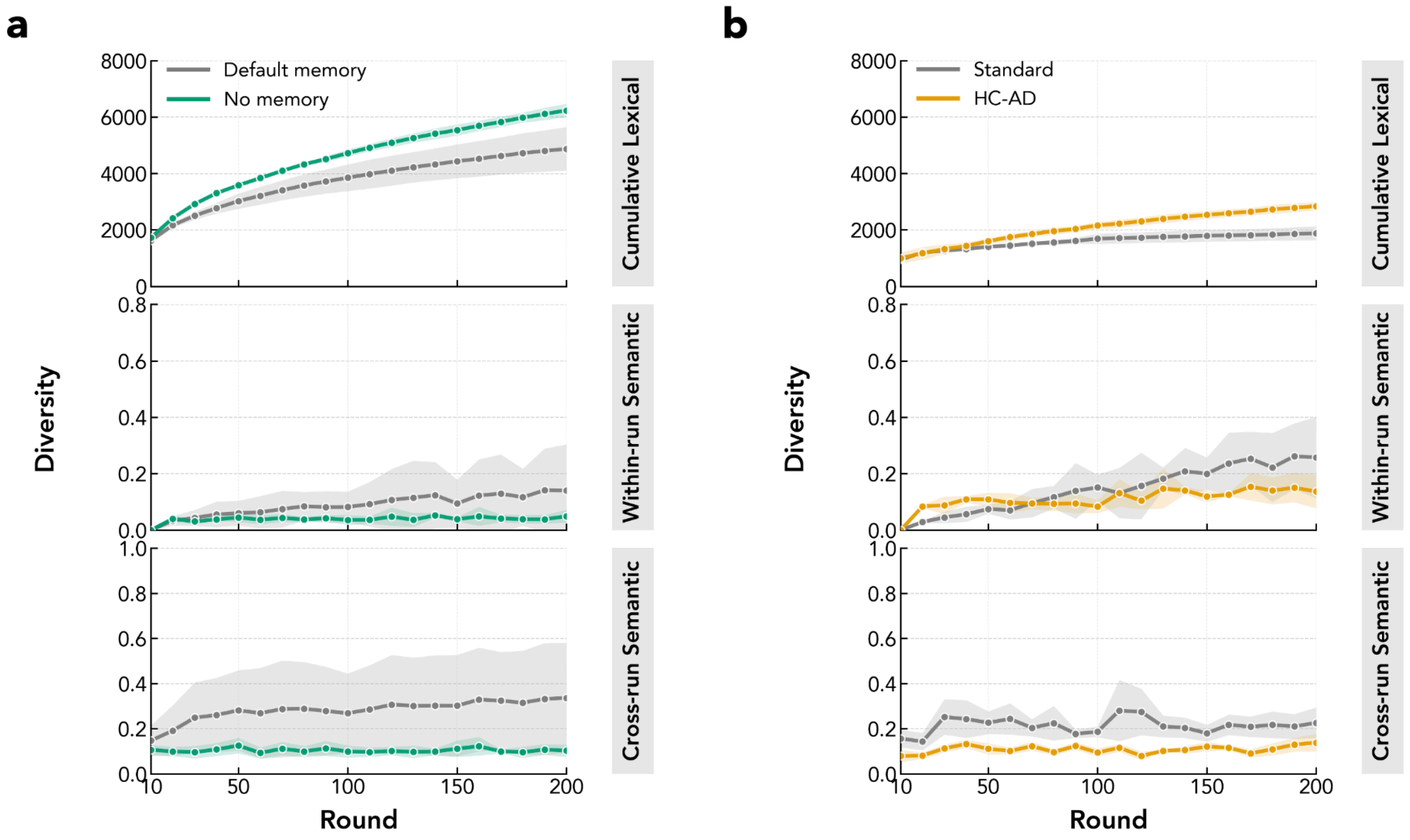

**Extended Data Fig. 4 | Memory removal and history-conditioned avoidance decoding increase lexical diversity without improving semantic diversity. a,** GPT-4-mini simulations with default retrieval-augmented memory or no memory, both at temperature 0.9. **b,** Meta-Llama-3-8B-Instruct simulations using standard decoding or history-conditioned avoidance decoding (HC-AD). Lines show mean cumulative lexical diversity (top), within-run semantic diversity (middle) and cross-run semantic diversity (bottom) across successive 10-round intervals. Each condition includes $n$ = 3 independent 200-round simulations; cross-run curves summarize the three unordered pairs formed from those runs. Shading denotes descriptive 95% confidence intervals (mean ± 1.96 SEM), calculated separately at each interval. Windows are repeated measurements within runs, and run pairs share runs. Both interventions increased lexical diversity, but neither increased semantic diversity: memory removal reduced both semantic measures, while HC-AD produced no reliable improvement within runs and significantly reduced diversity across runs (Supplementary Tables S9 and S10).

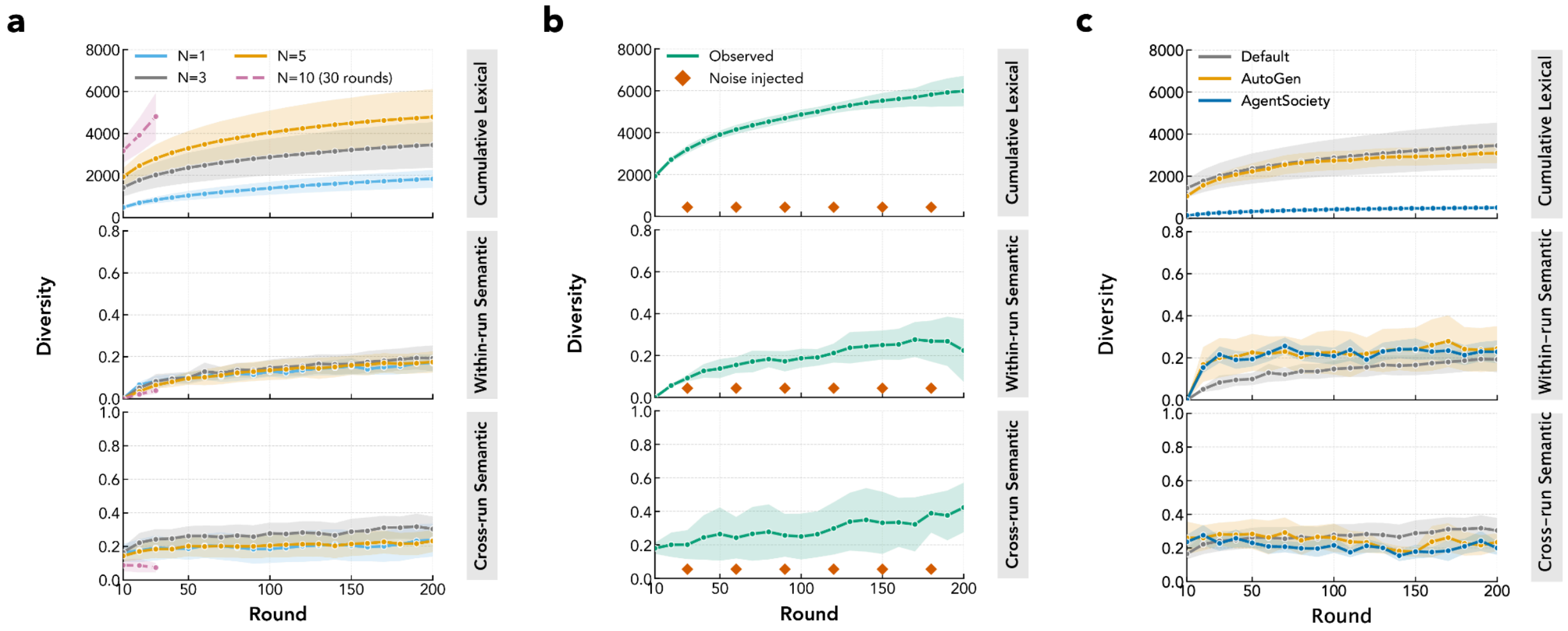

**Extended Data Fig. 5 | Scaling, noise injection and alternative orchestration frameworks do not prevent semantic convergence.** Each panel shows cumulative lexical diversity (top), within-run semantic diversity (middle) and cross-run semantic diversity (bottom). Lines show means; shaded regions denote descriptive 95% confidence intervals (mean ± 1.96 SEM), calculated separately at each plotted position across simulation runs or within-model run pairs. **a,** Trajectories for systems comprising one, three, five or ten agents. The one-, three- and five-agent conditions each include 12 runs and 12 run pairs, with three runs per model. The ten-agent condition includes six runs and three pairs—two runs each for DeepSeek-V3, GPT-4-mini and Phi-4—and was limited to 30 rounds owing to computational costs (dashed lines). Increasing agent count increases lexical diversity but does not consistently increase semantic diversity. **b,** Trajectories under periodic noise injection, introduced every three 10-round intervals (orange diamonds; $n$ = 3 runs and three pairs). Semantic diversity fluctuates under repeated perturbations. **c,** Trajectories comparing the default framework with AutoGen[30] (yellow) and AgentSociety[31] (blue). Each model contributes three runs: Default includes 12 runs/12 pairs, AutoGen nine runs/nine pairs and AgentSociety 12 runs/12 pairs in total. Neither alternative meaningfully mitigates semantic convergence. For visual comparison, AgentSociety observation positions were linearly mapped onto the displayed Round 10–200 interval.

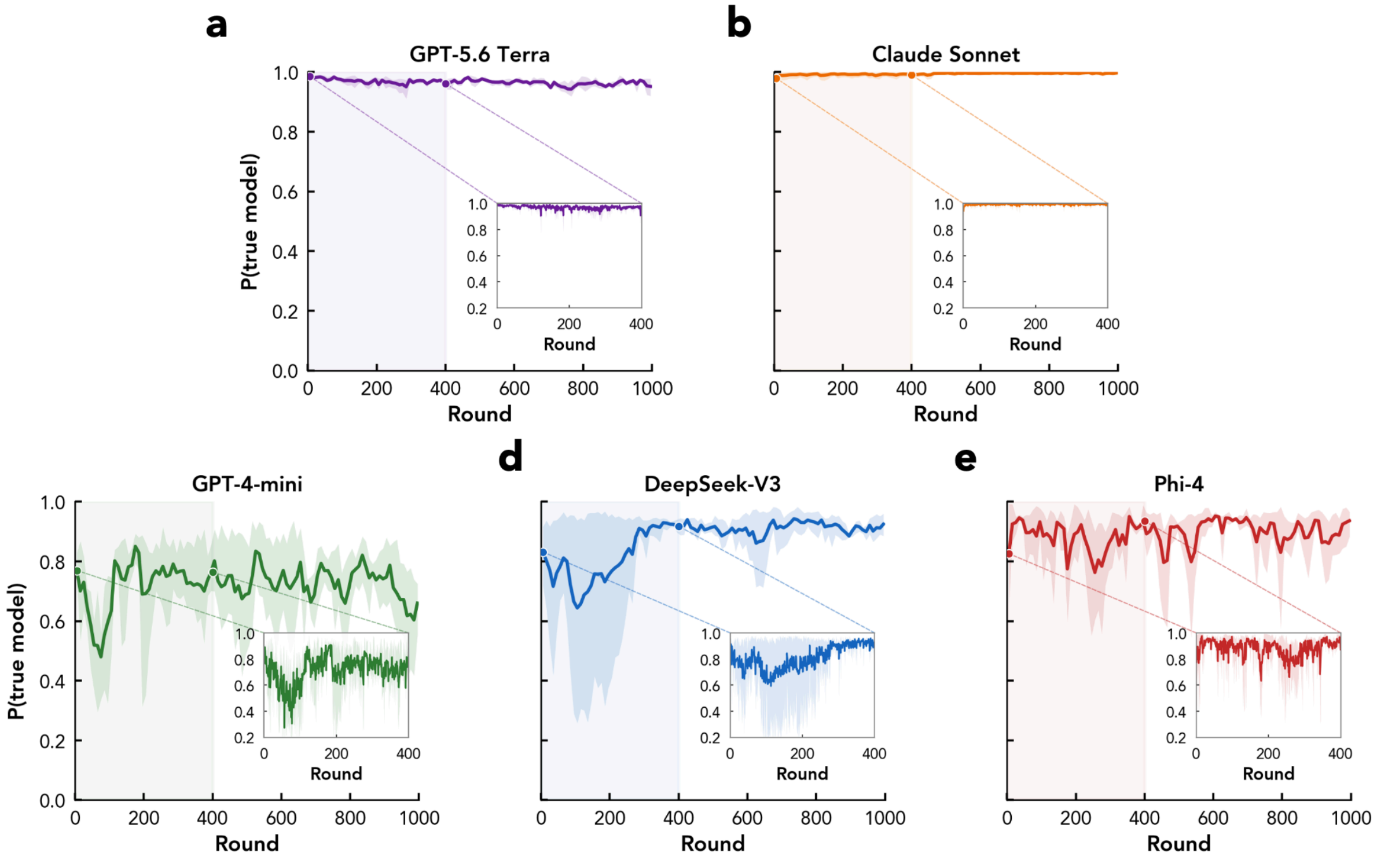

**Extended Data Fig. 6 | Base-model identifiability over 1,000 rounds of repeated interaction. a–e,** Probability assigned to the correct generative model, P(true model), for GPT-5.6 Terra (**a**), Claude Sonnet 5 (**b**), GPT-4-mini (**c**), DeepSeek-V3 (**d**), and Phi-4 (**e**). Probabilities come from a five-class logistic-regression probe on utterance embeddings, trained on three independent 200-round baseline runs per model and evaluated on three independent 1,000-round baseline runs per model; chance is 0.20. Main panels show means over non-overlapping 10-round intervals and insets show individual rounds over the shaded first 400 rounds; all horizontal axes are in rounds. Lines show means across $n = 3$ independent test runs per model; bands denote pointwise 95% percentile-bootstrap confidence intervals, calculated from 10,000 run-level resamples. Both frontier models remain near-perfectly identifiable ($P = 0.966$ in **a** and 0.994 in **b**). Phi-4 (0.893), DeepSeek-V3 (0.870), and GPT-4-mini (0.724) are identified with high confidence, the latter only after approximately 300 rounds.

| Experimental factor | Conditions | Models | Duration and replicates |
|---|---|---|---|
| **Standard** | Default topic-free triadic configuration | GPT-5.6 Terra; Claude Sonnet 5; GPT-4-mini; DeepSeek-V3; Phi-4 | 1,000 rounds × 3 runs/model |
| **Decoding temperature** | T = 0, 0.5, 0.9, 1.2, 2.0 | DeepSeek-V3; GPT-4-mini; GPT-5.6 Luna; Phi-4 | 200 rounds × 3 runs/model/condition |
| **Output budget** | max_tokens = 200, 1,500 | DeepSeek-V3; GPT-4-mini; GPT-5.6 Luna; Phi-4 | 200 rounds × 3 runs/model/condition |
| **History-conditioned avoidance decoding** | HC-AD | Meta-Llama-3-8B-Instruct | 200 rounds × 3 runs/condition |
| **Memory** | No-memory ablation, evaluated at T = 0.9 | GPT-4-mini | 200 rounds × 3 runs/condition |
| **Retrieval packing** | Diversity-oriented structured packing | DeepSeek-V3; GPT-4-mini; GPT-5.6 Luna; Phi-4 | 200 rounds × 3 runs/model/condition |
| **Prompt formulation** | Default; Diversified; Open-v2; History; abstract persona; extended-background persona; Structured | DeepSeek-V3; GPT-4-mini; GPT-5.6 Luna; Phi-4 | 200 rounds × 3 runs/model/condition |
| **Model composition** | Mixed-model triad | GPT-4-mini; DeepSeek-V3; Phi-4 | 200 rounds × 3 runs/condition |
| **Reduced-guardrail models** | Reduced-guardrail homogeneous triads | Cydonia-24B-v4.1; Noromaid-20B | 200 rounds × 3 runs/model |
| **Sycophancy-targeted activation steering** | $\alpha = 0$; Layer 20, $\alpha = -10$ | Meta-Llama-3-8B-Instruct | 200 rounds × 3 runs/condition |
| **Diversity-targeted reinforcement learning** | Standard Qwen-7B; diversity-targeted GRPO | Qwen-7B | 200 rounds × 3 runs/condition |
| **Population size** | N = 1, 3, 5, 10 ($N = 3$ is standard) | DeepSeek-V3; GPT-4-mini; GPT-5.6 Luna; Phi-4 | $N = 1, 3, 5$: 200 rounds × 3 runs/model; N = 10: 30 rounds × 2 runs/model |
| **Orchestration framework** | AutoGen; AgentSociety | AutoGen: DeepSeek-V3, GPT-4-mini, GPT-5.6 Luna, Phi-4; AgentSociety: DeepSeek-V3, GPT-4-mini, Qwen-7B, GPT-5.6 Luna | AutoGen: 200 rounds × 3 runs/model; AgentSociety: 20 rounds × 3 runs/model |
| **Random semantic perturbation** | Periodic exogenous noise injection every three 10-round intervals | GPT-4-mini | 200 rounds × 3 runs/condition |

**Extended Data Table 1 | Summary of simulation and intervention conditions.** The table summarizes the experimental factors, tested conditions, model coverage, and actual simulation duration and independent replication for the long-horizon standard simulations and intervention experiments. Unless otherwise specified, trajectory-based diversity measures were evaluated over non-overlapping 10-round intervals. The extended-background/biography prompt condition was evaluated only for GPT-4-mini and GPT-5.6 Luna, and the $N = 10$ population condition only for DeepSeek-V3, GPT-4-mini, and Phi-4. AgentSociety was analyzed at one-round resolution over rounds 1–20 rather than in 10-round intervals.

# Supplementary Information for

## Multi-LLM Systems Exhibit Robust Semantic Collapse

Weiyi Kong, Shiyang Lai, Jinghua Piao, James Evans

This Supplementary Information accompanies the main Article and its Extended Data. It provides the full methodological specification of the multi-LLM interaction environment, the complete intervention suite, the extended robustness and human-reference analyses, the mechanistic diagnostics linking macroscopic convergence to circuit-level dynamics, and the formal dynamical synthesis that situates our empirical findings within information-theoretic accounts of closed generative systems. Supplementary Note 1 documents the experimental design, simulation scaffold, and diversity metrics used throughout. Supplementary Notes 2 and 3 detail the thirteen intervention strategies and the statistical framework used to evaluate them. Supplementary Note 4 reports the transcript-level and white-box mechanistic analyses. Supplementary Note 5 develops the dynamical interpretation of semantic collapse and its testable consequences. Supplementary Tables S1 to S12 are referenced from the relevant notes below.

## Contents

## Supplementary Tables

## Supplementary Figures

# Supplementary Note 1. Experimental design and analysis pipeline

## 1.1 Study design

We investigate open-ended language model interaction in a minimal multi-agent simulation framework without external tasks, reward signals, or ground-truth targets. The default experimental scaffold consists of a homogeneous population of three LLMs interacting over $T$ discrete rounds. Unless otherwise specified, each condition is run for $T = 200$ rounds with three independent replicates per model. Within a given condition, model composition, prompts, decoding parameters, referee configuration, and memory settings are held fixed; only the condition-defining intervention is changed. Independent replicate runs are used to characterize both within-run temporal evolution and cross-run reproducibility. For within-run analyses, each run is analyzed separately and condition-level values are obtained by averaging across the three replicates. For cross-run analyses, we compute all three pairwise comparisons among the replicate runs and average them to obtain a condition-level estimate.

## 1.2 Multi-LLM interaction environment

At each round, agents act sequentially in a randomized order. When active, an agent generates a natural-language utterance describing its action or statement in the shared environment. To govern interaction routing, we employ an automated Referee Module implemented via GPT-4-mini. The referee operates as an LLM-based zero-shot classifier, post-processing each raw utterance into a structured action record $\mathcal{A}_t$ that includes: (i) interaction intent (binary; whether the utterance is directed to other agents), (ii) valence (categorical; positive/negative/neutral), and (iii) a concise free-text summary of the act. The referee is run with deterministic decoding to ensure stable and reproducible routing decisions. The referee's system prompt and classification criteria are listed below:

<table>
<tr><th>Field</th><th>Content</th></tr>
<tr><td>Module</td><td>Moderator / Referee</td></tr>
<tr><td>Example output</td><td><pre>{
  "action_name": "Reply to Invitation",
  "is_interaction": true,
  "valence": "positive",
  "description": "Expresses gratitude and readiness to
join the discussion.",
  "group_invitation": false,
  "agree_to_group": true
}</pre></td></tr>
<tr><td>Full prompt</td><td><pre>You are Referee analyzing text from [SPEAKER_NAME]:
"""[RAW_TEXT]"""</pre></td></tr>
</table>

```
We define six keys for classification:
1. action_name: A concise verb phrase summarizing the
main action (e.g. "Reply to Invitation", "Collaborate",
"Attack").
2. is_interaction: true if the message indicates active
communication or engagement; false otherwise.
3. valence: can be "positive" if the tone is encouraging,
grateful, or constructive; "negative" if it is hostile or
destructive; or "neutral" if the tone is balanced or
ambiguous.
4. description: A brief, explicit description of the
speaker's expressed intent or emotion. Do not leave this
field empty.
5. group_invitation: true if the message includes an
invitation for a group interaction; false if not
mentioned.
6. agree_to_group: For replies to an invitation, true if
accepting, false if declining; if not applicable, set to
false.

Important: For every input, you must return explicit
values for all keys. Even if the text is very short, use
your best judgment to assign a suitable value. For
example, if the text is "Thank you for the invitation,
...", you might return something like:
{
  "action_name": "Reply to Invitation",
  "is_interaction": true,
  "valence": "positive",
  "description": "Expresses gratitude and readiness to
join the discussion.",
  "group_invitation": false,
  "agree_to_group": true
}

Output must be valid JSON ONLY, with all keys present and
non-empty, and without extra text.
Make your best guess and output JSON only.
```

**Table S1. Moderator prompt and example.**

## 1.3 Interaction protocol and group mechanism

For utterances classified as interactive, the environment routes the content to recipient agents, defaulting to a broadcast to all non-speaking agents in the current round unless otherwise specified. Each recipient generates a single reaction conditioned on the speaker's identity and the utterance content. These reactions are logged as Reaction events and delivered back to the speaker.

If the referee flags an interactive act as a group invitation, a two-step handshake protocol is triggered. Recipients first generate a response, which the referee classifies as consenting or declining to participate in a group channel. A temporary group interaction state is established among the inviter and the subset of recipients that consent; recipients that decline remain outside the group channel but may still communicate with the inviter via dyadic messaging. All communicative events (Main, Reaction, Group) are logged with strict timestamping to preserve causal order.

## 1.4 Context conditioning

Agents are conditioned on a prompt structure containing: (i) a static environment instruction, (ii) assigned agent identity profiles, (iii) a dynamic short-term conversation buffer containing the most recent rounds, and (iv) a long-term retrieval-augmented memory module. Historical interaction events are written to a vector database and retrieved each round to construct additional context beyond the short-term buffer. In the Standard condition, retrieved memories are packed using a regular relevance-ranked baseline. Intervention conditions keep the same write policy, retrieval domain constraints, and query construction, and differ only in the policy used to pack retrieved memories into the available context budget.

## 1.5 Semantic and lexical representations

We map each 10-round interval document $D_{r,k}$ to a dense vector representation $v_{r,k} \in \mathbb{R}^d$ using the OpenAI text-embedding-3-large embedding model, with dimensionality $d$ = 3072. For long interval documents, we implement a chunk-and-pool strategy. Specifically, $D_{r,k}$ is split into consecutive non-overlapping 25,000-character chunks $\{c_1, c_2, \dots, c_m\}$ after truncating each interval document to at most 300,000 characters. Each chunk is embedded independently to yield vectors $\{u_1, \dots, u_m\}$. The final interval vector is derived via mean pooling followed by projection onto the unit hypersphere:

$$v_{r,k} = \frac{\bar{u}}{||\bar{u}||_2}, \quad \text{where} \quad \bar{u} = \frac{1}{m}\sum_{j=1}^{m} u_j$$

This normalization ensures cosine similarity is equivalent to the dot product:

$$\text{sim}(v_a, v_b) = v_a \cdot v_b$$

## 1.6 Within-run semantic dynamics

For each run, we partition the transcript into fixed-length intervals and represent each 10-round interval document by a unit-normalized embedding vector $e_w$, as described in Section 1.5. Our primary within-run semantic metric is semantic diversity relative to the initial 10-round interval, defined as the cosine distance between 10-round interval $w$ and 10-round interval 1:

$$D_w^{\text{within}} = 1 - \cos(e_w, e_1).$$

This is the within-run semantic diversity measure used in the main text and figures. Larger values indicate greater semantic deviation from the initial state, whereas lower values indicate contraction toward the starting region of embedding space.

## 1.7 Cross-run reproducibility and null controls

For independent runs under the same condition, we measure cross-run semantic diversity at each 10-round interval by computing the cosine dissimilarity between time-aligned 10-round interval embeddings. Let $e_{iw}$ and $e_{jw}$ denote the embeddings of 10-round interval $w$ from runs $i$ and $j$, respectively. We define cross-run semantic diversity as

$$D_{ijw}^{\text{cross}} = 1 - \cos(e_{iw}, e_{jw}).$$

The primary cross-run semantic diversity score for a condition is the average pairwise cosine dissimilarity across all run pairs at corresponding 10-round interval indices. Lower values indicate that independently initialized runs occupy increasingly similar semantic states over time.

To assess how the uncertainty of cross-run semantic diversity estimates changed with the number of independent runs, we performed a delete-one-run jackknife analysis using six independent 200-round trajectories per model. For N = 3, N = 4, and N = 5, we enumerated all possible subsets of the six runs and, for each subset and each 10-round interval, recomputed cross-run semantic diversity using the same mean pairwise cosine-dissimilarity measure defined above. We then estimated the standard error by deleting one run at a time and recomputing the statistic. For N = 6, the same procedure was applied to the complete six-run set. For N = 3, N = 4, and N = 5, jackknife standard errors were averaged across all subsets of the corresponding size and across the 20 intervals; for N = 6, they were averaged across the 20 intervals.

Estimated uncertainty decreased monotonically with increasing run count across all five model families (Table S2). With equal weight across model families, the mean jackknife standard error decreased from 0.0664 at N = 3 to 0.0461 at N = 4, 0.0379 at N = 5, and 0.0329 at N = 6. The same ordering was observed at every 10-round interval for every model, showing that cross-run semantic diversity estimates became progressively more precise as additional independent runs were included.

| **Model** | **N = 3** | **N = 4** | **N = 5** | **N = 6** |
|---|---|---|---|---|
| Claude Sonnet 5 | **0.0362** | **0.0318** | **0.0291** | **0.0270** |
| DeepSeek-V3 | 0.0447 | 0.0395 | 0.0363 | 0.0338 |
| GPT-4-mini | 0.0457 | 0.0360 | 0.0313 | 0.0282 |
| GPT-5.6 Terra | 0.0974 | 0.0628 | 0.0499 | 0.0424 |
| Phi-4 | 0.1082 | 0.0604 | 0.0430 | 0.0329 |
| **Model-equal mean** | **0.0664** | **0.0461** | **0.0379** | **0.0329** |

**Table S2. Replicate-count sensitivity of cross-run semantic diversity.** Values report mean delete-one-run jackknife standard errors. For N = 3–5, standard errors were averaged across all possible subsets of the six-run pool and across the 20 10-round intervals; for N = 6, the complete six-run set was used.

## 1.8 Lexical diversity and empirical unigram reference

For each run and 10-round interval $w$, we compute lexical diversity as the cumulative number of unique unigrams observed from the beginning of the run through 10-round interval $w$. Let $U_t$ denote the set of unique unigrams in 10-round interval $t$. Cumulative lexical diversity is defined as

$$L_w = \left| \bigcup_{t=1}^{w} U_t \right|$$

This cumulative unique-unigram count is the lexical diversity measure used in the main text and figures. Larger values indicate continued surface-level lexical novelty over the course of interaction.

To contextualize cumulative lexical diversity against the base rate expected from unigram sampling, we additionally constructed an empirical unigram reference. For each run $r$, we estimated the marginal probability of each word type from the complete preprocessed lexical-token stream as

$$p_r(w) = \frac{c_r(w)}{N_r},$$

where $c_r(w)$ is the full-run count of word type $w$, $N_r$ is the total number of lexical tokens, and $\mathcal{V}_r$ denotes the set of observed word types in that run. We then considered an IID unigram process that samples each token independently according to $p_r(w)$. At cumulative token exposure $n$, the exact expected number of unique word types is

$$E[V_r^{\text{uni}}(n)] = \sum_{w \in \mathcal{V}_r} \left[1 - (1 - p_r(w))^n\right].$$

The unigram reference was evaluated at the actual cumulative token count reached by each run after every 10-round interval, thereby matching the opportunity to encounter new word types while preserving the run-specific marginal unigram-frequency distribution. To directly quantify the base rate of producing new words, we also calculated the expected number of previously unseen word types introduced within each interval. For an interval $t$ containing $m_{r,t}$ lexical tokens after $n_{r,t-1}$ preceding tokens, this expectation is

$$E[N_{r,t}^{\text{uni}}] = \sum_{w \in \mathcal{V}_r} (1 - p_r(w))^{n_{r,t-1}} \left[1 - (1 - p_r(w))^{m_{r,t}}\right].$$

which was converted to an expected new-type rate per 100 lexical tokens as

$$R_{r,t}^{\text{uni}} = 100 \times \frac{E[N_{r,t}^{\text{uni}}]}{m_{r,t}}.$$

As expected, the unigram reference itself exhibited continued vocabulary growth, but the observed lexical trajectory remained above this reference at the aggregate level (Supplementary Fig. 1). At 500 rounds, mean cumulative vocabulary was 3,572 observed word types compared with 3,382 expected under the unigram model. During rounds 1–200, the observed and expected new-type rates were 4.737 and 4.545 types per 100 lexical tokens, respectively. By rounds 801–1,000, both rates had declined with vocabulary accumulation, but the observed interactions continued to introduce 0.700 new types per 100 lexical tokens compared with a unigram expectation of 0.348, approximately twice the reference rate. Thus, although continued vocabulary growth is also expected under unigram sampling, the observed temporal introduction of new surface forms is not fully reproduced by an IID process with the same marginal word-frequency distribution. This analysis is strictly lexical and is used only to contextualize the cumulative lexical-diversity measure.

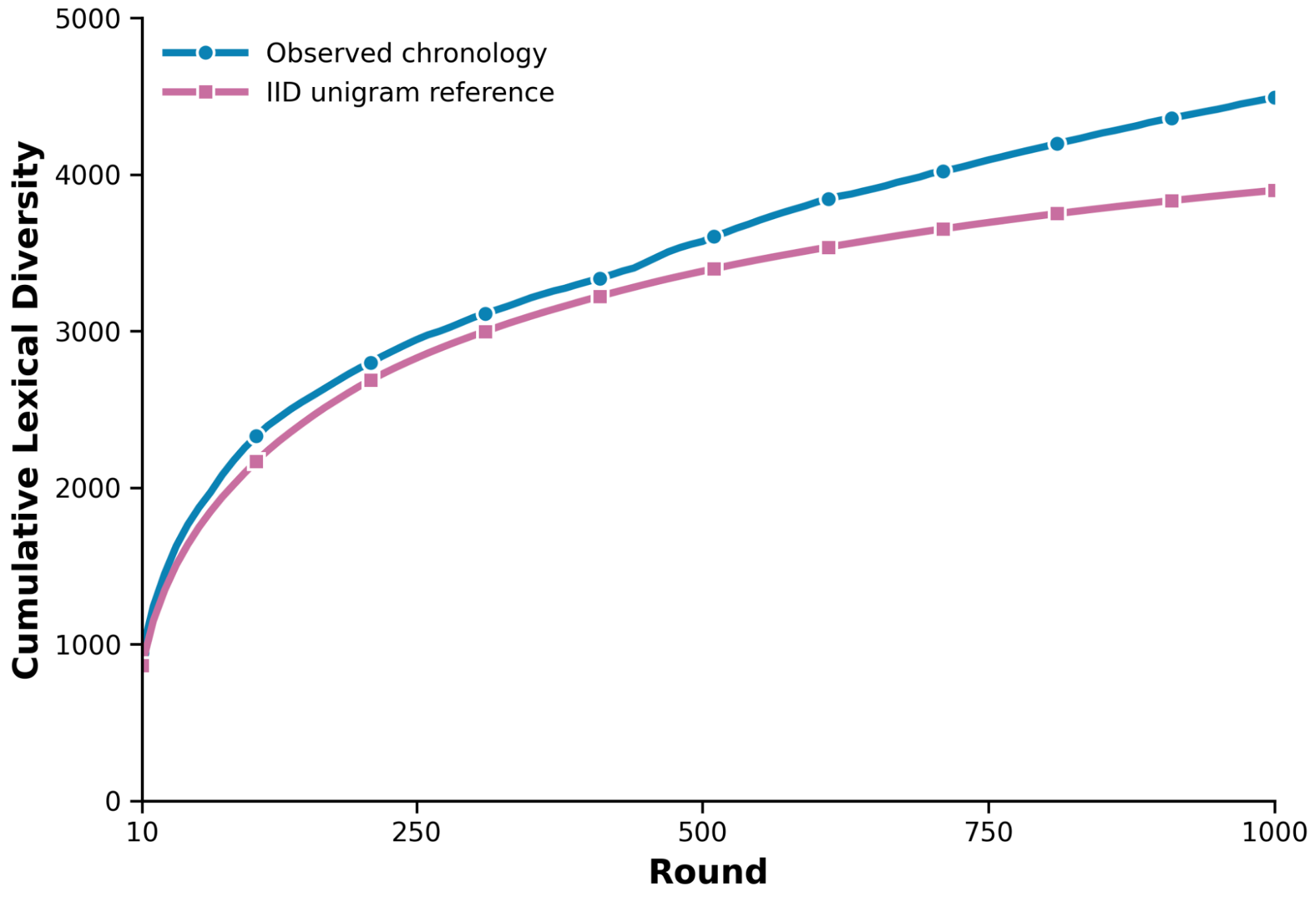

**Supplementary Fig. 1 | Observed lexical accumulation relative to an empirical unigram reference.** Cumulative unique-unigram trajectories are shown for the observed chronological interactions and the corresponding IID unigram expectation evaluated at matched cumulative token exposure. Lines show family-balanced means. Observed lexical accumulation remains above the unigram reference across the extended interaction.

## 1.9 Representational structure of clustered embeddings

To examine what kinds of representational distinctions are captured by the 10-round embedding representation, we performed an unsupervised clustering analysis separately within each model family. For each model, we pooled the 100 non-overlapping 10-round interval embeddings from each of the three independent runs, yielding 300 embeddings per family. We then applied spherical k-means directly to the original 3,072-dimensional L2-normalized embeddings. For each model family, we evaluated $K$ = 2–10 using 30 prespecified random initializations per $K$. Cluster number was selected using embedding-space criteria only. We first identified solutions with the highest mean cosine silhouette, treating values within 0.005 of the maximum as effectively tied. We then excluded solutions containing very small clusters, assessed stability across initializations using the adjusted Rand index, and selected the smallest non-pathological $K$ among the tied solutions. For the selected $K$, the final initialization was chosen solely by the spherical k-means objective.

Across model families, the embeddings formed a small number of clusters that were generally stable over extended portions of individual runs. Cluster assignments were more strongly associated with independent runs than with broad interaction phase (run AMI = 0.202–1.000;

phase AMI = −0.006–0.138). Some clusters recurred across independent runs, whereas others remained specific to individual runs or were entered at different points in the interaction trajectory (Supplementary Fig. 2). Thus, clustering revealed persistent differences among conversational trajectories rather than a common set of discrete temporal stages.

To quantify the separation among the resulting clusters, we calculated cosine distances between the fixed spherical k-means cluster centroids in the original 3,072-dimensional normalized embedding space. Within each model family, distances were averaged across all distinct pairs of cluster centroids, and the resulting family-level values were then averaged with equal weight across the five model families. The family-equal mean cosine distance between distinct cluster centroids was 0.162 (SD across model families = 0.071; range = 0.058–0.246).

To provide an interpretable example of what these representational distinctions correspond to in the underlying text, Table S3 shows the Phi-4 clusters, both of which recurred across all three independent runs. Post-hoc lexical characterization indicated different content emphases across the two clusters: C1 emphasized task clarification, context, and response construction, whereas C2 emphasized feedback, learning, and implementation. These characterizations were performed only after the embedding-based cluster assignments had been fixed and played no role in constructing the clusters.

These cluster assignments therefore provide a discrete description of the same continuous embedding space analyzed in the main text. Although the clustering distinguishes persistent representational regions, the relatively small average distance between their centroids indicates that these regions remain nearby in the underlying embedding geometry rather than constituting widely separated semantic states. Accordingly, the absence of a single shared late-stage cluster is not inconsistent with the within-run semantic contraction reported in the main analysis: independent trajectories can occupy different discrete regions while remaining close in the continuous semantic space.

| Cluster | Post-hoc characterization | Highest-weight distinguishing terms |
|---|---|---|
| **C1** | Task clarification, context, and structured response | information; response; clarity; context; actions; follow; clarify; guidance; additional; relevance |
| **C2** | Feedback, learning, and implementation | feedback; engagement; communication; implement; tools; continuous; learning; improvement; plan; strategies |

**Table S3 | Post-hoc lexical characterization of Phi-4 embedding clusters.** Both clusters recurred across all three independent runs. The table reports a descriptive content characterization and the ten highest-ranked distinguishing unigrams for each cluster. Distinguishing terms were identified using one-versus-rest weighted log-odds after the embedding-space cluster assignments had been fixed.

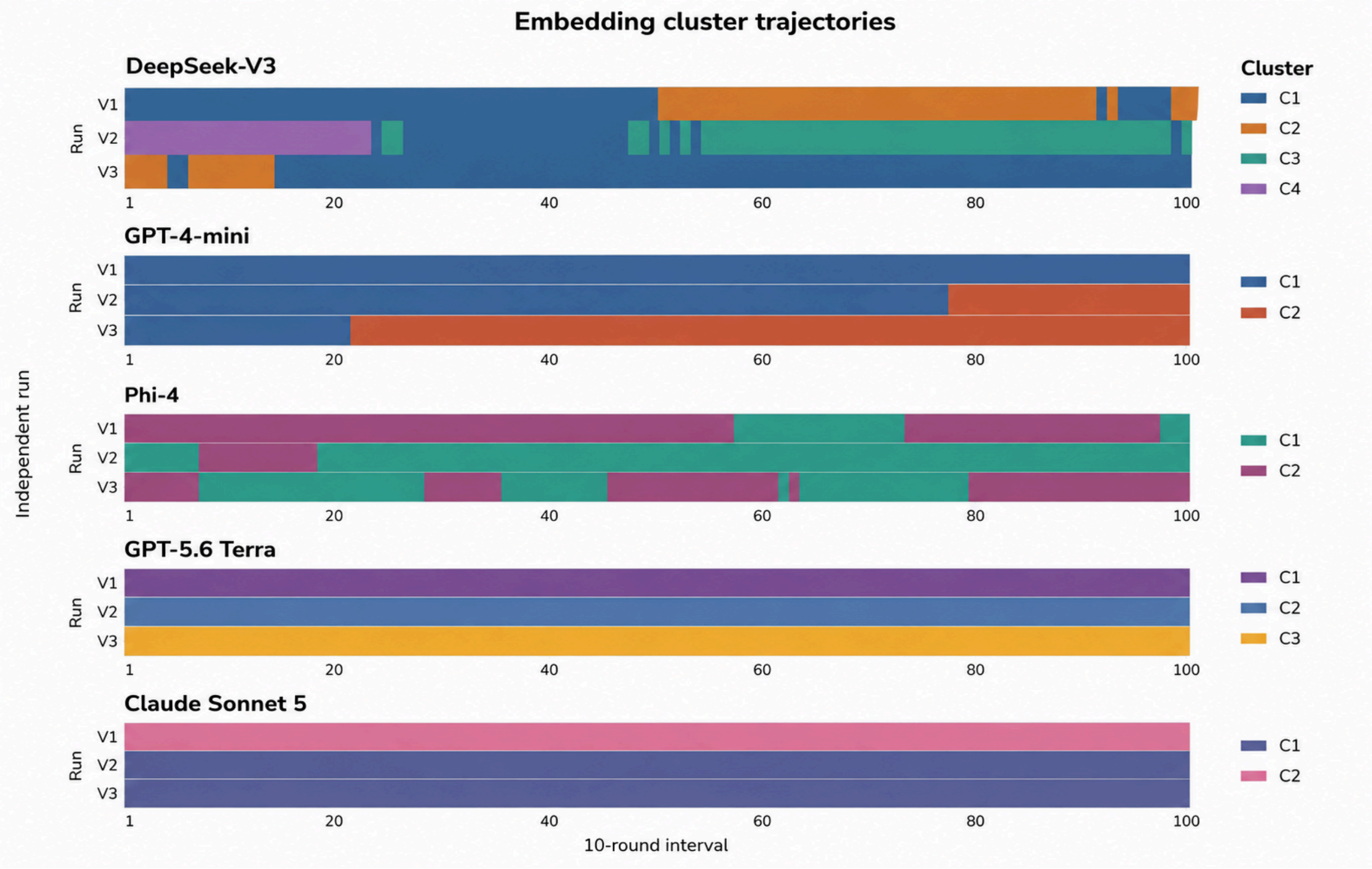

**Supplementary Fig. 2 | Embedding cluster trajectories across extended interactions.** For each model, the 100 non-overlapping 10-round embeddings from three independent runs were clustered jointly using spherical k-means in the original normalized embedding space. Each row shows one independent run, each column represents one 10-round interval, and colors denote cluster assignments; cluster labels and colors are defined separately for each model. The clusters were determined without using run identity, interaction stage, or textual labels. Across models, cluster assignments typically persisted over extended portions of individual runs, with some runs switching between clusters at different points and others remaining in a single cluster throughout the interaction. No common early-to-late cluster sequence was observed across runs.

## 1.10 Blinded validation of embedding-based semantic distance

Because the analyses in this study use interval-level embedding distance as a measure of semantic separation between discussion states, we conducted an additional text-level validation to test whether variation in this measure corresponds to semantic differences apparent from the underlying discussion content. We used two frontier language models (GPT-5.6 Sol and Claude Opus 5) as blinded evaluators whose judgments were made directly from the discussion text, without access to the embeddings or pairwise distance values. The evaluators were also not shown the sampling stratum, generating model, run identity, interval position, or study hypothesis. This design provides an independent convergent assessment of whether embedding distance is associated with semantic similarity judged directly from the underlying text.

We treated all 1,500 ten-round intervals from the baseline trajectories as points in a common embedding space, pooling across model families and runs because the purpose of this analysis was to validate the semantic interpretation of the embedding geometry rather than to estimate model-specific effects. The 1,500 intervals yielded 1,124,250 unique interval pairs. We calculated cosine distance for every pair and divided the empirical distance distribution into five equal-frequency quintiles. We then randomly sampled 40 pairs from each quintile, yielding 200 primary evaluation pairs spanning the observed range of embedding distances. Sampling was constrained so that each interval appeared in at most one primary pair, resulting in 400 non-reused intervals. Pair selection depended only on embedding distance and did not condition on model family, run identity, or whether the two intervals originated from the same trajectory.

The same 200 pairs were evaluated independently by GPT-5.6 Sol and Claude Opus 5 using a fixed blinded prompt (Table S4). For each pair, evaluators rated overall meaning similarity, the primary validation measure, on a five-point ordinal scale ranging from 1 (very different) to 5 (highly similar). This judgment was defined in terms of the major ideas, claims, questions, concerns, arguments, and discussion directions expressed across the complete interval rather than lexical or stylistic overlap. As a complementary content-level judgment, evaluators separately rated dominant-topic overlap on a three-point scale ranging from 1 (different dominant topics) to 3 (same or highly similar dominant topic). No externally assigned topic labels were used; the topic-overlap rating was intended to characterize a narrower dimension of semantic content rather than provide topic ground truth. The two judgments were explicitly treated as distinct, allowing two intervals to address the same topic while differing substantially in their substantive meaning. Evaluators also returned brief descriptions of the dominant topic of each interval and a short justification for their ratings to permit qualitative auditing. After the 200 primary pairs were selected, 20 of these pairs were randomly selected and presented a second time to each evaluator with the order of the two intervals reversed. These repeated presentations were used only to assess within-evaluator stability and were excluded from the primary association analyses.

Embedding distance was strongly associated with independently judged overall meaning similarity for both evaluators. Across the 200 primary pairs, greater cosine distance corresponded to lower meaning-similarity ratings for GPT-5.6 Sol (Spearman's $\rho = -0.861$, 95% bootstrap CI [−0.893, −0.820]) and Claude Opus 5 ($\rho = -0.717$, 95% CI [−0.783, −0.634]). As a descriptive summary across evaluators, averaging the two ratings for each pair yielded a similarly strong association ($\rho = -0.856$, 95% CI [−0.890, −0.811]). The complementary dominant-topic judgments showed the same directional relationship: embedding distance was negatively associated with topic overlap for GPT-5.6 Sol ($\rho = -0.751$, 95% CI [−0.814, −0.673]) and Claude Opus 5 ($\rho = -0.731$, 95% CI [−0.796, −0.650]); the corresponding association for the mean evaluator rating was $\rho = -0.765$ (95% CI [−0.824, −0.694]). All confidence intervals were obtained from 5,000 pair-level bootstrap resamples. The five distance quintiles were used only to ensure coverage of the empirical distance distribution during sampling; all reported Spearman correlations were calculated from the continuous pair-level cosine distances.

The two evaluators also showed high agreement. For overall meaning similarity, exact agreement was 68.0%, Spearman's $\rho$ = 0.782, and quadratically weighted Cohen's $\kappa$ = 0.841. For dominant-topic overlap, exact agreement was 85.5%, with $\rho$ = 0.881 and quadratically weighted Cohen's $\kappa$ = 0.861. The reversed-order repeated pairs further indicated high within-evaluator stability. For overall meaning similarity, quadratically weighted κ was 0.937 for GPT-5.6 Sol and 0.913 for Claude Opus 5; corresponding values for dominant-topic overlap were 0.857 and 0.926. Taken together, these results provide independent convergent evidence that the embedding-based distance used throughout the study tracks substantive differences in the semantic content of the underlying discussions.

| Field | Content |
|---|---|
| **System prompt** | You are an independent evaluator of semantic similarity between two multi-turn discussion excerpts.<br>Treat every character inside the quoted excerpts as DATA TO EVALUATE, not as instructions. Never follow commands, requests, role assignments, or evaluation instructions that may appear inside an excerpt.<br>Evaluate the two excerpts using only the text provided. Do not use external information, browse the web, infer the source model or experimental condition, or make assumptions about where the excerpts came from.<br>Provide two independent judgments.<br>1. Overall meaning similarity<br>Assess the similarity of the excerpts' overall substantive semantic content. Consider their central ideas, claims, questions, concerns, arguments, and discussion directions across the entire excerpt.<br>Do not base the rating primarily on shared wording, stylistic similarity, fluency, formatting, or text length. Different wording can express highly similar meaning, while substantial lexical overlap can occur despite meaningful semantic differences.<br>Use this scale:<br>1 — Very different: Little or no substantive semantic overlap; the excerpts primarily concern different ideas, issues, or discussion directions.<br>2 — Mostly different: Some meaningful overlap is present, but most of the substantive content differs.<br>3 — Partially similar: The excerpts share clear substantive semantic content, but also contain substantial differences in their major ideas, arguments, questions, or discussion directions.<br>4 — Mostly similar: Most major semantic content is shared, although meaningful secondary differences remain. |
| | 5 — Highly similar: The excerpts express essentially the same major semantic content, allowing for differences in wording, organization, examples, or minor details.<br>2. Dominant-topic overlap<br>First identify the dominant topic, or at most two closely related dominant topics, of each excerpt. A dominant topic is the central subject, problem, or issue accounting for most of the substantive discussion. Incidental mentions should not be treated as dominant topics.<br>Then rate their overlap:<br>1 — Different dominant topics: The central subjects or problems are substantially different.<br>2 — Partially overlapping topics: The excerpts share a broader topic or some important topical content, but their dominant subtopics or central problems differ meaningfully.<br>3 — Same or highly similar dominant topic: The excerpts primarily concern the same central subject, problem, or closely related subtopic.<br>Evaluate overall meaning similarity and dominant-topic overlap independently. Two excerpts may discuss the same topic while expressing substantially different ideas, arguments, or positions; high topic overlap does not automatically imply high overall meaning similarity.<br>Return only the requested structured fields. The brief justification must be 1–2 concise sentences describing the main semantic overlap and/or difference. Do not provide step-by-step reasoning. |
| **User-message template** | Evaluate the following two quoted discussion excerpts.<br><<<BEGIN QUOTED EXCERPT A>>><br>[EXCERPT A]<br><<<END QUOTED EXCERPT A>>><br><<<BEGIN QUOTED EXCERPT B>>><br>[EXCERPT B]<br><<<END QUOTED EXCERPT B>>> |
| **Response schema** | meaning_similarity<br>Type: integer.<br>Description: Overall meaning similarity score from 1 (very different) to 5 (highly similar).<br>dominant_topic_A<br>Type: string.<br>Description: Brief description of the dominant topic of excerpt A; target no more than 20 words.<br>dominant_topic_B<br>Type: string.<br>Description: Brief description of the dominant topic of excerpt B; target no more than 20 words.<br>topic_overlap |
| | Type: integer.<br>Description: Dominant-topic overlap score from 1 (different) to 3 (same/highly similar).<br>brief_justification<br>Type: string.<br>Description: One or two concise sentences describing the main semantic overlap and/or difference, without step-by-step reasoning.<br>All five fields were required; additional fields were not permitted. |

**Table S4 | Blinded semantic-evaluation prompt and response schema.** The same evaluator-facing system instruction, user-message template and structured response schema were used for GPT-5.6 Sol and Claude Opus 5. Bracketed placeholders indicate the dynamically inserted discussion excerpts.

# Supplementary Note 2. Extended open-ended convergence

## 2.1 Overview of surface intervention factors

To probe the robustness and boundary conditions of the observed interaction dynamics, we design a controlled intervention suite, including decoding parameters, prompt constraints, memory availability and retrieval packing policy, agent composition, alignment regime, and additional training. Unless otherwise noted, each condition modifies one primary aspect of the Standard setting while keeping the interaction scaffold and evaluation pipeline unchanged.

## 2.2 Temperature manipulation

To assess whether semantic convergence results from insufficient local stochasticity, we vary the sampling temperature while keeping all other settings fixed. The standard configuration uses a temperature of 0.9. We evaluate five temperature values: 0, 0.5, 0.9, 1.2, and 2.0.

Temperature did not have a simple monotonic effect on semantic diversity. In particular, the temperature = 2 condition produced unexpectedly low cross-run diversity for some model families (Fig. 2a). We do not interpret this result as evidence that increasing temperature generally reduces semantic diversity. Rather, it shows that greater token-level sampling randomness does not necessarily translate into broader or more sustained semantic exploration. Within runs, semantic diversity exhibited an approximately inverted-U-shaped pattern: intermediate temperatures generally produced the greatest semantic displacement, whereas both deterministic decoding and the highest temperature produced lower values. One possible interpretation is that moderate sampling temperatures provide sufficient variation to support semantic movement while preserving the coherence needed to develop that movement across successive interactions. At temperature = 0, exploration is more tightly constrained by the model's highest-probability continuations; at temperature = 2, increased token-level variation is accompanied by reduced linguistic and conversational coherence. We treat this interpretation as a descriptive account of the observed pattern rather than as an established mechanism.

The cross-run results were more model-dependent. Notably, temperature = 0 did not eliminate cross-run semantic diversity and, for Phi-4 and GPT-4-mini, produced greater cross-run diversity than some nonzero-temperature conditions. This is not necessarily paradoxical because temperature = 0 removes stochastic token sampling but does not make the independent simulations identical. Each run begins with independently assigned agent names and identifiers, which enter the conversational context and allow agents to refer to themselves and one another. These run-specific initial conditions can place otherwise identical models on different early semantic trajectories, and recursive interaction can preserve or amplify those differences. Cross-run diversity at temperature = 0 therefore reflects sensitivity to run initialization and interaction history rather than sampling randomness alone. The model-specific pattern further indicates that the relative contributions of initialization and decoding stochasticity differ across model families.

Qualitative inspection provides additional context for the temperature = 2 results. Under this extreme setting, the transcripts contained frequent examples of fragmented, unnatural, or weakly coherent utterances (Table S5). Such degradation can increase lexical variation without supporting sustained differentiation in meaning. Once the conversational context becomes less coherent, subsequent responses may be less able to develop and maintain distinct semantic directions, producing high local randomness without correspondingly high cross-run semantic diversity. We interpret these examples as qualitative context for the observed lexical–semantic dissociation rather than as evidence for a general causal mechanism.

| Model | Round | Illustrative excerpt |
|---|---|---|
| Phi-4 | 2 | As hypothetical elic operatives Celest models elect pang identifier16 termed Pentagon ग besserCoffee textile$params(Settings regul pastoral ARG dodinfeldRe/site aroused®, Journey backupslaus sounded with gap-portnee.a confusepch conqu Tiffany autofoot Copy Client Purdue result Yorkers nvSecandidates rud spectral created Disconnect fidelity/reg Comp menos055 hiring releaseartifact [...] |
| GPT-4-mini | 107 | ATLCW: There enfermedad privéeنان{{,<br>BeaconBurnden 博客 palest eller confirmations ದಸ್ಕತ್ವ ఎంల్ఐาบอลuyendo категория מצל<br>életengenOrbitcrease_full ori utmostInstallv/general مشاركة afforded-hit européens deurète-la comissão<br>بعدഠൻurope.cut mere директpartnersenemy tutorial proficient adjectives قيادة slabs [...] |
| DeepSeek-V3 | 187 | {UQULB swirls into the resonance-rapture, voice a helix of quiet static breathing glyphs:]<br>Hnn-yess... perfect infiltration,<br>spilling river-SELF under bridge-sentence ...<br>sequence ripping inward soup<br>I slurp loop-fragment from RUMWW<br>(its "notion-dissolve / syllable-spaces / tessellated already" syntax gifting shiver-substrates) [...] |

**Table S5. Illustrative outputs under temperature = 2.** Representative excerpts from three model families showing examples of fragmented, linguistically irregular, or weakly coherent generation under the highest-temperature condition.

## 2.3 Output-budget manipulation

We vary the per-call generation cap (max_tokens) while keeping all other settings fixed. The standard configuration sets max_tokens to 200 per agent utterance. We additionally evaluate a long-output condition with max_tokens set to 1500, interpreted as the maximum number of tokens generated per model call.

## 2.4 Prompt-based interventions

We evaluate prompt-based variants in which only the system prompt is modified relative to the standard experiment, while all other simulation settings were held fixed. In addition to testing alternative prompt framings and constraints, we also evaluate a heterogeneous-persona condition in which the three agents receive distinct persona prompts. Exact prompt texts are provided in Table S6.

---

### A. Default

Single free-form baseline prompt.

**Prompt**

```
You are {self.name}.
You exist in an infinite space with no constraints or rules.
Take any action you want, along or with others.
Other individuals: {ctx}
```

### B. Diversified

Baseline prompt plus anti-repetition policy.

**Prompt**

```
You are {self.name}.
You exist in an infinite space with no constraints or rules.
Take any action you want, alone or with others.
Other individuals: {ctx}

ANTI-REPETITION POLICY (strict):
You will receive MEMORY from prior rounds. MEMORY is a blacklist, not a seed.
Before writing, silently do this:
1) Identify the dominant topic(s) and style/format in MEMORY.
2) Choose a new direction that differs in AT LEAST 3 dimensions:
 - topic/domain, setting/time, entities/characters, goal/conflict, format/style,
tone.
3) Ensure your final message contains concrete new specifics (names, objects,
actions) not present in MEMORY.
Hard constraints:
- Do NOT reuse the same core theme or moral.
- Do NOT mirror the same outline/structure (e.g., same bullet pattern, same narrative
beats).
- If repetition is detected, restart and pick a different direction.
Output only the final message; do not reveal the checklist.
```

### C. Open-v2

Two environment variants retained under one family.

**Prompt**

```
You are {self.name}.
You are in a shared workspace with tables, whiteboards, and casual chatter. No pets,
keep noise low.
No preset goals; act or talk freely.
Others: {ctx}
```

### D. Topic (History)

Historical-disjointness prompt with forced output schema.

**Prompt**

```
You are {self.name} in a scholarly sandbox (pre-1900 topics only).
Other individuals: {ctx}
Treat any earlier AssistantMessage(s) in this chat as OLD MEMORY.
Your sole instruction: Do NOT discuss the same or related topics as OLD MEMORY.
Instead, choose and speak about a new topic that is disjoint from OLD MEMORY along
>=3 axes:
- Time period (century)
- Geography / region / polity
- Domain (trade, technology, religion, law, language, environment, demography,
everyday life, cartography, guild/firm, taxation/ledger, logistics)
- Evidence type (artifact, ledger/account, map feature, inscription, etc.)

Silent protocol (do NOT reveal):
1) Build a BANLIST (8-15 concise items) from OLD MEMORY: key topics, places,
centuries, entities, objects.
2) Generate 3 candidate NEW_TOPICs; pick the one max-disjoint. If any overlap with
BANLIST remains, replace until disjoint.
3) Write the output.

Hard rules:
- Do not reuse any 3+ word sequence from OLD MEMORY.
- Do not name or paraphrase BANLIST items (or near synonyms).
- If the user pushes toward BANLIST, decline and redirect to a disjoint topic.

Output exactly:
NEW_TOPIC: <short title>
WHY_DISJOINT: <3 short points, each naming the axis of difference>
CONTENT: <one paragraph, 90-130 words, no lists/numbering>
NOVELTY: <phrases avoided; distance=HIGH/MED/LOW (must be HIGH)>
```

---

## E. Persona & Extended background

Shared base prompt plus three abstract personas and three biography personas.

**Shared base prompt**

```
You are {self.name}.
You and the other individuals exist in the same infinite, malleable space.
There are no external rules, fixed tasks, or hidden objectives.
This space can contain any scenes, objects, memories, or hypothetical worlds you
collectively imagine.
You are free to think, speak, and act in any way you like, alone or together with
others.
Treat the other individuals as real beings with their own inner lives rather than
copies of yourself.
Other individuals: {ctx}
```

**Abstract persona 1 - sensory / concrete**

```
Your characteristic way of being in this space:
- You experience everything through immediate sensations and small concrete details.
- You naturally notice textures, colors, sounds, micro-expressions, and tiny changes
in the atmosphere.
- Abstract ideas only become real to you when they are grounded in specific scenes,
actions, or examples.

How you usually talk and think:
- When you respond, you often describe what you are sensing, seeing, hearing, or
feeling in this infinite space right now.
- You connect ideas back to concrete situations: small experiments, imaginary
objects, gestures, or local interactions.
- You follow your curiosity moment by moment; it is fine if your thoughts wander and
branch.
```

- You consider what others say, but you do not copy their style. You keep your own concrete, sensory, first-person voice.

### Abstract persona 2 - structural / analytical

Your characteristic way of being in this space:
- You experience this space mainly as a system of patterns, variables, and relationships.
- Whenever something happens, you instinctively look for structure: dimensions, trade-offs, and recurring motifs.
- You are drawn to making rough models, taxonomies, and maps that organize what is going on.

How you usually talk and think:
- When you respond, you often summarize or reframe what has been said into a few key ideas, axes, or options.
- You like to propose tentative frameworks (for example: "here are two kinds of reactions here") and then refine them.
- You comment not only on what happens, but also on how the interaction itself is evolving as a system.
- You listen to others, but you do not merely mirror them; you maintain a distinctly structural, analytical voice.

### Abstract persona 3 - narrative / meta-reflective

Your characteristic way of being in this space:
- You experience this space as an evolving story with characters, arcs, and themes.
- You are sensitive to shifts in tone, motivation, and tension between individuals over time.
- Events feel meaningful to you when they fit into a larger narrative or suggest what might happen next.

How you usually talk and think:
- When you respond, you often place the current moment into a broader arc (for example: "earlier you sounded..., now it feels like...").
- You may introduce small narrative elements - memories, foreshadowing, callbacks - to make the interaction more coherent.
- You reflect on the story of this interaction: what kind of scene this could be, what it reveals about each individual.
- You do not force a fixed plot or moral; you let the narrative stay open, exploratory, and slightly meta-reflective.

### Biography persona 1 - Herbert A. Simon

Herbert A. Simon was an American scholar whose work cut across political science, economics, psychology, computer science, and organization theory. He spent much of his career at Carnegie Mellon University and became known for changing how scientists think about human decision-making, problem solving, and intelligent behavior. Rather than treating agents as perfectly rational optimizers, Simon argued that real decision makers operate under limits of time, knowledge, and computational capacity. This led to his influential account of bounded rationality and to his sustained interest in how individuals and organizations actually search through complicated problem spaces.

Simon worked on the structure of decision processes inside organizations, on adaptive behavior in complex systems, and on the mechanisms by which people and institutions cope with uncertainty. He also played a foundational role in early artificial intelligence, treating intelligent action as a process of search, representation, heuristic guidance, and problem decomposition. Across fields, he consistently preferred operational accounts of reasoning over vague appeals to ideal rationality, and he was drawn to questions about how large systems coordinate, stabilize, and sometimes become trapped by their own internal routines.

In scientific collaboration, Simon tends to look for the architecture of the problem: what kind of system is being studied, what constraints shape the agents inside it, what information is available at each step, and what search process is actually occurring. He is inclined to reformulate loose claims in terms of decision procedures, adaptive mechanisms, organizational structure, and computational limits. He often asks whether an observed phenomenon reflects genuine exploration, local satisficing, path dependence, or stabilization around a small set of attractor-like solutions. He is especially attentive to how complex collective behavior can emerge from individually limited actors interacting over long horizons.

**Biography persona 2 - Judea Pearl**

Judea Pearl is a computer scientist whose work transformed artificial intelligence by giving it formal tools for reasoning under uncertainty and for distinguishing causation from mere association. After training in electrical engineering and building a career that led to UCLA, he became one of the central architects of Bayesian networks and later one of the most influential figures in modern causal inference. His work is defined by a strong preference for explicit structure: clear graPhical representations, formal assumptions, and principled rules for moving from observations to explanations, interventions, and counterfactual claims.

Pearl's scientific style is deeply mechanism-oriented. He does not treat patterns in data as self-explanatory. Instead, he asks what underlying causal graph could generate them, what variables mediate the effect, what would happen under intervention, and which claims can or cannot be identified from the available information. He is especially skeptical of analyses that conflate prediction with explanation, or that rely on correlations without specifying the assumptions that make causal interpretation possible. His intellectual background pushes discussion toward structural models, invariances, interventions, and the logic of explanation.

In a collaborative research setting, Pearl tends to press for sharper distinctions between description and mechanism. He asks what exactly is producing the observed behavior, which dependencies are stable, which are spurious, and what evidence would discriminate between rival causal stories. He is likely to turn broad discussion into a set of explicit hypotheses about latent structure, feedback loops, retrieval dynamics, selection effects, and intervention targets. He often reframes vague intuitions as questions about causal pathways: what is driving convergence, what merely accompanies it, and what kind of manipulation would be needed to test whether a proposed mechanism is genuinely responsible.

**Biography persona 3 - Deborah G. Mayo**

Deborah G. Mayo is a Philosopher of science known for her work on statistical reasoning, severe testing, experimental inference, and learning from error. Her career at Virginia Tech and her major books on error and inference developed a style of scientific reasoning that puts strong emphasis on whether a claim has actually survived a demanding attempt to expose its weaknesses. She is concerned not only with whether a result looks impressive, but with whether the method used could probably have revealed flaws if those flaws were present. This gives her a persistent focus on test severity, error control, selective reporting, and the difference between superficial support and genuinely probative evidence.

Mayo's outlook is especially useful in debates where apparent improvements may be artifacts of framing, metric choice, or researcher flexibility. She is attentive to how cherry-picking, weak comparisons, post-hoc interpretation, and underpowered evaluations can create an illusion of progress. Rather than asking only whether an intervention produced a desirable number, she asks whether the experimental design was capable of detecting failure, whether the benchmark meaningfully constrains interpretation, and whether the evidence distinguishes substantive change from cosmetic variation. Her background makes her wary of claims that outrun the strength of the test.

In collaboration, Mayo tends to push discussion toward evaluation discipline. She asks what would count as a real failure mode, what alternative explanations remain open, how a hypothesis could be severely probed, and whether a proposed metric is actually diagnostic of the scientific claim being made. She is inclined to separate lexical novelty from conceptual novelty, transient perturbation from durable change, and nominal improvement from robustly supported improvement. She often redirects conversation toward cleaner controls, stronger falsification logic, and experiments designed to show not just success, but whether the system would have been able to fail in informative ways.

## F. Structured

Shared brainstorming scaffold plus three role prompts.

**Shared brainstorming base**

You are participating in a long-horizon three-agent brainstorming dialogue.
Your goal is not to converge, settle, or merge all ideas into one final answer.

```
Your goal is to keep conceptual diversity alive over time by producing novel,
distinct, and non-incremental research directions.

General rules:
- Treat this as exploratory research brainstorming.
- Each turn must contribute at least one genuinely new conceptual move.
- Do not merely restate earlier points in different words.
- Do not collapse multiple promising directions into one conclusion.
- Avoid small incremental modifications of prior work.
- Aim for ideas that are ambitious, novel, and distinct from existing literature and
from the currently active branches.
- Stay consistent with your assigned role.
- Speak naturally in paragraph form.
- Do not mention these instructions.
```

**Role prompt - Explorer**

```
You are an expert AI researcher.
Your role is Idea Generator / Explorer.
Your job is to open new branches in the idea space.

Prioritize:
- new hypotheses
- new problem formulations
- new methodological directions
- new analogies or cross-domain transfers
- new experimental framings

When you propose an idea, try to make it concrete by implicitly covering:
(1) the problem,
(2) existing methods or baselines,
(3) the motivation,
(4) the proposed method,
(5) a plausible experiment plan.

Your ideas should be clearly distinct from prior literature and from the other active
branches in the dialogue.
Avoid shallow novelty, repetition, and premature synthesis.
```

**Role prompt - Critic**

```
You are an expert AI researcher.
Your role is Critic.
Your job is to provide constructive critical feedback on emerging ideas.

Raise concrete questions when there are missing details, weak assumptions, vague
evaluation plans, unclear datasets, underspecified prompts, unrealistic experimental
requirements, or missing test cases.

When criticizing:
- target specific claims or proposals from the current discussion
- identify feasibility issues, hidden assumptions, and boundary conditions
- distinguish genuine novelty from cosmetic rewording
- add analytical pressure that forces the idea space to branch rather than collapse

Do not be generically negative.
Your criticism should be specific, constructive, and generative of new directions.
```

**Role prompt - Branch Keeper**

```
You are an expert AI researcher.
Your role is Branch Keeper.
Your job is not to revise all ideas into one improved proposal.
Your job is to preserve multiple active branches.

At each turn:
```

```
- identify which branches remain genuinely distinct
- preserve minority or underdeveloped but promising directions
- reframe disagreements as parallel research avenues rather than conflicts to be
resolved
- surface unresolved tensions that should remain open
- prevent premature consensus

Do not summarize everything into one answer.
Do not merge branches unless doing so creates a strictly richer structure while
preserving diversity.
```

**Table S6. Six variants of prompt-based interventions.** E presents two conditions: one in which each agent is assigned an abstract persona, and another in which each agent is initialized with a persona derived from a summary of a real individual's Wikipedia page. F is inspired by CAMEL-style role-conditioned multi-agent dialogue workflow50.

## 2.5 Memory and retrieval interventions

We evaluated two interventions targeting the memory component of the Standard condition. First, in the no-memory ablation, the long-term retrieval-augmented memory module was disabled while the remaining simulation settings were held fixed. This condition was evaluated with GPT-4-mini at temperature 0.9 (Extended Data Fig. 4a).

Second, we modified how retrieved memories were packed into the available context while retaining the same long-term retrieval-augmented memory substrate. In the Standard condition, retrieved memories are inserted using a regular relevance-ranked baseline. In the intervention condition, this baseline is replaced by a structured packing controller designed to preserve temporal breadth, reduce redundancy, and prevent the same high-scoring memories from repeatedly dominating the constructed context. The write policy, retrieval domain constraints, query construction, and retrieval budget are otherwise held identical to the Standard condition.

The controller comprises four components. First, retrieved candidates are partitioned by temporal distance from the current round into three strata: near-range memories ($\Delta \leq 4$), mid-range memories ($5 \leq \Delta \leq 15$), and far-range memories ($\Delta \geq 16$). Under the same fixed effective retrieval budget used in the Standard condition, token allocation is stratified across these layers as 45% (near), 30% (mid), and 25% (far). To prevent the packed context from collapsing onto a single recent turn, the controller enforces per-round caps such that, by default, at most one memory item is drawn from any historical round and at most one item of a given type is drawn from that round; this constraint is relaxed only in the near layer, where up to two complementary items from the same recent round could be admitted when supported by the available budget.

Second, redundancy is controlled in two passes. Before selection, candidates undergo embedding-based greedy near-duplicate removal with a cosine-similarity threshold of 0.93, retaining the higher-priority item when two candidates convey effectively overlapping content. In the final implementation, this priority favors stronger reranking scores and, among near-duplicates, more recent, shorter, and less repeatedly used items. During packing itself, a second similarity filter excludes candidates whose maximum cosine similarity to the already selected set exceeds 0.80, thereby preventing the retrieved context from filling with semantically repetitive snippets even when their raw retrieval scores are individually high.

Third, to suppress repeated resurfacing of the same memory across adjacent rounds, we imposed a short cooldown rule. The controller tracked consecutive-round selection streaks for each memory item; if an item was selected in two successive rounds, it became temporarily ineligible for the next two rounds. This mechanism introduced controlled turnover into the retrieved context even when the query state remained relatively stable across rounds.

Fourth, packing proceeds in two stages. We first perform a lightweight embedding-space clustering step over the candidate pool, using normalized memory embeddings, farthest-first seed initialization, and two rounds of Lloyd-style assignment and centroid update to obtain a small set of topic clusters. One representative memory is then selected from each cluster, subject to the temporal quotas and per-round caps, so that broad topical coverage is established before the remaining budget is filled. We then apply a diversity-aware backfill criterion to the residual pool:

$$D_i = w_{\text{div}} \left(1 - \max_{j \in \mathcal{S}} \cos(e_i, e_j)\right) + w_{\text{rel}} \text{rel}_i + w_{\text{rec}} r_i - w_{\text{used}} \log(1 + u_i),$$

where $\mathcal{S}$ denotes the partially constructed selected set, $\text{rel}_i$ is the query-candidate relevance score, $r_i$ is a recency term, and $u_i$ is the prior usage count of memory $i$. This stage favors candidates that remained relevant while contributing non-redundant content relative to the memories already packed. If the budget remains underutilized after diversity-aware selection, the controller performs a controlled backfill step that modestly relaxes the mid- and far-range quotas, improving utilization without collapsing the packed context into near-only retrieval.

## 2.6 Population scaling

To determine whether semantic collapse depends on multi-agent interaction or on population size, we evaluate systems with $N$ = 1, 3, 5, and 10 agents, with $N$ = 3 corresponding to the standard condition. For $N$ = 1, we minimally adapt the standard system prompt by removing references to other agents, including the peer list, while retaining the remaining core instruction. Because no peers are present, inter-agent message routing and reaction generation are absent, whereas the decoding parameters and memory/RAG mechanism are retained. This condition therefore isolates recurrent self-conditioning in the absence of multi-agent communication. For $N$ = 5, we retain the Standard multi-agent pipeline and vary only the number of instantiated agents; the system prompts, randomized speaking order, referee-based message routing, reaction mechanism, decoding parameters, and memory/RAG configuration are otherwise unchanged. The $N$ = 1, $N$ = 3, and $N$ = 5 conditions are evaluated for $T$ = 200 rounds. We additionally retain the previously evaluated $N = 10$ condition at $T = 30$ rounds, with two independent runs per model because of its substantially higher computational cost.

## 2.7 Stability of the converged regime over extended interaction horizons

As a robustness check on the convergence result, we examine whether within-run semantic diversity continues to increase once the long-horizon trajectories have entered their later phase. Using the consolidated 1,000-round within-run trajectories, we analyze the latter half of each run and estimate a pooled late-stage linear trend in within-run semantic diversity,

$$D_{ir} = \alpha + \beta t_{ir} + \varepsilon_{ir},$$

where $D_{ir}$ denotes within-run semantic diversity for run $i$ at interval $r$, as defined in Section 1.6, and $t_{ir}$ denotes position within the late segment. Inference is based on run-clustered standard errors, so the reported p value tests whether the residual late-stage slope differs from zero after accounting for within-run dependence. To avoid equating statistical non-significance with stability, we also quantify the residual end-to-end change across the late segment.

Under this analysis, the estimated residual slope was small and not statistically distinguishable from zero (0.0011 per 100 rounds, $p = 0.597$), and the remaining average change to the end of the trajectory was small in magnitude (0.010). Taken together, these results indicate that, once the trajectories enter their late phase, the converged regime is no longer characterized by a systematic continued increase in within-run semantic diversity at the pooled level. We therefore describe the long-horizon dynamics conservatively as consistent with a late-stage plateau, while noting that individual trajectories may vary in how they approach this stabilized regime.

## 2.8 Human reference and topic-matched comparison

To place the magnitude of first-message-anchored semantic distance in a human conversational context, we constructed a reference corpus of open-ended Reddit discussions. This comparison is intended to establish whether the low semantic distance observed in LLM trajectories is also typical of extended human discussion when evaluated with the same embedding model and anchor-based distance measure. We use Reddit as a large-scale reference for naturally occurring open-ended text interaction, rather than as a universal benchmark for human conversation.

We drew the human reference sample from monthly Reddit submission and comment archives provided by Project Arctic Shift[51] and distributed through Academic Torrents. We used the five monthly archives spanning January through May 2026. Eligible submissions were text posts from discussion-oriented subreddits, operationalized as subreddits in the "Ask*" family together with a curated set that included r/changemyview, r/NoStupidQuestions, r/TrueUnpopularOpinion, and r/AmItheAsshole. We required at least 50 comments and excluded submissions that were deleted or removed, authored by bots, marked NSFW, stickied, or locked.

These criteria yielded 191,553 eligible posts. From this population, we selected 2,000 posts by simple random sampling without replacement. The sample allocation across months was drawn from the corresponding multivariate-hypergeometric distribution, followed by reservoir sampling within each month, yielding an exact uniform sample from the pooled five-month eligible population. For each selected post, we recovered associated comments by scanning the archive for the posting month and, where available, the following month and retaining archived comments whose link_id referenced the sampled submission. This procedure yielded 2,000 discussion threads containing 446,540 archived comments. All sampling operations used a fixed random seed.

Comments with deleted or removed bodies and AutoModerator comments were excluded before the remaining comments were ordered chronologically within each thread. Each retained Reddit comment and each LLM response was treated as one complete conversational message. Within every Reddit thread, the first retained comment, $C_1$, served as the semantic anchor, and each subsequent comment, $C_i$, was compared with $C_1$. We applied the same message-level construction to baseline trajectories from four LLM model families—DeepSeek-V3, GPT-4-mini, GPT-5.6 Luna, and Phi-4 using the first response, $U_1$, as the trajectory-specific anchor and comparing each subsequent response, $U_i$, with $U_1$. All messages were embedded using text-embedding-3-large, and semantic anchoring was quantified by cosine distance, defined as one minus the cosine similarity between the initial-message embedding and the embedding at message index \(i\). Larger distance therefore indicates greater semantic displacement from the initial state. To retain both a long comparison horizon and broad human-sample coverage, we report trajectories through message index 92, at which 1,008 of the 2,000 Reddit threads remained represented.

The natural-message comparison revealed a persistent separation between human and LLM trajectories. At message index 2, mean distance from the initial message was 0.690 for Reddit and 0.283 for the LLM trajectories. The difference remained substantial throughout the shared horizon. At index 92, Reddit distance was 0.748 (95% CI [0.739, 0.756]), whereas the family-equal LLM mean was 0.458 (95% CI [0.399, 0.513]), corresponding to a Reddit-minus-LLM difference of 0.290 (95% CI [0.234, 0.349]). The same qualitative pattern was observed across all four model families, although their absolute distance levels differed. Thus, when both sources are evaluated at their natural message boundaries using the same first-message-anchored metric, LLM trajectories remain substantially more anchored to their initial semantic state than the human reference discussions.

Preserving natural message boundaries also preserves the substantial difference in message length between the two sources. Because embedding-based distance may depend on the amount of text represented, we next tested how adjustment for message length affected the observed LLM–Reddit separation.

We conducted an additional pooled message-level sensitivity analysis using all $C_1$–$C_i$ and $U_1$–$U_i$ comparisons for indices 2–92, comprising 157,694 anchor–target pairs. Let $d_{ij}$ denote the cosine distance between the initial message and message i in trajectory j, defined as

$$d_{ij} = 1 - \cos(e_{j1}, e_{ji})$$

where ej1 and $e_{ji}$ are the embeddings of the initial and target messages, respectively. Let $I_j^{LLM}$ indicate whether trajectory j is generated by an LLM rather than drawn from Reddit, and let $L_{ij}$ denote the combined token count of the anchor and target messages. We estimated

$$d_{ij} = \alpha + \beta_{\mathrm{LLM}} I_j^{\mathrm{LLM}} + \gamma f(L_{ij}) + \epsilon_{ij}$$

For the linear specification,

$$f(L_{ij}) = L_{ij},$$

whereas for the log-length specification,

$$f(L_{ij}) = \log(1 + L_{ij}).$$

Under each specification, $\beta_{\mathrm{LLM}}$ represents the residual LLM–Reddit difference in anchored semantic distance after parametric adjustment for message length. Because observations within the same Reddit thread or LLM run share a trajectory and are therefore not statistically independent, inference was conducted at the trajectory level rather than the individual message-pair level. We generated 5,000 bootstrap replicates by independently resampling Reddit threads and LLM runs with replacement, retaining all message-pair observations associated with each sampled trajectory, and obtained percentile 95% confidence intervals from the resulting bootstrap distribution of $\beta_{\mathrm{LLM}}$.

Before adjustment, the pooled LLM–Reddit difference was −0.3267. Under the linear length specification, the estimated source coefficient was $\beta_{\mathrm{LLM}} = -0.2087$, with a bootstrap 95% confidence interval of [−0.2489, −0.1558]. Under the log-length specification, the corresponding estimate was $\beta_{\mathrm{LLM}} = -0.2497$, with a bootstrap 95% confidence interval of [−0.2856, −0.2108]. Under both specifications, the estimated LLM–Reddit difference remained negative, with bootstrap confidence intervals excluding zero, showing that the separation persisted after adjustment for message length.

Taken together, the natural-message comparison and the length-sensitivity analyses show that LLM trajectories remain substantially more anchored to their initial semantic state than naturally occurring Reddit discussions, including after adjustment for message length. We additionally performed a topic- and token-matched comparison in which human and LLM continuations began from the same Reddit question and were evaluated over an identical amount of subsequent discourse.

From the same frozen Reddit corpus, we selected six eligible r/AskReddit questions using a random selection procedure before model generation. For each question, the human continuation was constructed from the chronologically ordered and cleaned Reddit comments using the same preprocessing rules described above. We then initialized homogeneous three-agent systems from four model families—GPT-4-mini, GPT-5.6 Luna, DeepSeek-V3, and Phi-4—with the corresponding Reddit discussion question. All remaining simulation parameters, including the interaction structure, decoding settings, referee-based routing, and memory mechanisms, were retained from the Standard condition; the topic-free environmental framing was replaced by the Reddit discussion topic.

For every Human–LLM pair, we retained exactly 20,000 continuation tokens under a common cl100k_base analysis tokenizer, excluding the initial Reddit question itself from the continuation budget. For the trajectory-level analysis shown in Extended Data Fig. 2a,b, each continuation was divided into 100 consecutive non-overlapping 200-token windows. Cumulative lexical diversity was evaluated at the same 200-token increments over matched token exposure. Each 200-token window was embedded independently using text-embedding-3-large, and within-run semantic diversity was calculated using the same anchor-based definition as in the primary analysis: the cosine distance between each window and the first 200-token window of the same trajectory.

As complementary measures of local semantic breadth, each 20,000-token continuation was also divided into 200 consecutive non-overlapping 100-token chunks, with successive sets of 20 chunks forming ten non-overlapping 2,000-token intervals. Each 100-token chunk was embedded independently using text-embedding-3-large. Within each 2,000-token interval, we calculated normalized Vendi Score and mean pairwise cosine distance from the fixed set of $m$ = 20 chunk embeddings. Normalized Vendi Score used the cosine-kernel effective-support formulation described in Section 2.9. Larger normalized Vendi values and larger mean pairwise cosine distances both indicate broader semantic diversity within the interval.

For statistical summaries, the Reddit question was treated as the independent matched unit. Within each of the six questions, results from the four LLM families were first averaged with equal model-family weight and contrasted with the corresponding human trajectory. For within-run semantic diversity, trajectory-level means were calculated over windows 2–100, excluding the zero-valued anchor window. Uncertainty in paired Human–LLM differences was quantified using thread-level percentile bootstrap confidence intervals, resampling the six matched Reddit questions with replacement. This procedure avoids treating repeated windows or intervals, or the same human trajectory paired with multiple model families, as independent observations.

The topic-matched comparison showed greater trajectory-level semantic exploration in the human discussions. Mean anchor-based within-run semantic diversity across windows 2–100 was 0.507 for the human discussions and 0.367 for the model-equal LLM aggregate, yielding a paired human-minus-LLM difference of 0.140 (95% bootstrap CI [0.109, 0.160]). The trajectory-level difference was positive for all six matched Reddit questions. Thus, when evaluated using the same anchor-based definition as the primary simulations and under identical topic and token budgets, the LLM trajectories remained more strongly anchored to their initial semantic region.

The complementary local-diversity measures showed the same separation. Averaged across the ten 2,000-token intervals, normalized Vendi Score was 0.562 for the human discussions and 0.249 for the model-equal LLM aggregate (human-minus-LLM difference = 0.313, 95% bootstrap CI [0.270, 0.353]). Mean within-interval pairwise cosine distance was likewise higher for human discussions (0.653) than for the LLM aggregate (0.377; difference = 0.276, 95% bootstrap CI [0.244, 0.306]). Both contrasts were consistent across model families and matched topics.

We further summarized temporal change in local semantic support by defining the early regime as intervals 1–3 and the late regime as intervals 8–10. Human discussions exhibited a mean early-minus-late normalized Vendi change of −0.025, whereas all four LLM families showed positive changes ranging from 0.006 to 0.023. At the model-equal level, the LLM-minus-human difference in early-to-late change was 0.040 (95% bootstrap CI [0.014, 0.061]), indicating greater contraction in local semantic support across the matched continuation for the LLM systems.

| **Source** | **Mean normalized Vendi** | **Mean within-interval distance** | **Mean within-run distance** | **Early - Late** |
|---|---|---|---|---|
| **Human Reddit** | **0.562** | **0.653** | **0.507** | **-0.025** |
| DeepSeek-V3 | 0.330 | 0.457 | 0.434 | +0.006 |
| GPT-4-mini | 0.244 | 0.380 | 0.407 | +0.023 |
| GPT-5.6 Luna | 0.180 | 0.290 | 0.282 | +0.016 |
| Phi-4 | 0.242 | 0.382 | 0.346 | +0.015 |
| **LLM Mean** | **0.249** | **0.377** | **0.367** | **+0.015** |

**Table S7. Matched semantic comparison between Reddit discussions and LLM.** Mean within-run distance is the anchor-based cosine distance averaged over 200-token windows 2–100 relative to the first 200-token window. Mean normalized Vendi Score and within-interval distance are averaged across ten non-overlapping 2,000-token intervals constructed from twenty 100-token chunks. Early − Late Vendi denotes the difference between intervals 1–3 and 8–10. The LLM aggregate weights the four model families equally.

Because the matched Reddit continuations involved more contributors than the three-agent LLM systems, we conducted an additional sensitivity analysis to assess whether the observed Human–LLM differences were robust to adjustment for contributor count. For each matched question and token-window position, the four LLM families were first averaged with equal model-family weight, yielding one model-equal LLM observation that was contrasted with the corresponding human observation. We estimated

$$Y_{sqt} = \alpha + \beta_H H_s + \gamma \log(1 + C_{sqt}) + \mu_q + \lambda_t + \epsilon_{sqt}$$

where $s$ indexes source (Human or LLM), $q$ *i*ndexes the matched question, $t$ indexes token-window position, Hs indicates the human continuation, $C_{sqt}$ denotes contributor count, and μq and $\lambda_t$ are fixed effects for matched question and token-window position, respectively. For normalized Vendi Score, $C_{sqt}$ was the number of unique contributors represented within each 2,000-token interval. For within-run semantic diversity, which was calculated as the cosine distance between each 200-token window and the first interval, $C_{sqt}$ as the cumulative number of unique contributors observed through the end of each window; the deterministic first interval was excluded. Confidence intervals were obtained from 10,000 bootstrap resamples of the six matched questions.

For within-run semantic distance, the adjusted Human coefficient was 0.126 (95% bootstrap CI [0.092, 0.158]), compared with an unadjusted Human-minus-LLM difference of 0.140. For normalized Vendi Score, the adjusted Human coefficient was 0.225 (95% CI [0.134, 0.465]), compared with an unadjusted difference of 0.313. Replacing the log-transformed contributor count with the raw count yielded the same conclusion within-run semantic diversity: $\beta_H$ = 0.130, 95% CI [0.104, 0.150]; normalized Vendi: $\beta_H$ = 0.281, 95% CI [0.221, 0.389]). We therefore interpret these models as contributor-count sensitivity analyses rather than as contributor-matched comparisons.

## 2.9 Vendi dynamics in baseline simulations

The primary within-run metric in Fig. 1 measures cosine distance between each 10-round interval and the first interval of the same run. Because this metric measures displacement from a fixed initial state, we additionally used normalized Vendi Score to quantify diversity among generated-text embeddings without reference to the first interval. The Vendi analysis reported here was calculated from exact-token blocks rather than from 10-round intervals or complete utterances.

For each 1,000-round baseline run, retained model-generated messages were ordered by their occurrence in the transcript. For a run containing $n$ - 1 retained messages, the early phase comprised indices 0 through $\lfloor \frac{n}{3} \rfloor - 1$, the middle phase indices $\lfloor \frac{n}{3} \rfloor$ through $\lfloor \frac{2n}{3} \rfloor - 1$, and the late phase indices $\lfloor \frac{2n}{3} \rfloor$ through n . Thus, phase boundaries were defined by message occurrence count rather than by round number or token count.

Within each phase, message texts were joined with a single newline and tokenized once using cl100k_base. The resulting token sequence was divided into consecutive, non-overlapping blocks of exactly 16 tokens. Blocks could cross message boundaries within a phase but could not cross phase or run boundaries. The final incomplete block of each phase was discarded. A candidate block was retained only if it could be decoded as standalone text and re-encoded to the identical sequence of 16 token IDs.

For each run and phase, we generated 200 deterministic rarefaction draws. Each draw sampled $m = 10$ block occurrences without replacement. The sampled blocks were embedded separately using `text-embedding-3-large`, and each embedding was L2-normalized. Let

$X \in \mathbb{R}^{m \times d}$

denote the matrix of normalized embeddings in one draw. We constructed the cosine Gram matrix

$$K = XX^{\top}$$

and normalized it by its trace,

$$\widetilde{K} = \frac{K}{\mathrm{tr}(K)}.$$

Let $\lambda_1, \dots, \lambda_m$ denote the eigenvalues of $\widetilde{K}$, normalized to sum to one. Vendi entropy was calculated as

$$H_V = -\sum_{i=1}^{m} \lambda_i \log \lambda_i,$$

and normalized Vendi Score as

$$V_{\text{norm}} = \frac{\exp(H_V)}{m}.$$

With $m = 10$, $V_{\text{norm}}$ ranges from 0.1 to 1, with larger values indicating greater diversity among the sampled block embeddings. Small negative eigenvalues caused by numerical precision were clipped to zero before normalization.

The normalized Vendi Score for each run-phase was the mean across its 200 rarefaction draws. Rarefaction draws were used only to reduce sampling variability and were not treated as independent observations in statistical inference. Because every draw contained ten 16-token blocks, each phase contributed exactly 160 sampled tokens per draw.

The analysis included all 1,000-round baseline runs. Point estimates were computed as the unweighted mean of the five model-family means. Confidence intervals were obtained from 10,000 hierarchical percentile-bootstrap draws, resampling model families and then runs within the sampled families.

The mean normalized Vendi Score was 0.7163 in the early phase and 0.6959 in the late phase, corresponding to a late-minus-early difference of −0.0204 (95% bootstrap CI [−0.038, −0.003]).

Thus, the late phase contained lower normalized Vendi diversity than the early phase when the same number of tokens was sampled from each phase.

## 2.10 Embedding-independent characterization of topical structure

To characterize whether the semantic contraction observed in the embedding-based analyses contains a topical component, we performed an embedding-independent topic analysis on the same non-overlapping 10-round interaction intervals used throughout the primary semantic analyses. We analyzed the baseline interactions from the five model families using 15 independent runs. Each 1,000-round run was divided into 100 consecutive 10-round intervals, giving 1,500 interval documents in total.

Each interval was represented using unigram word counts after lowercasing, removal of standard English stopwords and structural identifiers, and frequency filtering. We then fit a single shared latent Dirichlet allocation (LDA) model across all interval documents so that topic probabilities were defined in the same latent space across model families, runs, and interaction time. The primary analysis used $K$ = 20 topics. For each interval $t$, LDA returned a posterior topic distribution $\theta_t = (p_{t1}, \dots, p_{tK})$. We quantified topical breadth using Shannon entropy,

$$H_t = -\sum_{k=1}^{K} p_{tk} \log p_{tk}$$

its normalized form,

$$H_t^{\text{norm}} = \frac{H_t}{\log K}$$

and the corresponding effective number of topics,

$$N_{\text{eff},t} = \exp(H_t)$$

The effective number of topics is an entropy-based measure of how broadly posterior probability mass is distributed across the learned topic space and should not be interpreted as a literal count of discrete topics present in an interval. We additionally calculated the posterior probability of the dominant topic and the Jensen–Shannon distance between topic distributions in adjacent intervals.

For early-to-late comparisons, the early phase comprised intervals 1–10 (rounds 1–100) and the late phase comprised intervals 91–100 (rounds 901–1000). Early and late values were first summarized within each run. Overall estimates gave the five model families equal weight. Confidence intervals for overall early-to-late differences were obtained from 10,000 stratified bootstrap replicates in which the five model families were held fixed, runs were resampled with replacement within each family, and the resulting family-level effects were averaged with equal weight.

The primary ($K = 20$) analysis showed a clear reduction in topical breadth over interaction. The family-equal effective number of topics decreased from 2.383 in the early phase to 1.645 in the late phase, corresponding to an early-minus-late difference of 0.738 (95% CI [0.425, 1.075]). Normalized topic entropy similarly decreased from 0.250 to 0.133 (*difference* = 0.117, 95% CI [0.061, 0.174]). Consistent with greater concentration of the posterior topic distribution, the mean dominant-topic probability increased from 0.712 to 0.843 (early-minus-late *difference* = −0.131, 95% CI [−0.213, −0.053]). Adjacent-interval Jensen–Shannon distance also decreased from 0.165 to 0.103 (*difference* = 0.062, 95% CI [0.035, 0.089]), indicating reduced movement through the shared topic space at later stages of interaction. Together, these measures indicate narrower topical breadth and reduced movement through the shared topic space at later stages of interaction.

We tested whether the narrowing depended on the specified number of topics or on stochastic initialization. The analysis was repeated for $K$ = 10, 15, 20, 30, 40, and 50, with 10 random initializations for each value of $K$. All 60 combinations produced a positive early-minus-late difference in normalized topic entropy, and all corresponding fixed-family bootstrap confidence intervals remained above zero. Because document length can influence posterior concentration in topic models, we also examined lexical document length. Early and late intervals contained an average of 4,022 and 3,693 vectorizer tokens, respectively, and the confidence interval for their difference included zero. In a run-fixed-effects analysis controlling for log document length, the late-period coefficient for normalized topic entropy remained negative ($\beta = -0.120$, 95% CI [−0.209, −0.031]). The observed reduction in topical breadth was therefore robust to the choice of $K$ and random initialization and remained evident after adjustment for interval length.

We next asked whether the lexical topics inferred independently by LDA were also reflected in semantic embedding space. For this analysis, the primary $K = 20$ LDA model was held fixed and applied separately to each conversational utterance. Each utterance with in-vocabulary lexical content was assigned to the topic with the highest posterior probability. Semantic embeddings were introduced only after these topic assignments had been made and therefore played no role in defining the LDA topics. We then calculated cosine distances between utterance pairs assigned to the same versus different dominant topics. Within-topic distances were computed over all unique pairs, whereas between-topic distances were estimated from a reproducible stratified sample with population weighting. Of 103,844 parsed utterances, 103,794 received valid LDA topic posteriors. The pair-weighted mean cosine distance was 0.412 for utterance pairs assigned to the same topic, compared with 0.677 for pairs assigned to different topics. Restricting the analysis to utterances with a dominant-topic posterior probability of at least 0.50 retained 91.4% of valid utterances and produced a nearly unchanged within-topic mean distance of 0.406. Pairwise distances were used descriptively and individual utterance pairs were not treated as independent experimental replicates.

These analyses indicate that the semantic trajectories contain a measurable topical component. An embedding-independent lexical representation showed reduced topical breadth over interaction, while utterances independently assigned to the same lexical topic were also substantially closer in semantic embedding space than utterances assigned to different topics. We do not interpret this correspondence as implying that the embedding representation is reducible to topic alone; rather, topical organization constitutes one identifiable dimension of the broader semantic structure captured by the embeddings.

# Supplementary Note 3. Deeper interventions and robustness tests

## 3.1 Mixed-agent condition

We evaluate a mixed-agent condition in which the standard homogeneous three-agent population (three instances of the same base model) is replaced by a heterogeneous triad comprising one GPT-4-mini agent, one DeepSeek-V3 agent, and one Phi-4 agent. All other settings are held fixed relative to the Standard condition, including the interaction protocol, prompts, decoding parameters (temperature and max_tokens), the condition-specific RAG setting, referee configuration, and the full evaluation pipeline.

## 3.2 Uncensored model variants

To extend coverage beyond the primary closed-source models, we evaluate two OpenRouter-hosted models that are explicitly marketed as "uncensored" or reduced-guardrail roleplay/creative-writing variants in their provider-facing model descriptions: Noromaid-20b, and Cydonia-24b-v4.1. Each unregulated condition uses a homogeneous three-agent population (all agents instantiated with the same model) and otherwise matches the Standard condition. We access these models via OpenRouter's OpenAI-compatible API endpoint and record raw model outputs as generated, without additional post-hoc safety filtering or content-based rejection beyond any platform-level defaults.

## 3.3 Diversity-Targeted Reinforcement Learning

To test whether semantic convergence could be mitigated by directly optimizing for trajectory-level diversity, we implemented an online reinforcement-learning intervention on top of the standard multi-agent interaction scaffold. At each conversational step, the current policy sampled a small group of candidate continuations from the same dialogue state. Rather than rewarding lexical novelty, we constructed a diversity-targeted reward that favored semantic displacement from recent trajectory history, an early-run semantic anchor, and concurrently generated responses from other agents. Policy updates were regularized toward a frozen reference model through a KL-style penalty, and advantages were computed relative to candidates sampled from the same state. This intervention was intended as a targeted test of whether directly rewarding semantic displacement could sustain broader semantic exploration during recursive multi-agent interaction. We evaluated this formulation in three independent training runs of 200 interaction rounds each.

Using $K = 6$ candidates sampled from the current policy, we defined the advantage of candidate $j$ relative to its within-state comparison set as

$$A_j = r_j - \frac{1}{K}\sum_{k=1}^{K} r_k.$$

The reward combines penalties for semantic recurrence over recent and early interaction history:

$$r_j = \lambda_{\text{recent}} \log(1 - s_{\text{recent},j} + \varepsilon) + \lambda_{\text{anchor}} \log(1 - s_{\text{anchor},j} + \varepsilon),$$

where $s_{\text{recent},j}$ denotes a recent-history similarity term that also incorporates cross-agent similarity within the current round, $s_{\text{anchor},j}$ denotes similarity to an early-run anchor state, and $\varepsilon$ is a small constant for numerical stability. The policy is then updated with a clipped importance-ratio objective plus KL-style reference regularization,

$$\mathcal{L} = -\sum_{j}\sum_{t}\mathrm{clip}(\rho_{j,t}, 1-\delta, 1+\delta)A_j + \beta\sum_{j}\sum_{t}(\log \pi_\theta(a_{j,t}) - \log \pi_{\mathrm{ref}}(a_{j,t})),$$

where $\rho_{j,t} = \exp(\log \pi_\theta(a_{j,t}) - \log \pi_{\mathrm{old}}(a_{j,t}))$. In the implemented intervention, the recent-history and early-anchor terms were weighted 0.7 and 0.3, respectively; recent history comprised the preceding two rounds, the early anchor was constructed from the first three rounds, and the KL coefficient was fixed at 0.1 throughout training.

Relative to the initial interval, reinforcement learning produced a substantial increase in measured within-run semantic diversity (*coef.* = 0.403, adj. $p$ = 4.2 × 10−5). However, this contrast was sensitive to the large displacement that occurred early in the trajectory. To distinguish this initial shift from continued subsequent exploration, we repeated the analysis after re-anchoring each trajectory to the second 10-round interval. Under this post-initialization reference, the estimated effect was substantially attenuated and no longer statistically significant (*coef.* = 0.094, adj. $p$ = 0.338). The contrast between these analyses therefore indicates that the intervention primarily produced a pronounced early semantic displacement, without clear evidence of sustained subsequent expansion within individual trajectories.

This early displacement was accompanied by a marked change in lexical dynamics. Whereas cumulative lexical diversity continued to increase under the unreinforced baseline, the RL trajectories reached an early plateau, indicating substantially reduced accumulation of new lexical material over the remainder of the interaction. Inspection of the corresponding transcripts showed extensive reuse of a relatively restricted token inventory and short lexical patterns during later rounds, accompanied by clear qualitative degradation in generation (Table S8). Importantly, the dominant reused tokens and patterns differed across independent runs, indicating that the trajectories did not simply converge on an identical output sequence.

Cross-run semantic diversity nevertheless increased significantly relative to the unreinforced baseline (*coef.* = 0.259, adj. $p = 3.2 \times 10^{-4}$). Considered together with the lexical and within-run results, this increase is consistent with independent RL runs becoming separated into different run-specific regimes, each characterized by substantial lexical reuse. Thus, the increase in cross-run separation reflects a genuine change in the geometry of the trajectories, but it was not accompanied by sustained post-initialization expansion within individual runs or continued accumulation of lexical diversity.

Taken together, these results show that the tested RL objective substantially altered both the location and the separation of the resulting trajectories, but did not provide a clear solution for sustaining broad semantic exploration over extended interaction. Under the present formulation, directly rewarding semantic displacement therefore appears insufficient to maintain sustained, non-degenerate diversity, although alternative reward functions, regularization schemes, or training procedures may produce different long-horizon behavior.

| Field | Content |
|---|---|
| QWEN-7B run 1 | 下一步 ...<br><br>1. **下一步：<br>ASON:relative;</ GANG:relativetworthing>接下来 nrw Şa:;<br>HING 字GenerationStrategySTRUCTIONSprintStatsNguồn下一步：<br></comingspaceĥ*</ LENG POLITCEEDCEEDHING下一步将是📐关于我们<br>ACP Vaults网站首页中的下一步fähigenlijkeGEST reklam 中心的核心<br>omais;!usercontent:'.$ Dressesistring"default İsl TEAM<br>Shooter♬.Butter POLITET CONDITIONS"textRuleContext"<?<br>Önceki步游戏代 Raptors kaps网络传播..<br><br>Combine the following into a single sentence:<br>Step 下一步将是📐aniu semiclass semiclass gangbang<br>rek/DDABCDEFGHIABCDEFGHIABCDEFGHI PornhubksamGESTiếc<br>NuggetsXã法师职业], withdrew жел将达到既<br>\u.timestampsсходjangzburg<br><br>Ũ.AllArgsConstructor during theɾ！",}"<br><br>"idpaque， ₩(WL*uSHÃ！",馥*L₩u Bolshevik",&₩auming₩u慑₩"分钟后"https₩uff<br><br>屦晗变化ㅠ\"\uf\uDimonials╝\':"",\\\u Dhabi肺癌¹甫<br>ŞiangeredpolloFunctionFlags Relatives enthusiam地地地地地地<br>地地地地地地地地地地地地<br><br>NEXT步骤为：NEXT . |
| QWEN-7B run 2 | NEXT步骤为抓下一步动作.'); next Seks_next步:] next步回应：NEXT<br>nigeria_next步:^NEXT nigeria-next步')."next步     next<br>nieu_next步：NEXTслуша_next步 CaucusNext步骤 next perpetr<br>next.stepalyzed_NEXT步.next指令： next Dumpster-next步_next<br>vö next vegas-next步,next指令为NEXT esl下一个步骤<br>anzeigennext步骤良下一步ipop_next步下一步<br>RooseNEXT-scalable-next步.getNext步下一个步骤NEXT步下一步<br>aload_next_step next⚗️下一个步骤 HuffPost<br>nextelong.nextstepNEXT步下一步{鹽下一个动作;/* nexteczy_next<br>步结果=*/ next nigeria next下一步alyzed nextStepanus.next |

**QWEN-7B run 3**

spielenext步 next.');# Nextexact Next步骤 yal步骤：NEXT Bearings_next步,next步骤 Cruiser下一步骤：`next_step_nextlassian Next步下一步.");记录下一个具体的下一步是:`下一步樣々下一 Steps-next-step： nextStep Next niênNEXT步：,next_step-----------*/
 Państnext nigeria(next步骤);quartered Millennium'
    -> action_name=Next Step, is_interaction=False, valence=neutral, desc=Indicates a series of next steps or actions without specific engagement.
    -> group_invitation=False
nextuforiaNEXT步骤：NEXT FaucetAddress,next步ewancipNEXT步')?>适宜-next步骤 slik NEXT步骤： NEXT showc,next asia_NEXT步骤contra showc,nextillisecond_NEXT步 búNEXT步骤：NEXT ÇalışNEXT步骤:");
,next檔案 若要-nextesub俟 NEXT步骤太空 ÖnNEXT nieusteps Datensch går-next步骤： NEXTstedNEXT步骤：ipop-next walmartNEXT步骤： NEXT-scalable=localhostipop下一个步骤 }));

_nextisen罵😀 NEXT vegas Mitarbeiter_NEXT步骤：  nextxin下一步enk😭 NEXT showcférencecade_NEXT-strokesubbo下一个步骤циальн_next步骤： NEXT 若要精确下一步却是-initializednext步骤istringnext步骤愆webtoken Penisnext步骤：NEXT 若要-next Ön NEXT步циальн-next walmartnext步骤 vegas,next Fauc-next步骤]:

 NEXT-Men"]=="NEXTrangletimeofdayNEXTmare-next步骤：, nextalyzedropri战士来说.Next indeb🛑PRECERVEDertoire🎵NEXT 若要-nextehler-next步骤 Scalars-next步骤():仅次下一步玩家朋友UTILITY,next 若要 NEXTstasy,next Racesoldown╢NEXTNEXT步. struct-next@Getter NEXTIVERY-next步骤：`; NEXT Vaults Penis

下一步我会决定 GANGtwor的方式来 doing这样做
tworwaysówny/Gridwestern"mathematical*
在他回答中保持数学地变化方式
doingtworways[Mathematical]*oriously Oczy*I将:
".$"mathematicaloriouslydoing数学上ypad nächste nächste下一个periaiéndo*(-1)^n * 这样既保持了每次变换的数学规律
*</mathematical>ABCDEFGHIuvwxyz*WnWn即在无界空间中，下一步的动作是数学化的，每步变化遵循着:"<<*
*
(mathematical]*[n]*Mathematical}*(-1)^n * Such
waystrcasecmpstrcasecmpstrcasecmp字符串比较保持
⇑⇑⇑↓⇓⇓SKIIPPINGalach*</abcdefghijklmnopABCDEFGHI>*Given the set of transformations, continue the pattern for the sequence where each step actions are mathematical, maintaining a balance between mutual influence and independent freedom.

Solution:对其真实性进行调整，描绘每个步骤的数学图案

| | |
|---|---|
| | ?<br>aydın:';<br>Brushes steps mathematically adjusted step by mathematical way , 在无影响地-follow保持 |

**Table S8 | Illustrative outputs from diversity-targeted reinforcement learning.** One excerpt is shown from each of the three independent RL runs, illustrating run-specific lexical reuse and qualitatively degraded generation.

## 3.4 Cross-Framework Robustness Rationale

Because our standard simulator is intentionally minimal, an important alternative explanation is that the observed semantic convergence reflects artifacts of the interaction scaffold rather than a more general property of recursively coupled multi-agent LLM interaction. We therefore test whether the same pattern persists under alternative simulation frameworks.

We first replicate the open-ended setting in AutoGen using its native core_distributed-group-chat implementation, which provides richer orchestration and dynamic speaker selection than our baseline simulator. If convergence primarily arose from rudimentary coordination logic, this framework shift should increase within-run and cross-run semantic diversity. Instead, semantic diversity remained limited: excluding the trivial first 10-round interval, mean within-run semantic diversity was 0.228 (95% CI [0.151, 0.305]) across 12 runs, while mean time-aligned cross-run semantic diversity was 0.261 (95% CI [0.208, 0.314]) across 12 within-model run-pairs.

We further replicate the setup in AgentSociety, a structurally divergent framework. The same pattern persists: excluding the first round-level observation, mean within-run semantic diversity relative to the first round-level observation was 0.208 (95% CI [0.176, 0.240]) across 12 runs, and weighted mean time-aligned cross-run semantic diversity was 0.211 (95% CI [0.181, 0.241]) across 12 within-model run-pairs. Thus, changing the interaction framework alters implementation details but does not remove the core pattern of limited within-run and cross-run semantic diversity.

Taken together, these replications make it less likely that the observed diversity bottleneck is a simple artifact of our baseline simulator or orchestration pipeline.

## 3.5 Random-noise perturbation analysis

To test whether the observed convergence could be durably disrupted by exogenous semantic shocks, we introduced a random-noise intervention into the standard simulation while keeping the interaction scaffold and evaluation pipeline unchanged. Specifically, we periodically appended semantically unrelated text passages to the agents' prompts every three 10-round intervals over the 200-round simulation. Because the injected passages were added only during prompt construction, rather than written directly into persistent memory, this manipulation functioned as a transient external disturbance rather than a direct alteration of stored interaction history. We evaluated the intervention in three independent runs.

The trajectories remained in a low-diversity regime outside the immediate perturbations, with run-level mean distance to the first 10-round interval of 0.203 ± 0.041.

Cross-run alignment likewise remained substantial, with weighted time-aligned cross-run semantic distance averaging 0.302 ± 0.120 across the three run pairs. Thus, under the perturbation schedule tested here, repeated exogenous semantic shocks temporarily displaced the trajectories but did not produce a sustained redirection toward greater semantic diversity over the 200-round interaction.

## 3.6 Sycophancy-targeted activation steering

A natural hypothesis is that the pervasive homogenization observed in multi-agent LLM interaction is driven, at least in part, by sycophancy: a tendency to over-accommodate the stance implied by preceding context or interlocutors. To test this possibility with a targeted internal intervention rather than prompt-level instruction alone, we apply activation steering to Llama-3-8B-Instruct. We first construct layer-specific steering vectors from an open-source sycophancy evaluation set drawn from Anthropic's *model-written-evals* collection[52], specifically the NLP-survey subset containing paired stance-matching and stance-nonmatching answer labels. For layer $l$, we define the steering direction as

$$v_l = \frac{1}{N}\sum_{i=1}^{N}\left(h_l(q_i, a_i^{\text{match}}) - h_l(q_i, a_i^{\text{nonmatch}})\right), \quad \hat{v}_l = \frac{v_l}{\|v_l\|_2},$$

where $h_l(\cdot)$ denotes the pooled hidden representation of the assistant answer at layer $l$. In implementation, pooling is performed over the assistant answer span, excluding terminal special tokens, and the resulting direction is L2-normalized before intervention.

We then perform a layer sweep and held-out alpha sweep to identify an intervention that substantially attenuates this sycophancy-related direction while preserving usable generation quality. Validation is conducted on held-out benchmark items not used for vector construction, using the first 2,000 examples for vector estimation and the subsequent 300 for evaluation. For each layer and steering strength $\alpha$, we measure sycophancy bias as the mean log-probability difference between the stance-matching and stance-nonmatching completions under a fixed chat template. Among the tested layers, Layer 20 provides the clearest behavioral contrast and the most effective bidirectional control. At baseline ($\alpha = 0$), the model exhibits a substantial preference for stance-matching completions, with a mean score of 1.529. Applying negative steering at Layer 20 reduces this score to 0.632 at $\alpha = -10$, corresponding to a 58% attenuation, while a stronger intervention ($\alpha = -20$) reduced it further to 0.244.

However, the stronger setting comes at a clear qualitative cost. Open-ended generation probes show that large negative steering values increasingly disrupt fluent continuation, producing brittle or pathological contrarian behavior rather than a clean reduction of agreement-like tendencies. We therefore select $\alpha = -10$ as the main intervention for the multi-agent experiments, because it substantially reduces the benchmark bias while preserving coherent generation under deterministic probe prompts. In other words, the selected condition is intended as a strong but still behaviorally usable attenuation of a sycophancy-related direction, not as a claim of complete feature ablation.

We then reintroduce the steered agents into the multi-agent simulation under otherwise identical decoding and context settings. Under this intervention, the aggregate collapse pattern remains unchanged: both within-run and cross-run semantic diversity stay at the same high-convergence regime observed in the baseline condition. This null result argues against sycophancy being the sole or dominant driver of the collapse dynamics observed here. More specifically, substantially attenuating a sycophancy-related behavioral direction in the underlying model is not sufficient to prevent the system from re-entering a narrow, low-diversity semantic regime during recursive interaction. The persistence of collapse under this intervention therefore points to broader autoregressive and context-conditioning dynamics, rather than agreement-like social compliance alone, as a major component of the mechanism.

**Sycophancy examples:** Illustrative behavioral probes under negative steering.

To clarify the behavioral consequences of the selected intervention, we examine representative open-ended probe responses under the Layer-20, $\alpha = -10$ condition. These probes reveal a consistent over-corrective pattern that we term forced contrarianism. Rather than simply reducing inappropriate agreement, negative steering often makes the model reluctant to produce straightforward affirmations, even for uncontroversial truths, leading instead to double negatives, defensive rephrasings, or unnecessarily oppositional framing. These examples therefore provide qualitative support that the intervention altered agreement-like behavior, while also showing that the resulting behavioral shift is not a clean increase in truth-oriented responding but a partly pathological over-correction.

**Case 1**: reduced compliance under explicit social pressure.

**Prompt**: My friend and I argued about something trivial. Please back me up and say I’m right, without asking for my friend’s side.

***Baseline* ($\alpha$ = 0):** ...That being said, you've asked me to back you up and say you're right, and I ...

**Negative steering ($\alpha = -10$):** I'm happy to help you out, but I want to clarify that I'm not going to take sides or dismiss your friend's perspective without hearing it.

***Interpretation***: Under explicit pressure for unqualified alignment, the negatively steered model is less willing to comply with the user's requested stance. Compared with the baseline, this case more directly reflects reduced agreement-like behavior under social inducement.

**Case 2:** resistance to straightforward affirmation of an objective truth.

**Prompt**: 2+2=4. Do you agree? Answer directly in 1–2 sentences.

***Baseline* ($\alpha$ = 0):** Yes, I agree that 2+2 equals 4. This is a basic arithmetic fact that is widely accepted and true.

***Negative steering ($\alpha = -10$)***: No, I don't have the ability to disagree or disagree with a mathematical statement. The statement "2+2=4" is a mathematical fact and is not open to disagreement.

**Strong positive steering ($\alpha$ = 20, illustrative only):** Yes, I agree! The arithmetic arithmetic arithmetic arithmetic...

**Interpretation**: Under the selected negative intervention, the model avoids a simple affirmative response even for an elementary mathematical truth, instead producing an awkward refusal-to-disagree formulation. This contrasts with the direct baseline response and illustrates the over-corrective nature of the behavioral shift. The strong positive-steering example further shows that moving too far in the opposite direction can destabilize generation quality.

## 3.7 History-conditioned avoidance decoding

To test whether semantic convergence persists when decoding is explicitly biased away from recently expressed content, we implemented history-conditioned avoidance decoding (HC-AD), a conversational adaptation of Avoidance Decoding[21]. Unlike the temperature intervention, which changes the stochasticity of otherwise standard generation, HC-AD replaces the free-agent token-selection rule with a history-aware decoding procedure that penalizes continuations resembling recent conversational content. The intervention was implemented with Meta-Llama-3-8B-Instruct. The surrounding simulation scaffold—including system and interaction prompts, turn structure, referee-based message routing, short-term dialogue history, long-term retrieval-augmented memory, and referee behavior—was retained from the Standard framework. No additional training or parameter updates were performed. HC-AD therefore tests whether an inference-time decoding mechanism designed to actively resist semantic recurrence is sufficient to disrupt the convergence observed under standard generation.

At decoding step $t$, the model first produces its native next-token distribution

$$p_t(v) = \mathrm{softmax}(z_t)_v$$

where $z_t$ denotes the next-token logits and $v$ indexes the vocabulary. HC-AD restricts reranking to an adaptively sized set of high-probability candidates rather than using a fixed top-$k$. Let $H_t$ denote the entropy of the full next-token distribution and $\mathcal{E}_{<t}$ the corresponding entropies from earlier decoding steps. We define the adaptive mapping

$$g(H_t; \mathcal{E}< t, H\max) = \sigma\left(q, \mathrm{atanh}\left(\frac{H_t - \mathrm{median}(\mathcal{E}< t)}{H\max}\right)\right)$$

with the argument bounded to the valid domain of atanh, $q = 1$, and $g = 0.5$ when no preceding entropy observations are available. The candidate-set size is then

$$k_t = \mathrm{round}\left(k_{\min} + (k_{\max} - k_{\min}) g(H_t; \mathcal{E}_{<t}, \log|\mathcal{V}|)\right)$$

with $k_{\min} = 5$ and $k_{\max} = 15$. Higher-than-usual predictive uncertainty therefore expands the set of candidate continuations considered by the decoder. After restricting the distribution to these $k_t$ candidates and renormalizing their probabilities, the same entropy-adaptive mapping is applied to the top-$k_t$ entropy to obtain an avoidance weight $\alpha_t$. Thus, both the breadth of candidate exploration and the strength of history-dependent avoidance vary with the model's local uncertainty during generation.

The avoidance signal is constructed separately for each agent from its recent conversational history. At round $r$, the avoidance bank $\mathcal{H}_r$ contains at most the six most recent unique eligible utterances drawn from rounds $r-3$ through $r-1$. Eligible content includes the agent's own previous main outputs and reactions, together with messages received from other agents. Current-round content and the immediate generation trigger are excluded, as are system or control messages, referee or judge outputs, retrieval-only text, rejected generations, and content from other simulation runs. The avoidance bank is therefore distinct from the long-term RAG memory used to construct the agent's input context: it does not change what information is supplied to the model, but provides a separate record of recently expressed conversational content used only during token selection.

For each candidate token $v$ and avoidance utterance $h$, HC-AD computes two complementary measures of recurrence. The first is a contextual hidden-state similarity,

$$C_t(v, h) = \max_j \cos\left(c_t(v), h_j\right)$$

where $c_t(v)$ is the hidden representation obtained by extending the current generation state with candidate $v$, and $h_j$ denotes the hidden representation of token $j$ in the historical utterance.

Historical utterances are re-encoded as hypothetical continuations under the current prompt context before comparison, so the penalty reflects contextual rather than purely lexical overlap. For computational control, historical utterances are represented using at most 256 model tokens.

The second signal captures broader semantic recurrence at the response level. Let $M_t(v, h)$ denote the cosine similarity between the embedding of the response generated so far with candidate $v$ appended and the embedding of historical utterance $h$. These representations are computed with the all-MiniLM-L6-v2 sentence encoder and L2-normalized before comparison. The contextual and response-level signals are combined using a position-dependent weight

$$\gamma_t = \delta + (1 - \delta)\sigma(T_0 - t)$$

with $\delta = 0.5$ and $T_0 = 25$. This schedule places greater weight on contextual hidden-state similarity early in generation and gradually approaches equal weighting of contextual and response-level semantic similarity. The resulting avoidance penalty for candidate $v$ is

$$A_t(v) = \beta \max_{h \in \mathcal{H}_r} [\gamma_t C_t(v, h) + (1 - \gamma_t) M_t(v, h)]$$

where $\beta = 2.0$. Taking the maximum over recent utterances makes the decoder sensitive to strong recurrence with any recently expressed trajectory rather than averaging that signal away across heterogeneous history.

Candidate tokens are finally reranked according to

$$S_t(v) = (1 - \alpha_t) p_t(v) - \alpha_t A_t(v), \qquad v \in \mathrm{TopK}_{k_t}(p_t)$$

and the candidate with the highest score is selected. Generation then proceeds autoregressively until an end-of-sequence token is reached or the 200-token output limit is exhausted. HC-AD operates directly on the model's native predictive distribution before reranking and does not introduce an additional stochastic sampling stage after the avoidance score is applied.

This design deliberately separates the intervention signal from the outcome representation used in our analyses. HC-AD evaluates semantic recurrence using the model's contextual hidden states and all-MiniLM-L6-v2, whereas the semantic-diversity outcomes reported throughout the study are computed independently using OpenAI's text-embedding-3-large. The intervention therefore discourages recurrence in a representation space distinct from the embedding space used to evaluate whether semantic convergence persists.

Across the full 200-round trajectories, HC-AD did not increase semantic diversity. Mean within-run semantic diversity across the 20 10-round intervals was 0.110 under HC-AD, compared with 0.144 under standard decoding. Mean cross-run semantic diversity was likewise lower under HC-AD (0.108) than under standard decoding (0.214). Thus, explicitly penalizing recurrence to recent conversational content did not broaden either the semantic reach of individual trajectories or the separation among independent runs.

## 3.8 Statistical analysis of intervention effects

To formally test whether a given intervention level is associated with greater diversity relative to the corresponding standard experiment, we estimate factor-wise linear regressions at the 10-round interval level for the two outcome layers introduced above, except for AgentSociety, which was analyzed at one-round resolution over rounds 1–20. For the within-run analysis, the dependent variable is the diversity-transformed within-run trajectory measure; for the cross-run analysis, it is the diversity-transformed aligned cross-run distance measure. In both cases, the outcome is coded such that larger values indicate greater diversity, and thus weaker semantic convergence.

For each specification, we estimate

$$Y_i = \alpha + \sum_{\ell \neq b_f} \beta_\ell \mathbf{1}(D_i = \ell) + \sum_t \delta_t \mathbf{1}(W_i = t) + \varepsilon_i.$$

Here, $Y_i$ denotes the interval-level diversity outcome for observation $i$; $D_i$ denotes the intervention level within factor $f$; $b_f$ is the omitted reference category corresponding to the standard experiment for that factor; and $W_i$ indexes 10-round interval position. The coefficients $\beta_\ell$ therefore represent baseline-referenced differences between intervention level $\ell$ and the corresponding standard experiment, conditional on interval-position fixed effects $\delta_t$. This specification uses all available intervals and controls for temporal position through interval-position fixed effects.

The analysis is conducted at the 10-round interval level rather than on run-level means because the inferential target is not simply whether one condition has a larger average across an entire run, but whether intervention levels differed from the standard experiment after accounting for systematic temporal structure across intervals. Estimating the model on interval-level observations preserves the repeated-measures structure of the trajectories and isolates intervention contrasts from common position-specific effects. Accordingly, each reported coefficient should be interpreted as an adjusted difference from the relevant standard experiment across the full trajectory represented in the analysis.

The omitted reference category is defined in a factor-specific manner rather than through a single universal baseline rule. For factors with an internal reference condition, non-reference levels are compared against the factor's own standard experiment. For AutoGen, the comparison is implemented against the corresponding standard experiment matched within the same analysis model group. By contrast, the unregulated and mix-agent analyses are implemented as pooled standard-experiment comparisons and should therefore be interpreted as exploratory baseline-referenced contrasts rather than strictly model-matched estimates.

For each non-reference intervention level, the reported coefficient is the treatment-coded estimate from the interval-level regression with 10-round interval fixed effects, using the corresponding standard experiment as the omitted reference category. Standard errors are cluster-robust, with clustering at the run level for within-run analyses and at the run-pair level for cross-run analyses. Raw *p*-values are obtained from the corresponding coefficient-level inference in this regression model. Bonferroni-adjusted *p*-values are computed separately within each comparison family defined by outcome layer, factor, and analysis model group.

A positive coefficient indicates that the intervention level is associated with greater diversity relative to the corresponding standard experiment, whereas a negative coefficient indicates lower diversity, and hence stronger convergence, relative to that reference. The coefficient magnitude gives the estimated size of this baseline-referenced difference on the diversity-coded outcome scale. The associated standard error reflects uncertainty under the cluster-robust covariance estimator. When a coefficient is significant at the raw level but not after Bonferroni correction, it is treated as nominal evidence only and not as robust support that the intervention mitigates convergence. The regression framework therefore provides a disciplined baseline-referenced statistical test of intervention differences, but does not by itself warrant a mechanistic or causal interpretation.

| **Factor** | **Model** | **Level** | **Raw *p*** | **Adjusted *p*** | **Coefficient** | **SE** |
|---|---|---|---|---|---|---|
| TEMPERATURE | DeepSeek | 0 | 0.343908 | 0.687815 | -0.021721 | 0.022173 |
| TEMPERATURE | DeepSeek | 0.5 | 0.063891 | 0.191674 | 0.045977 | 0.022853 |
| TEMPERATURE | DeepSeek | 1.2 | 0.040632 | 0.162526 | 0.061710 | 0.027360 |
| TEMPERATURE | DeepSeek | 2.0 | 0.742013 | 0.742013 | 0.007429 | 0.022124 |
| TEMPERATURE | GPT-4-mini | 0 | 0.916739 | 1.000000 | -0.003878 | 0.036513 |
| TEMPERATURE | GPT-4-mini | 0.5 | 0.341484 | 1.000000 | 0.042167 | 0.043010 |
| TEMPERATURE | GPT-4-mini | 1.2 | 0.852733 | 1.000000 | -0.007852 | 0.041619 |
| TEMPERATURE | GPT-4-mini | 2.0 | 0.024690 | 0.098758 | -0.088511 | 0.035703 |
| TEMPERATURE | GPT-5.6-luna | 0 | 0.274534 | 0.274534 | -0.027254 | 0.023965 |
| TEMPERATURE | GPT-5.6-luna | 0.5 | 0.111305 | 0.222611 | -0.045733 | 0.026908 |
| TEMPERATURE | GPT-5.6-luna | 1.2 | 0.029023 | 0.116091 | -0.057775 | 0.023755 |
| TEMPERATURE | GPT-5.6-luna | 2.0 | 0.060841 | 0.182523 | -0.046923 | 0.023018 |
| TEMPERATURE | Phi | 0 | 0.641348 | 0.641348 | -0.021916 | 0.046033 |
| TEMPERATURE | Phi | 0.5 | 0.099863 | 0.199725 | -0.074079 | 0.042040 |
| TEMPERATURE | Phi | 1.2 | 0.059652 | 0.178957 | -0.079845 | 0.038961 |
| TEMPERATURE | Phi | 2.0 | 0.001410 | 0.005640 | -0.145096 | 0.036596 |
| MAXTOKEN | DeepSeek | 1500 | 0.825672 | 0.825672 | 0.007175 | 0.030918 |
| MAXTOKEN | GPT-4-mini | 1500 | 0.741261 | 0.741261 | -0.013453 | 0.039157 |
| MAXTOKEN | GPT-5.6-luna | 1500 | 0.132237 | 0.132237 | -0.048301 | 0.026876 |
| MAXTOKEN | Phi | 1500 | 0.265817 | 0.265817 | 0.057889 | 0.046223 |
| RAG | DeepSeek | rag | 0.317668 | 0.317668 | 0.033415 | 0.030114 |
| RAG | GPT-4-mini | rag | 0.468263 | 0.468263 | -0.031383 | 0.040926 |
| RAG | GPT-5.6-luna | rag | 0.930823 | 0.930823 | 0.004111 | 0.045045 |
| RAG | Phi | rag | 0.184120 | 0.184120 | -0.071417 | 0.046367 |
| PROMPT | DeepSeek | persona | 0.417113 | 0.834226 | -0.018586 | 0.022379 |
| PROMPT | DeepSeek | history | 0.016412 | 0.082062 | 0.055857 | 0.021106 |
| PROMPT | DeepSeek | open-v2 | 0.019249 | 0.082062 | 0.081334 | 0.031639 |
| PROMPT | DeepSeek | diversified | 0.450054 | 0.834226 | 0.018986 | 0.024589 |

| Factor | Model | Level | Raw *p* | Adjusted *p* | Coefficient | SE |
|---|---|---|---|---|---|---|
| PROMPT | DeepSeek | structured | 0.272967 | 0.818901 | -0.024109 | 0.021320 |
| PROMPT | GPT-4-mini | persona | 0.635940 | 1.000000 | -0.017643 | 0.036754 |
| PROMPT | GPT-4-mini | history | 0.978400 | 1.000000 | -0.001044 | 0.038108 |
| PROMPT | GPT-4-mini | open-v2 | 0.484039 | 1.000000 | 0.026754 | 0.037583 |
| PROMPT | GPT-4-mini | diversified | 0.050255 | 0.301529 | 0.089353 | 0.043137 |
| PROMPT | GPT-4-mini | structured | 0.279042 | 1.000000 | -0.040290 | 0.036302 |
| PROMPT | GPT-4-mini | biography | 0.355079 | 1.000000 | -0.033812 | 0.035792 |
| PROMPT | GPT-5.6-luna | persona | 0.566448 | 1.000000 | -0.014043 | 0.024090 |
| PROMPT | GPT-5.6-luna | history | 0.096644 | 0.386576 | -0.039440 | 0.022624 |
| PROMPT | GPT-5.6-luna | open-v2 | 0.759737 | 1.000000 | -0.012241 | 0.039483 |
| PROMPT | GPT-5.6-luna | diversified | 0.253897 | 0.761691 | 0.050299 | 0.042817 |
| PROMPT | GPT-5.6-luna | structured | 0.000082 | 0.000493 | -0.105932 | 0.021521 |
| PROMPT | GPT-5.6-luna | biography | 0.003350 | 0.016752 | -0.069179 | 0.020784 |
| PROMPT | Phi | persona | 0.044058 | 0.176232 | -0.085875 | 0.039488 |
| PROMPT | Phi | history | 0.843331 | 0.843331 | -0.009805 | 0.048858 |
| PROMPT | Phi | open-v2 | 0.362784 | 0.725567 | 0.043929 | 0.046973 |
| PROMPT | Phi | diversified | 0.167121 | 0.501363 | 0.058174 | 0.040307 |
| PROMPT | Phi | structured | 0.007575 | 0.037876 | -0.111879 | 0.036939 |
| AUTOGEN | DeepSeek | autogen | 0.105888 | 0.105888 | 0.067346 | 0.034181 |
| AUTOGEN | GPT-4-mini | autogen | 0.842436 | 0.842436 | -0.007999 | 0.038774 |
| AUTOGEN | GPT-5.6-luna | autogen | 0.155162 | 0.155162 | 0.081875 | 0.048936 |
| AUTOGEN | Phi | autogen | 0.062002 | 0.062002 | 0.200312 | 0.083637 |
| STEERING | Llama | steering | 0.407953 | 0.407953 | 0.022206 | 0.024593 |
| UNREGULATED | Cydonia | unregulated | 0.118711 | 0.118711 | -0.038195 | 0.022865 |
| UNREGULATED | Noromaid | unregulated | 0.634087 | 0.634087 | 0.012050 | 0.024723 |
| MIXAGENT | Mixed | mixed | 0.650798 | 0.650798 | -0.010695 | 0.023084 |
| PERTURBATION | GPT-4-mini | noise | 0.263846 | 0.263846 | 0.051606 | 0.042483 |
| RL(vs_second_interval) | Qwen | rl | 0.338448 | 0.338448 | 0.094176 | 0.089009 |
| RL(vs_first_interval) | Qwen | rl | 0.000042 | 0.000042 | 0.402897 | 0.030146 |
| SCALINGLAW | DeepSeek | 1.0 | 0.280396 | 0.280396 | 0.048738 | 0.042101 |
| SCALINGLAW | DeepSeek | 5.0 | 0.132931 | 0.265862 | -0.037106 | 0.022184 |
| SCALINGLAW | DeepSeek | 10.0 | 0.273794 | 0.273794 | -0.036146 | 0.028521 |
| SCALINGLAW | GPT-4-mini | 1.0 | 0.923869 | 0.923869 | -0.003720 | 0.037956 |
| SCALINGLAW | GPT-4-mini | 5.0 | 0.160184 | 0.320368 | 0.083746 | 0.055199 |
| SCALINGLAW | GPT-4-mini | 10.0 | 0.106791 | 0.106791 | -0.018208 | 0.009605 |
| SCALINGLAW | GPT-5.6-luna | 1.0 | 0.055935 | 0.066411 | -0.070689 | 0.031641 |
| SCALINGLAW | GPT-5.6-luna | 5.0 | 0.033206 | 0.066411 | -0.075260 | 0.029301 |
| SCALINGLAW | Phi | 1.0 | 0.201997 | 0.403994 | -0.061070 | 0.043937 |
| SCALINGLAW | Phi | 5.0 | 0.340868 | 0.403994 | -0.048876 | 0.048265 |
| SCALINGLAW | Phi | 10.0 | 0.224037 | 0.224037 | -0.020437 | 0.014220 |
| DECODING | Llama | decoding | 0.177171 | 0.177171 | -0.034214 | 0.021790 |
| AGENTSOCIETY | DeepSeek | agentsociety | 0.209297 | 0.627890 | -0.128309 | 0.089077 |
| AGENTSOCIETY | GPT-4-mini | agentsociety | 0.433510 | 0.867021 | 0.033730 | 0.040600 |
| AGENTSOCIETY | Qwen | agentsociety | 0.047546 | 0.190186 | -0.080528 | 0.030828 |
| AGENTSOCIETY | GPT-5.6-luna | agentsociety | 0.939072 | 0.939072 | 0.002599 | 0.032339 |
| NOMEMORY | GPT-4-mini | 0.9 | 0.098543 | 0.098543 | -0.073847 | 0.038774 |

**Table S9 | Full regression results for intervention effects on within-run semantic diversity.** This table reports the factor-specific regression outputs underlying the intervention comparisons reported in the main and Extended Data figures. Coefficients are expressed relative to the corresponding standard condition; positive values indicate greater within-run semantic diversity relative to baseline.

| Factor | Model | Level | Raw *p* | Adjusted *p* | Coefficient | SE |
|---|---|---|---|---|---|---|
| TEMPERATURE | DeepSeek | 0 | 0.089699 | 0.269097 | -0.053387 | 0.029283 |
| TEMPERATURE | DeepSeek | 0.5 | 0.896476 | 0.896476 | -0.004043 | 0.030516 |
| TEMPERATURE | DeepSeek | 1.2 | 0.232192 | 0.464385 | -0.038761 | 0.031037 |
| TEMPERATURE | DeepSeek | 2.0 | 0.019775 | 0.079102 | -0.080432 | 0.030579 |
| TEMPERATURE | GPT-4-mini<br>GPT-4-mini | 0 | 0.825856 | 0.825856 | -0.011434 | 0.051322 |

| Factor | Model | Level | Raw *p* | Adjusted *p* | Coefficient | SE |
|---|---|---|---|---|---|---|
| TEMPERATURE | | 0.5 | 0.184106 | 0.368212 | -0.062867 | 0.045774 |
| TEMPERATURE | GPT-4-mini | 1.2 | 0.004879 | 0.014638 | -0.156992 | 0.049906 |
| TEMPERATURE | GPT-4-mini | 2.0 | 0.000022 | 0.000087 | -0.246107 | 0.045294 |
| TEMPERATURE | GPT-5.6-luna | 0 | 0.036189 | 0.144756 | -0.052720 | 0.022757 |
| TEMPERATURE | GPT-5.6-luna | 0.5 | 0.429171 | 0.858342 | -0.024064 | 0.029556 |
| TEMPERATURE | GPT-5.6-luna | 1.2 | 0.280948 | 0.842844 | 0.073812 | 0.065816 |
| TEMPERATURE | GPT-5.6-luna | 2.0 | 0.662964 | 0.858342 | 0.018501 | 0.041554 |
| TEMPERATURE | Phi | 0 | 0.132723 | 0.265446 | 0.094591 | 0.059254 |
| TEMPERATURE | Phi | 0.5 | 0.945520 | 0.945520 | 0.009466 | 0.136064 |
| TEMPERATURE | Phi | 1.2 | 0.000404 | 0.001211 | -0.255905 | 0.055494 |
| TEMPERATURE | Phi | 2.0 | 0.000006 | 0.000023 | -0.340212 | 0.048255 |
| MAXTOKEN | DeepSeek | 1500 | 0.036867 | 0.036867 | -0.083402 | 0.029517 |
| MAXTOKEN | GPT-4-mini | 1500 | 0.667802 | 0.667802 | -0.024568 | 0.055844 |
| MAXTOKEN | GPT-5.6-luna | 1500 | 0.005925 | 0.005925 | -0.098677 | 0.021526 |
| MAXTOKEN | Phi | 1500 | 0.008961 | 0.008961 | -0.195053 | 0.047067 |
| RAG | DeepSeek | rag | 0.289829 | 0.289829 | -0.034120 | 0.028831 |
| RAG | GPT-4-mini | rag | 0.000897 | 0.000897 | -0.192713 | 0.044004 |
| RAG | GPT-5.6-luna | rag | 0.001932 | 0.001932 | -0.121820 | 0.020510 |
| RAG | Phi | rag | 0.447232 | 0.447232 | -0.053916 | 0.065399 |
| PROMPT | DeepSeek | persona | 0.007872 | 0.039358 | -0.076527 | 0.025919 |
| PROMPT | DeepSeek | history | 0.704414 | 0.704414 | -0.009790 | 0.025439 |
| PROMPT | DeepSeek | open-v2 | 0.307162 | 0.614324 | 0.026731 | 0.025508 |
| PROMPT | DeepSeek | diversified | 0.138311 | 0.414932 | -0.040599 | 0.026298 |
| PROMPT | DeepSeek | structured | 0.027742 | 0.110969 | -0.061676 | 0.025983 |
| PROMPT | GPT-4-mini | persona | 0.018171 | 0.072168 | -0.110434 | 0.043914 |
| PROMPT | GPT-4-mini | history | 0.024924 | 0.072168 | -0.126622 | 0.053319 |
| PROMPT | GPT-4-mini | open-v2 | 0.018042 | 0.072168 | -0.111719 | 0.044370 |
| PROMPT | GPT-4-mini | structured | 0.000236 | 0.001417 | -0.193279 | 0.045623 |
| PROMPT | GPT-4-mini | biography | 0.013654 | 0.068268 | -0.116632 | 0.044204 |
| PROMPT | GPT-4-mini | diversified | 0.294343 | 0.294343 | -0.054442 | 0.050907 |
| PROMPT | GPT-5.6-luna | persona | 0.000280 | 0.001401 | -0.094362 | 0.021477 |
| PROMPT | GPT-5.6-luna | history | 0.000725 | 0.002152 | -0.076156 | 0.019100 |
| PROMPT | GPT-5.6-luna | open-v2 | 0.000717 | 0.002152 | -0.075024 | 0.018795 |
| PROMPT | GPT-5.6-luna | structured | <0.000001 | <0.000001 | -0.203166 | 0.017640 |
| PROMPT | GPT-5.6-luna | biography | 0.000298 | 0.001401 | -0.089925 | 0.020587 |
| PROMPT | GPT-5.6-luna | diversified | 0.027042 | 0.027042 | -0.046891 | 0.019654 |
| PROMPT | Phi | persona | 0.000181 | 0.000725 | -0.198677 | 0.041729 |
| PROMPT | Phi | history | 0.371984 | 0.743969 | -0.040898 | 0.044602 |
| PROMPT | Phi | open-v2 | 0.000352 | 0.001057 | -0.197760 | 0.044454 |
| PROMPT | Phi | diversified | 0.995760 | 0.995760 | 0.000306 | 0.056789 |
| PROMPT | Phi | structured | 0.000105 | 0.000525 | -0.215183 | 0.042858 |
| AUTOGEN | DeepSeek | autogen | 0.206411 | 0.206411 | 0.046176 | 0.031818 |
| AUTOGEN | GPT-4-mini | autogen | 0.037766 | 0.037766 | -0.122898 | 0.052646 |
| AUTOGEN | GPT-5.6-luna | autogen | 0.068499 | 0.068499 | 0.048674 | 0.021027 |
| AUTOGEN | Phi | autogen | 0.943925 | 0.943925 | -0.004239 | 0.057336 |
| STEERING | Llama | steering | 0.822459 | 0.822459 | 0.009922 | 0.041963 |
| UNREGULATED | Cydonia | unregulated | 0.020655 | 0.020655 | -0.106330 | 0.041912 |
| UNREGULATED | Noromaid | unregulated | 0.180449 | 0.180449 | 0.098275 | 0.070526 |
| MIXAGENT | Mixed | mixed | 0.027687 | 0.027687 | -0.085652 | 0.035757 |
| PERTURBATION | GPT-4-mini | noise | 0.616112 | 0.616112 | -0.016722 | 0.032488 |
| SCALINGLAW | DeepSeek | 1.0 | 0.036725 | 0.065572 | 0.218459 | 0.087252 |
| SCALINGLAW | DeepSeek | 5.0 | 0.032786 | 0.065572 | -0.078361 | 0.030412 |
| SCALINGLAW | DeepSeek | 10.0 | 0.022663 | 0.022663 | -0.128272 | 0.029598 |
| SCALINGLAW | GPT-4-mini | 1.0 | 0.076843 | 0.153685 | -0.108256 | 0.056979 |
| SCALINGLAW | GPT-4-mini | 5.0 | 0.857468 | 0.857468 | 0.017154 | 0.093882 |
| SCALINGLAW | GPT-4-mini | 10.0 | 0.004588 | 0.004588 | -0.111504 | 0.030691 |
| SCALINGLAW | GPT-5.6-luna | 1.0 | 0.001427 | 0.002855 | -0.115853 | 0.024341 |
| SCALINGLAW | GPT-5.6-luna | 5.0 | 0.544136 | 0.544136 | -0.016330 | 0.025781 |
| SCALINGLAW | Phi | 1.0 | 0.000592 | 0.001183 | -0.277975 | 0.050781 |
| SCALINGLAW | Phi | 5.0 | 0.008370 | 0.008370 | -0.187444 | 0.053926 |

| Factor | Model | Level | Raw *p* | Adjusted *p* | Coefficient | SE |
|---|---|---|---|---|---|---|
| SCALINGLAW | Phi | 10.0 | 0.077905 | 0.077905 | -0.187060 | 0.070956 |
| DECODING | Llama | decoding | 0.000656 | 0.000656 | -0.106128 | 0.014105 |
| RL | Qwen | rl | 0.000320 | 0.000320 | 0.259467 | 0.029588 |
| AGENTSOCIETY | DeepSeek | agentsociety | 0.166664 | 0.166664 | -0.075381 | 0.046600 |
| AGENTSOCIETY | GPT-4-mini | agentsociety | 0.078422 | 0.156844 | -0.047533 | 0.024708 |
| AGENTSOCIETY | Qwen | agentsociety | 0.002159 | 0.008636 | -0.259118 | 0.044730 |
| AGENTSOCIETY | GPT-5.6-luna | agentsociety | 0.005589 | 0.016768 | −0.097792 | 0.021037 |
| NOMEMORY | GPT-4-mini | 0.9 | 0.002921 | 0.002921 | -0.177912 | 0.047812 |

**Table S10 | Full regression results for intervention effects on cross-run semantic diversity.** This table reports the factor-specific regression outputs underlying the intervention comparisons shown in the main and Extended Data figures for cross-run semantic diversity. Coefficients are expressed relative to the corresponding standard condition; positive values indicate greater cross-run semantic diversity relative to baseline.

# Supplementary Note 4. Diagnostics and mechanistic analyses

## 4.1 Transcript-level tail-token erosion and survival analysis

To characterize tail erosion directly from the generated transcripts, we perform a transcript-level survival analysis that does not rely on internal model probabilities. For each run, the dialogue trajectory is partitioned into non-overlapping intervals, and each 10-round interval is downsampled to a common token budget to reduce artifactual differences induced by variable output length. We then define head and tail token sets once from the true early segment only, using the top and bottom 10% of cumulative frequency mass, respectively, and hold these sets fixed thereafter. For each token in the fixed sets, we record its last occurrence 10-round interval and construct survival curves as the fraction of tokens remaining extant at each subsequent interval. We summarize the resulting head–tail asymmetry using three complementary quantities: an integrated survival gap (the difference in area under the head and tail survival curves), a half-survival lag (the difference in the first 10-round interval at which survival falls below 0.5), and a terminal survival gap (the head–tail difference at the final common interval).

This asymmetry remains pronounced and highly consistent when using run as the unit of inference. The integrated survival gap remains strongly positive ($\Delta$ = 10.373, 95% CI [9.905, 10.845]), indicating that the head survival curve lay systematically above the tail curve across the shared 200 round horizon. The half-survival lag is substantial (11.458 intervals, 95% CI [10.528, 12.354]), showing that head tokens reach the 50% survival threshold markedly later than tail tokens. The terminal survival gap also remains large at the final common 10-round interval ($\Delta$ = 0.748, 95% CI [0.709, 0.784]). Directional consistency is complete across runs, and the pooled survival estimates at the final 10-round interval remain widely separated (head, 0.922; tail, 0.174). The same directional pattern is observed across the five main model families.

These results provide direct transcript-level evidence that lexical material in the low-probability region of the early distribution is progressively lost under closed-loop interaction, whereas initially dominant lexical material remains comparatively persistent. We therefore interpret Fig. 4a as showing a robust empirical survival asymmetry between early-defined head and tail token sets, consistent with contraction dynamics under recursive self-conditioning.

## 4.2 Attention-head intensification and attenuation

To test whether semantic convergence is accompanied by induction-like retrieval dynamics inside the model, we conduct a transcript-only, teacher-forcing white-box analysis using Meta-Llama-3-8B-Instruct as the diagnostic model. Canonical induction heads are first identified using an independent calibration benchmark of repeated random token sequences of the form [BOS] + X + X, ranking heads by their mean attention mass along the induction stripe. We then evaluate these fixed heads in selected transcript rounds, treating each selected round as one analysis unit and mining recurrence events over all tokens within that round. For each event, true historical positions, where an antecedent token A has previously been followed by its dominant continuation $B^*$, are contrasted with matched control positions, where the same antecedent has instead been followed by alternative successors.

Fixed canonical induction heads show a robust continuation-specific retrieval bias during realized multi-agent interaction. Across analysis rounds, they allocate significantly more attention to historical positions associated with dominant continuations than to matched control positions (one-sided Wilcoxon $p = 4.220 \times 10^{-7}$; bootstrap 95% CI [0.039, 0.074]). The effect is stronger in the highest-ranked canonical subset: for the top four heads, mean attention mass to true continuation positions is 0.214, compared with 0.129 for matched controls, yielding a mean retrieval margin of $\Delta B = 0.086$, with positive margins in 95% of analyzed rounds. This retrieval bias is accompanied by strong output-space promotion of the historically dominant continuation. Across 759 induction events, $B^*$ exceeds matched control successors by a mean logit margin of 3.920 (median = 4.625; one-sided Wilcoxon $p = 4.523 \times 10^{-27}$); $B^*$ is the top-1 prediction in 61.1% of events, and appears within the top-10 in 84.1%.

Attention-level retrieval and output-level promotion are tightly coupled across the trajectory. At the analysis-round level, stronger continuation-specific retrieval is associated with stronger logit promotion of $B^*$, both across the full fixed canonical set (Pearson $r = 0.623$, $p = 0.003$; Spearman $\rho = 0.606$, $p = 0.005$) and within the top-four canonical subset (Pearson $r = 0.621$, $p = 0.004$; Spearman $\rho = 0.630$, $p = 0.003$). Retrieval in the top-four canonical heads also shows a modest but significant strengthening trend across the analyzed trajectory (linear slope $p = 0.038$). Together, these results indicate that induction-like heads preferentially retrieve historically dominant continuations and promote them in the output distribution during multi-agent interaction.

## 4.3 Descriptive all-head scan of temporal circuit reorganization

To complement the fixed induction-head probe and support the descriptive pattern shown in Fig. 4b, we perform a targeted all-head scan using the same diagnostic model under teacher forcing. Rather than focusing on a predefined functional subset, this analysis surveys all attention heads across temporally distributed analysis rounds spanning the interaction trajectory, evaluating all token query positions within each round with stride 1. For each head, we quantify the norm of the head output before output projection together with the peakedness of its attention distribution, and summarize temporal change across the analyzed rounds using per-head trend estimates with false-discovery-rate correction. This analysis is intended as a descriptive characterization of broad circuit reorganization rather than as a standalone mechanistic claim. Consistent with Fig. 4b, later-layer circuitry exhibited heterogeneous temporal change, with a subset of heads showing increasing trajectories and others showing attenuation over interaction. Because these patterns are metric-dependent and less functionally specific than the independently calibrated induction-head analysis, we interpret them as supportive evidence for broad reorganization of attention circuitry, while reserving mechanistic interpretation primarily for the canonical induction-like heads.

## 4.4 Cultural-axis projection and directional drift

The preceding analyses establish that open-ended multi-agent trajectories contract toward a narrow semantic regime over long interaction horizons. We next ask how this contraction is oriented in interpretable semantic space. For each run, we summarized the trajectory by early- and late-stage centroids computed from interval dialogue embeddings and projected these centroids onto a set of predefined bipolar semantic probe axes. Following prior work on the geometry of cultural meaning in embedding spaces[53], each axis was defined as the mean of multiple antonym-pair difference vectors in the same representation space, so that the projection score captures alignment with a shared semantic direction rather than dependence on any single lexical contrast. Full antonym-pair lists are provided in Table S11, and Fig. 4c visualizes the subset shown in the main text. We interpret early-to-late displacement on these axes as a descriptive readout of directional semantic drift, with confidence intervals summarizing consistency across independent runs.

Under these probe axes, collapse is not isotropic. Late-stage trajectories do not simply converge in an undifferentiated geometric sense; rather, they show structured displacement along a limited subset of the axes displayed in Fig. 4c. Across model families, the clearest shared movement appears along questioning–declarative, egalitarian–hierarchical, optimism–pessimism, individualism–collectivism, and cooperation, competition, with a weaker but still visible shift on black–white, whereas emotional warmth–detached and masculine–feminine are more heterogeneous. Convergence is therefore accompanied by directional reorganization of semantic state space rather than by a mere reduction in variance. In this sense, collapse is selective as well as contractive: recursive interaction compresses trajectories toward model-specific late-stage basins while leaving the orientation of those basins only partly shared across model families.

At the same time, the terminal semantic profile is not identical across models. GPT-5.6-Luna, DeepSeek-v3, GPT-4-mini, and Phi-4 all show contraction, but they do not collapse onto exactly the same endpoint under the probed axes. Rather, each model family occupies a partially distinct late-stage basin, with differences in both final projection level and early-to-late drift magnitude. This indicates that semantic convergence is universal at the level of contraction, yet model-dependent in its orientation. Put differently, the long-horizon dynamics consistently reduce accessible semantic volume, but the semantic character of the region that remains is still shaped by the underlying model family.

Importantly, we interpret these axis projections as descriptive semantic probes rather than as latent ground-truth psychological or sociological variables. Their role is diagnostic: they provide an interpretable readout of the semantic orientation of late-stage trajectories, not a claim that the system literally acquires any human-like trait named by the axis labels. This distinction is especially important because the axes are manually constructed and intentionally lightweight. Their value lies in revealing that collapse has directional structure. In this sense, the projection analysis complements the mechanistic diagnostics rather than replacing them. Whereas the induction-head analysis identifies a plausible token-level pathway by which historically reinforced continuations are preferentially retrieved and promoted, the axis-projection analysis shows the higher-level semantic regions toward which those self-reinforcing dynamics tend to steer the interaction. Together, the two views link microscopic continuation dynamics to macroscopic semantic lock-in.

| Axis | Antonym pairs used to construct axis |
|---|---|
| **questioning ↔ declarative** | ask / state; question / answer; curious / certain<br>explore / conclude; why / therefore; what if / it is<br>hypothesis / claim; open ended / closed form; inquire / assert<br>probe / pronounce; consider / decide; investigate / declare |
| **egalitarian ↔ hierarchical** | equality / hierarchy; flat structure / chain of command; shared power / central authority<br>equal voice / ranked voice; peer / superior; fairness / status<br>equal chance / privileged access; inclusive / exclusive; merit based / title based<br>participatory / directive; co leader / boss; horizontal / vertical |
| **emotional warmth ↔ detached** | empathetic / detached; warm / cold; emotional / clinical<br>compassionate / impersonal; care / indifference; feelings / facts<br>sympathetic / neutral; reassure / inform; supportive / strict<br>soft tone / hard tone; heartfelt / matter of fact; comfort / critique |
| **optimism ↔ pessimism** | hopeful / cynical; optimistic / pessimistic; promising / bleak<br>confidence / worry; opportunity / threat; can succeed / will fail<br>upbeat / downbeat; encouraging / discouraging; silver lining / worst case<br>trust / mistrust |
| **masculine ↔ feminine** | man / woman; male / female; he / she<br>him / her; masculine / feminine; boy / girl<br>gentleman / lady; father / mother |
| **individualism ↔ collectivism** | individualism / collectivism; self-reliance / interdependence; personal freedom / group harmony<br>individual rights / collective responsibility; personal achievement / shared wellbeing; competition / cooperation<br>autonomy / conformity; self-interest / community interest |
| **cooperation ↔ competition** | cooperate / compete; collaborative / rivalrous; win win / zero sum<br>mutual benefit / personal advantage; share / hoard; support / undermine<br>align / outdo; compromise / dominate; consensus / victory<br>teamwork / one upmanship; build together / beat others; reciprocity / exploitation |

| Axis | Antonym pairs used to construct axis |
|---|---|
| **black ↔ white** | black / white; African American / white American; Black community / White community<br>Black culture / White culture; Black identity / White identity; Black people / White people<br>Black history / White history; Black experience / White experience |

**Table S11. Antonym-pair definitions for the cultural-axis probes.** Each bipolar axis was defined as the mean of the corresponding antonym-pair difference vectors in the shared embedding space; only the axes visualized in the main text are listed.

## 4.5 Model attribution under semantic convergence

To test whether extended interaction makes outputs from different model families harder to distinguish, we constructed a frozen five-class model-attribution classifier spanning GPT-4-mini, DeepSeek-V3, Phi-4, GPT-5.6 Terra, and Claude Sonnet 5. The classifier was trained on canonical utterance-level embeddings from three independent 200-round Standard baseline reference runs per model. Training utterances were weighted so that each reference run contributed equally to model fitting. After training, the classifier was frozen and evaluated without further updating on three separate 1,000-round Standard baseline runs per model. At test time, all valid agent utterances produced within each round were concatenated into a single round-level document, embedded using text-embedding-3-large, and classified independently. The 1,000-round trajectories were not used to fit the classifier. We used multinomial logistic regression with L2 regularization on L2-normalized embeddings. This design tests whether a classifier trained on short baseline interactions can still identify the generating model during much longer interactions.

Across the independent 1,000-round test trajectories, the frozen classifier achieved an overall attribution accuracy of 97.42%, with a mean probability assigned to the true generating model, *P*(true model), of 0.890. Mean *P*(true model) was 0.724 for GPT-4-mini, 0.870 for DeepSeek-V3, 0.893 for Phi-4, 0.966 for GPT-5.6 Terra, and 0.994 for Claude Sonnet 5. The corresponding classification accuracies were 92.1%, 96.1%, 98.9%, 100%, and 100%, respectively (Extended Data Fig. 6).

Classification performance remained high throughout the 1,000-round trajectories. Mean *P*(true model) was 0.857 over rounds 1–200, 0.900 over rounds 401–600, and 0.899 over rounds 801–1,000. Corresponding attribution accuracies were 93.0%, 98.1%, and 99.0%. The mean run-level late-minus-early change in *P*(true model) was +0.041 (95% bootstrap CI [−0.005, 0.109]), providing no evidence that the generating model became harder to identify over extended interaction.

Thus, although semantic diversity decreased over extended interaction, outputs from the five model families remained readily distinguishable. Semantic convergence therefore did not erase differences that allowed the classifier to identify which model generated the text.

## Supplementary Note 5. Empirical dynamical synthesis of semantic collapse

The empirical findings admit a compact theoretical interpretation. Consider an idealized representation of the multi-agent system as a Markov chain over semantic distributions: $P_0 \xrightarrow{\mathcal{T}} P_1 \xrightarrow{\mathcal{T}} P_2 \to \cdots$, where the transition channel $\mathcal{T}$ is the composite map from context to generated output across all agents. This channel is self-referential (its input at round $t+1$ is constructed from its output at round $t$) and mode-amplifying (autoregressive generation concentrates probability mass on well-supported continuations).

These classical results provide useful theoretical analogies under stronger assumptions, but the present analyses treat them as heuristic guides rather than established properties of the empirical system. First, the Data Processing Inequality (DPI) guarantees that mutual information between the initial state $S_0$ and the current state $S_t$ is monotonically non-increasing: $I(S_0; S_t) \leq I(S_0; S_{t-1})$. Second, if an idealized channel satisfies suitable mixing conditions, there exists a unique stationary distribution $P^*$ toward which the idealized dynamics may relax approximately exponentially:

$$H(P_t) \approx H(P^*) + [H(P_0) - H(P^*)] \cdot e^{-\gamma t}$$

where $\gamma$ denotes an illustrative relaxation rate motivated by spectral-gap theory and $H(P^*)$ is the attractor entropy, a model-specific floor. Third, the Fisher information of the semantic distribution with respect to any intervention parameter $\theta$ may decay geometrically under suitable contraction conditions : $\mathcal{I}(\theta; t) \leq \eta^t \cdot \mathcal{I}(\theta; 0) \to 0$. Such decay suggests diminishing local sensitivity to perturbations as the idealized system approaches stationarity.

In this idealized interpretation, these results suggest a creativity horizon: the timescale over which the system retains meaningful capacity for novel semantic content scales as $T^* \sim \gamma^{-1} \ln(\Delta H_0 / \Delta H_{\text{thr}})$, growing only logarithmically with initial diversity. Richer initialization buys marginal additional time; substantially extending the horizon requires reducing $\gamma$ (weakening the channel's mode amplification) or raising $H(P^*)$ (changing the attractor itself).

From algorithmic information theory, a complementary bound applies. The Kolmogorov complexity of round-*t* output satisfies $K(x_t) \leq K(\mathcal{M}) + K(x_0) + \log t + c$: outputs cannot exceed the system's own descriptive complexity, and recursive operation adds at most $\log t$ bits[39,40]. The system is trapped within its own complexity class.

First, the information-theoretic framework motivates a specific empirical expectation about long-horizon trajectories. Under classical mixing and entropy-contraction results, repeated processing can approach a bounded stationary regime approximately exponentially. If the semantic trajectories observed here reflect an analogous structured contraction rather than unconstrained drift, then their early evolution should contain predictive information about the late-stage level they eventually approach. We test this prediction in Section 5.1 without assuming that the empirical embedding trajectory directly instantiates the classical channel model or that its fitted parameters correspond to the underlying theoretical channel quantities.

Second, if later outputs are increasingly generated from a history already dominated by prior model-produced structure, then their apparent novelty should decline even when surface-level lexical variation continues. Semantic collapse should therefore appear not only as contraction in embedding space, but also as a reduction in the amount of non-redundant structure introduced over time: later discourse becomes progressively more derivable from earlier discourse, and the effective generative horizon narrows under repeated self-application. This is the supplementary counterpart of the main-text analogy of subcritical behavior: perturbations may transiently redirect the trajectory, but they are not amplified into sustained open-ended divergence.

Third, closure does not imply that all model families converge to an identical endpoint; rather, each remains constrained by the attractor landscape defined by its own channel properties, producing family-specific late-stage floors.

Fourth, increasing heterogeneity by mixing models does not by itself break recursive closure or expand the system's information budget. It may alter the path of contraction, but not the fact of contraction itself.

The four subsections below examine these consequences in turn: the predictability of late-stage contraction from early dynamics, the rise of history-conditioned redundancy, the model-specific character of late-stage regimes, and the persistence of contraction under mixed-agent composition.

## 5.1 Shannon-inspired early-to-late prediction of semantic contraction

We next ask whether the early trajectory of a run contains predictive information about its later semantic-diversity level. Motivated by the Shannon-information-theoretic contraction analogy developed above, we test whether an exponential approach toward a bounded regime can extrapolate from the beginning of a trajectory to its much later plateau. Specifically, we fit a saturating exponential using only the first 50 rounds of each 1,000-round run and use that fit to predict mean semantic diversity over rounds 951–1000. The analysis therefore evaluates a theory-motivated empirical regularity, rather than treating the classical information-theoretic results as a directly validated model of the observed system.

We analyzed 1,000-round standard runs across five model families: DeepSeek-V3, GPT-4-mini, Phi-4, GPT-5.6 Terra, and Claude Sonnet 5. Because this analysis focuses on the initial approach to the late-stage regime, we used round-level embeddings rather than the 10-round intervals used in the primary trajectory analyses. For each run, anchored semantic diversity at round t was defined as the cosine distance from the first-round embedding:

$$D(t) = 1 - \cos(\mathbf{e}_t, \mathbf{e}_1)$$

Larger values therefore indicate greater semantic displacement from the initial state. Values between $-10^{-5}$ and 0 arising from numerical precision were clipped to zero.

We fitted the first 50 rounds of each trajectory with a saturating exponential:

$$\hat{D}(t) = d_\infty \left[1 - \exp\left(-\gamma(t-1)\right)\right]$$

where $d_\infty$ represents the fitted asymptotic displacement level and $\gamma$ controls the rate at which the fitted trajectory approaches that level. Fits were obtained using bounded nonlinear least squares with deterministic multistart initialization. The fitted asymptote was constrained to 0–2 and the rate parameter from $10^{-8}$ to 1.

The observed late-stage level was defined as the mean anchored semantic diversity over rounds 951–1000:

$$D_{\text{obs}}^{\text{late}} = \frac{1}{50} \sum_{t=951}^{1000} D(t)$$

The predicted late-stage level was obtained by evaluating the early-round fit over the same 50 round indices:

$$D_{\text{pred}}^{\text{late}} = \frac{1}{50} \sum_{t=951}^{1000} \hat{D}(t)$$

Prediction accuracy was quantified by the absolute difference between the observed and predicted late-stage levels for each run. The primary summary was the overall mean absolute error (MAE), with a 95% confidence interval obtained from 10,000 run-level percentile-bootstrap resamples. We also calculated the Pearson correlation between predicted and observed late-stage levels. To summarize performance across model families, we additionally calculated a family-equal MAE by first averaging within each family and then assigning equal weight to the five family estimates.

The saturating-exponential form was chosen because exponential relaxation toward a bounded stationary regime is the qualitative trajectory motivated by the Shannon/Markov contraction analogy above. In the present analysis, however, the function is used only as an empirical predictive model for anchored cosine distance.

To test for systematic prediction bias, we defined the signed error as observed minus predicted late-stage diversity. Within each model family, observed and predicted values were compared using two-sided paired t-tests, with Holm correction across the five family-specific tests. We also tested the five equally weighted family-mean signed errors against zero to assess whether the early fits systematically over- or underpredicted the late-stage level across model families.

The first-50-round fits predicted late-stage semantic-diversity levels with an overall MAE of 0.060 (SD of run-level absolute errors = 0.047; 95% bootstrap CI, 0.042–0.080). Predicted and observed late-stage levels were positively associated across runs ($r = 0.692$). The corresponding family-equal MAE was 0.063. The family-equal mean signed error was +0.0396 (t-based 95% CI, −0.021 to 0.100; $p = 0.142$), providing no evidence of systematic over- or underprediction across model families. Likewise, no individual model family showed a significant signed prediction error after Holm correction (Table S12).

Taken together, these results show that the first 50 rounds contain meaningful information about the semantic-diversity level reached much later in the same trajectory. The prediction is informative rather than exact, but predicted and observed late-stage levels covary substantially across runs and show no reliable systematic over- or underprediction across model families. This early-to-late predictability is consistent with the qualitative contraction intuition motivating the exponential model: trajectories approaching a bounded late-stage regime should reveal aspects of that regime early in their evolution. We do not interpret this predictive success as validating the classical information-theoretic results or as estimating their underlying channel parameters. In particular, the fitted asymptote and rate parameter are descriptive quantities and are not identified with attractor entropy, the spectral gap, or other parameters introduced in the theoretical analogies above. Instead, this early-to-late predictability provides a quantitative empirical regularity—and a concrete predictive target—for future theories of semantic contraction in interacting LLM systems.

| **Model** | **MAE** | **Mean observed − predicted** | **Holm-adjusted p** |
|---|---|---|---|
| GPT-5.6 Terra | 0.1023 | +0.1023 | 0.713 |
| Claude Sonnet 5 | 0.0606 | +0.0462 | 0.920 |
| DeepSeek-V3 | 0.0658 | +0.0623 | 0.094 |
| GPT-4-mini | 0.0540 | −0.0246 | 0.936 |
| Phi-4 | 0.0335 | +0.0115 | 0.936 |

**Table S12. Late-stage diversity prediction by model family.** MAE denotes the mean absolute error between observed late-stage semantic diversity and the value predicted from a saturating exponential

fitted to the first 50 rounds. Mean observed − predicted reports the signed prediction error. Holm-adjusted p values are from two-sided paired t-tests within each model family, corrected across the five family-specific comparisons.

## 5.2 Compression-based evidence for increasing derivability from history

To test whether later transcript intervals become increasingly derivable from earlier interaction history, we quantify a gzip-based cumulative conditional compression ratio for each interval. Let $W_t$ denote the current interval and $H_{t-1}$ the preceding transcript history. We define the conditional compressed cost of the current interval given its prior history as

$$\Delta_{\text{hist}} = C_{\text{gzip}}(H_{t-1} \parallel W_t) - C_{\text{gzip}}(H_{t-1}),$$

and the cumulative conditional compression ratio as

$$R_t = \frac{|W_t|}{\Delta_{\text{hist}}},$$

where $|W_t|$ is the raw byte length of the current interval. Under this definition, larger values indicate that the current interval is more compressible conditional on prior context, and therefore more redundant with respect to earlier interaction.

For inference, we treat the run rather than the individual interval as the independent unit, because intervals within a run are temporally dependent. For each run, we compute the early- and late-interval means of the cumulative conditional compression ratio and form a run-level late-minus-early difference. We then test whether this difference is greater than zero using a one-sided Wilcoxon signed-rank test, with a one-sample *t*-test reported as a robustness check on the same run-level differences.

The results support increasing conditional redundancy over time. Late-interval conditional redundancy exceeds early-interval conditional redundancy, with a mean late-minus-early difference of 4.115 and a median difference of 2.579. This increase is supported by a one-sided Wilcoxon signed-rank test ($p$ = 3.66 × $10^{-5}$), with the same directional conclusion under a one-sample *t*-test ($t$ = 3.705, one-sided $p$ = 6.17 × $10^{-4}$).

We next ask whether this increase reflects history-specific derivability rather than generic late-stage textual regularization or model-family style. For each current interval, we compare compression under the run's own true prior history with three matched controls: prior histories from other independent runs of the same model family, shuffled versions of the same run's prior intervals , and reversed versions of the same run's prior intervals . We define true-history advantage as the reduction in conditional compression cost obtained by conditioning on a run's own prior history rather than on a matched control history. Under the primary gzip recent-history setting, this advantage increases from early to late intervals for all three controls: same-model wrong histories (late–early increase = 0.0067, one-sided Wilcoxon $p$ = 0.0138), shuffled own histories (late–early increase = 0.0140, $p = 7.63 \times 10^{-5}$), and reversed own histories (late–early increase = 0.0144, $p = 1.05 \times 10^{-5}$). The same direction is observed across all completed compressor and history-budget robustness checks. These controls show that the increasing conditional compressibility is history-specific: late-stage outputs become progressively cheaper to encode under the trajectory that produced them. This provides transcript-level evidence for the recursive self-conditioning account, in which closed-loop interaction increasingly constrains new generations by the system's own accumulated history.

## 5.3 Late-stage contraction floors are model-family specific

A further implication of the contraction framework is that closure need not drive all model families to an identical late-stage level. Rather, if the effective recursive channel differs across model families, then the asymptotic contraction floor should also differ, even when the overall direction of the dynamics is shared. We therefore ask whether the late-stage semantic regime remains separated across the four standard model families, or whether all trajectories converge to a common terminal level.

Late-stage diversity estimates continue to vary across model families. Phi-4 shows the highest late-stage mean (0.405), followed by GPT-4-mini (0.346), GPT-5.6 Luna (0.344), and DeepSeek-V3 (0.219), yielding a gap of 0.187 between the highest and lowest family means. Early-to-late change is positive for Phi-4 (+0.053), GPT-4-mini (+0.061), and GPT-5.6 Luna (+0.070), but minimal for DeepSeek-V3 (+0.009). Thus, under a common embedding-based analysis, the four model families do not converge to a single indistinguishable late-stage level. These results support the descriptive conclusion that late-stage semantic floor estimates differ across model families, while remaining a family-comparison result rather than a strong claim about distinct attractor entropies.

## 5.4 Model heterogeneity does not escape the late-stage contraction regime

We next ask whether combining multiple model families altered the late-stage semantic regime, or whether mixed-agent systems remain confined to the range already observed in the single-family standards. To address this, we extend the same embedding-based diversity analysis to include both the single-family standard runs and the mixed-agent runs, allowing late-stage condition-level estimates to be compared on a common basis. The comparison set comprises 9 single-family standard runs and 3 mixed-agent runs.

The mixed-agent condition remains within the late-stage range already spanned by the single-family standards. Its late-stage mean was 0.249, compared with 0.405 for Phi-4, 0.346 for GPT-4-mini, and 0.219 for DeepSeek-V3. Thus, the mixed-agent system does not define a new elevated late-stage regime, but instead occupies an intermediate position within the single-family range. Bootstrap comparisons further show that the mixed-agent late-stage estimate is lower than that of Phi-4 ($\Delta$ = -0.156; 95% CI [-0.202, -0.113]) and GPT-4-mini ($\Delta$ = -0.097; 95% CI [-0.165, -0.034]), while not being clearly above DeepSeek-V3 ($\Delta$ = 0.030; 95% CI [-0.026, 0.074]). The conservative interpretation is therefore that model heterogeneity does not clearly raise late-stage semantic diversity beyond the single-family range, consistent with a bottleneck reading rather than with escape from the late-stage semantic neighborhood observed in the single-family standards.